%% file: main.tex
\documentclass[sigconf, nonacm]{acmart}
\usepackage{listings}

\input{common-includes}
\newcommand\vldbdoi{XX.XX/XXX.XX}

\newcommand\vldbvolume{14}
\newcommand\vldbissue{1}

\newcommand\vldbavailabilityurl{http://vldb.org/pvldb/format_vol14.html}

\begin{document}
\title{Making RDBMSs Efficient on Graph Workloads Through Predefined Joins}

\newif\iflong
\longtrue


\author{Guodong Jin}
\email{jinguodong@ruc.edu.cn}
\affiliation{%
  \institution{Renmin University of China}
}

\author{Semih Salihoglu}
\email{semih.salihoglu@uwaterloo.ca}
\affiliation{%
  \institution{University of Waterloo, Canada}
}

\input{abstract}

\maketitle

\begingroup
\renewcommand\thefootnote{}\footnote{\noindent
This work is licensed under the Creative Commons BY-NC-ND 4.0 International License. Visit \url{https://creativecommons.org/licenses/by-nc-nd/4.0/} to view a copy of this license. For any use beyond those covered by this license, obtain permission by emailing \href{mailto:info@vldb.org}{info@vldb.org}. Copyright is held by the owner/author(s). Publication rights licensed to the VLDB Endowment. \\
\raggedright Proceedings of the VLDB Endowment, Vol. \vldbvolume, No. \vldbissue\ %
ISSN 2150-8097. \\
\href{https://doi.org/\vldbdoi}{doi:\vldbdoi} \\
}\addtocounter{footnote}{-1}\endgroup


\input{introduction}
\input{related}
\input{rid-materiallization}
\input{sip-join}
\input{rid-index}
\input{implementation}
\input{evaluation}

\input{conclusion}

\bibliographystyle{ACM-Reference-Format}
\bibliography{references}

\iflong
\input{appendix}
\fi

\end{document}

%% file: common-includes.tex
\usepackage[T1]{fontenc}
\usepackage[utf8]{inputenc}
\usepackage{booktabs}
\usepackage{multirow}
\usepackage{balance}
\usepackage{xcolor}
\usepackage{xspace}
\usepackage{lscape}

\usepackage{verbatim}
\usepackage{graphicx}
\usepackage{balance}  
\usepackage{caption}
\usepackage{subcaption}
\usepackage{setspace}
\usepackage{float}
\usepackage{tikz}
\usepackage{amsmath}
\usepackage{breqn}
\usepackage{algorithm}
\usepackage[noend]{algpseudocode}
\usepackage{listings}
\definecolor{key-color}{rgb}{0.8, 0.47, 0.196}

\usepackage{enumitem}
\usepackage{hyphenat}

\usepackage{doi}
\usepackage{hyperref}
\expandafter\def\expandafter\UrlBreaks\expandafter{\UrlBreaks\do\/\do\-\do\.} 
\definecolor[named]{Purple}{cmyk}{0.55,1,0,0.15}
\definecolor[named]{DarkBlue}{cmyk}{1,0.58,0,0.21}
\hypersetup{bookmarksnumbered,unicode,naturalnames,colorlinks,breaklinks,
	linkcolor=Purple,
	citecolor=Purple,
	urlcolor=DarkBlue,
	filecolor=DarkBlue}

\graphicspath{ {images/} }

\newenvironment{squishedlist}
{
	\begin{list}{$\bullet$}
		{
			\setlength{\itemsep}{0pt}
			\setlength{\parsep}{1pt}
			\setlength{\topsep}{0.5pt}
			\setlength{\partopsep}{0pt}
			\setlength{\leftmargin}{01.0em}
			\setlength{\labelwidth}{1em}
			\setlength{\labelsep}{0.5em}
		}
	}
{
	\end{list}
}

\newenvironment{squishedenumerate}
{
	\begin{enumerate}[label=(\roman*),topsep=0.2em,itemsep=0.2em,leftmargin=1.2em] 
	}
	{
	\end{enumerate}
}

\newcommand{\Follows}{\texttt{Follows}\xspace}
\newcommand{\FollowsNS}{\texttt{Follows}}
\newcommand{\Person}{\texttt{Person}\xspace}
\newcommand{\PersonNS}{\texttt{Person}}
\newcommand{\HashJoin}{\texttt{HashJoin}\xspace}
\newcommand{\HashJoinNS}{\texttt{HashJoin}}
\newcommand{\SJoin}{\texttt{SJoin}\xspace}
\newcommand{\SJoinNS}{\texttt{SJoin}}
\newcommand{\SJoinIdxR}{\texttt{SJoinIdxR}\xspace}
\newcommand{\SJoinIdxRNS}{\texttt{SJoinIdxR}}
\newcommand{\SJoinIdxM}{\texttt{SJoinIdxM}\xspace}
\newcommand{\SJoinIdxMNS}{\texttt{SJoinIdxM}}
\newcommand{\Scan}{\texttt{Scan}\xspace}
\newcommand{\ScanNS}{\texttt{Scan}}
\newcommand{\ScanSJ}{\texttt{ScanSJ}\xspace}
\newcommand{\ScanSJNS}{\texttt{ScanSJ}}
\newcommand{\leftsemijoin}{\mbox{$\mathrel{\raise1pt\hbox{\vrule height5pt
depth0pt width0.6pt\hskip-1.5pt$>$\hskip -2.5pt$<$}}$}}
\newcommand{\rightsemijoin}{\mbox{$\mathrel{\raise1pt\hbox{\hskip-1.5pt$>$\hskip -2.5pt$<$\hskip -1.1pt\vrule height5pt
depth0pt width0.6pt}}$}}
\newtheorem{example}{Example}

\tikzstyle{io} = [trapezium, trapezium left angle=70, trapezium right angle=110, minimum width=3cm, minimum height=0.7cm, inner sep=1pt, text badly centered, text width=1.2cm, draw=black, fill=blue!10]
\tikzstyle{decision} = [diamond, minimum width=1cm, minimum height=1cm, text width=1.5cm, text badly centered, inner sep=1pt, draw=black, fill=green!10]
\tikzstyle{process} = [rectangle, minimum width=3cm, minimum height=1cm, text width=2.5cm, text centered, draw=black, fill=red!10]
\tikzstyle{arrow}=[draw, -latex]

\newtheorem{query}{}

%% file: abstract.tex
\begin{abstract}
\label{sec:abstract}
Joins in native graph database management systems (GDBMSs) 
are predefined to the system as edges, 
which are indexed in adjacency list indices and serve as pointers.
This contrasts with and can be more performant than value-based joins in RDBMSs and 
has lead researchers to investigate ways to integrate
predefined joins directly into RDBMSs.
Existing approaches adopt a strict
separation of graph and relational data and processors, where a graph-specific processor
uses left-deep and index nested loop joins for a subset of joins. 
This may be suboptimal, and may lead to non-sequential scans of data in some queries.
We propose a purely relational approach to integrate predefined joins in 
columnar RDBMSs that uses row IDs (RIDs) of tuples
as pointers. Users can predefine equality joins between any two tables,
which leads to
materializing RIDs in extended tables and optionally in RID indices. 
Instead of using the RID index to perform the join directly, we use it primarily in hash joins
to generate semi-join filters that can be passed to scans using sideways information passing, 
ensuring sequential scans. In some settings, we also use RID indices to reduce the number 
of joins in query plans.
Our approach does not introduce any graph-specific system components, can execute predefined joins on any join plan, and can improve
performance on any workload that contains equality joins that can be predefined. 
We integrated our approach to DuckDB and call the resulting system {\em GRainDB}.
We demonstrate that GRainDB far improves the performance of DuckDB on relational and 
graph workloads with large many-to-many joins, making it competitive
with a state-of-the-art GDBMS, and incurs no major overheads otherwise.
\end{abstract}

%% file: introduction.tex
\vspace{-30pt}
\section{Introduction}
\label{sec:introduction}

Perhaps the two most commonly used data structures to model data in enterprise database applications are tables, 
which are the core structures of relational database management systems (RDBMSs), and graphs, which
are the core structures of several classes of systems, most recently of property graph database
management systems (GDBMSs for short), such as Neo4j~\cite{neo4j}, TigerGraph~\cite{tigergraph}, 
DGraph~\cite{dgraph},  and GraphflowDB~\cite{kankanamge:graphflow, mhedhbi:sqs, mhedhbi:aplus, mhedhbi:optimizing, gupta:columnar}. Aside from
developer preference for using a graph-specific data model and query language,
GDBMSs target what are colloquially referred to as {\em graph workloads}, which refer to workloads
that contain large many-to-many joins. For example, these workloads appear in social networking applications
for finding long paths between two people over many-to-many friendship relationships or in financial fraud detection applications for finding fraudulent patterns across 
many-to-many money transfers across bank accounts.

At the same time, several economic and technical factors have lead researchers  
to investigate techniques to support efficient graph querying natively inside RDBMSs.
For example, it is recognized that the data stored in many specialized GDBMSs are extracted from 
RDBMSs~\cite{sahu:extended-survey, graphgen, tian:db2graph, anzum:graphwrangler}. In many enterprises, 
users replicate parts of the tabular data stored in RDBMSs to a 
GDBMS because their applications require  the fast join capabilities of GDBMSs. 
In addition, many applications require other processing on their graph workloads beyond evaluating large 
many-to-many joins, such as running predicates on node and edge properties or grouping and aggregations, 
for which RDBMSs already employ efficient techniques. 
Therefore leveraging mature RDBMS technology to support graph workloads 
natively is highly appealing to both users and vendors: users can avoid the challenges of duplicating data and keeping multiple systems in sync, while vendors can avoid the efforts to develop a new system from scratch.
We revisit this goal and research challenge in the context of columnar RDBMSs, which are similar to GDBMSs in that they also target read-heavy analytical workloads. Our specific goal is to 
extend a columnar RDBMS natively with the fast join capabilities of GDBMSs. 

Several prior approaches leverage RDBMS technology to evaluate graph workloads. 
One approach simply exposes a separate graph querying layer to users and implements 
a translation component that outputs SQL versions of queries, with no or minimal
modifications to the query processor of the RDBMS.
This approach is not focused on performance and is 
commonly employed in commercial products, such as IBM DB2 Graph~\cite{tian:db2graph}, 
SQLGraph~\cite{sun:sqlgraph}, SAP Hana's graph database extension~\cite{rudolf:hanagraph}.

A second approach introduces a new graph-specific query processor that co-exists with the existing processor of the RDBMS.
This has been most recently adopted by the GR-Fusion system~\cite{hassan:grfusion, hassan:grfusion-edbt}. 
Specifically, SQL is extended to contain graph-specific constructs, using which users create graphs.
The topologies of these views, i.e., the vertices and edges without properties, are stored  
in native adjacency list indexes, which are used during query processing for graph traversals/many-to-many joins, using new graph-specific operators, such as \texttt{EdgeScan} and \texttt{PathScan}. 
Parts of queries that refer to graph-specific constructs 
compile to these specialized graph operators, while
the non-graph parts of queries compile to existing operators of the RDBMS. 
GQ-Fast~\cite{lin:gq-fast} is another system that develops a separate query processor and 
storage sub-system specialized for graphs.
GQ-Fast is not integrated into an RDBMS but the authors' envisioned 
integration~\cite{lin:gq-fast} is similar to GR-Fusion's dual processor approach.  
Aside from being heavy-weight integration approaches that 
develop separate graph-specific components within an RDBMS, 
the strict separation of graph and non-graph data and operators can lead 
to inefficient accesses when a query accesses graph data and properties
or fail to apply efficient optimizations to the entire query.
For example, graph traversals in these systems, which are implemented in specialized traversal operators, are 
effectively left deep join plans that use index nested loop joins, so avoid
using efficient bushy plans. 
For many queries, these approaches can be amenable to 
significant performance improvements.

To motivate the key approaches of our solution, we begin by analyzing the primary differences 
between the join evaluation techniques in RDBMSs and GDBMSs. This question has been discussed 
since the birth of data management between proponents of DBMSs that adopted graph-based models, adopting 
Charles Bachmann's IDS system~\cite{bachman:ids}, and those that adopted Ted Codd's relational model~\cite{codd:relational-model}, such as System R~\cite{astrahan:systemr}. Perhaps 
Codd himself has best articulated the primary differences in his Turing  Award lecture~\cite{codd:lecture}. 
As a primary difference, Codd notes,  
joins in GDBMSs happen along {\em predefined} access paths, i.e., between existing records 
through predeclared {\em pointers} (or links). 
In contrast, joins in RDBMSs are {\em value-based}, so arbitrary tables can be joined on arbitrary columns as long as those 
columns have the same data types. Although much has changed since IDS and System R, this 
characterization is still accurate for contemporary GDBMSs and RDBMSs. 
Contemporary GDBMSs are indeed optimized to perform joins between node records along predefined edges and
use two common techniques to perform these joins efficiently: (i) dense integer ID-based joins, which serve as pointers to directly
look up records; and (ii) an adjacency list index that is used to quickly find joining edge and node records with a given node record during join evaluation.

Motivated by these observations, our approach integrates predefined joins into a columnar RDBMS by extending 
two components of the system:
(i) the physical storage and query processor; and (ii) the indexing sub-system, where
each integration progressively yields more performance benefits. Users perform two actions, the second of which is 
optional, to benefit from predefined joins: (i) predefinition of a primary-foreign key join to the system; and (ii) an index creation on these tables:

\vspace{2pt}
\noindent  {\bf $\bullet$ Physical Storage and Query Processor:} When a user predefines a primary-foreign key join from table $F$ to 
table $P$, where a column of $F$ has a foreign key to a column of $P$, this performs
an ALTER TABLE command that inserts an additional $RID_{p}$ column to $F$ that contains for each row $r_f$ 
in $F$ the row ID (RID) of row $r_p$ in $P$ that $r_f$ points to. RIDs are dense integer-based system-level IDs
in columnar RDBMS that are used to identify the physical locations of the column values of each row. They
are therefore system-level pointers, similar to node IDs in GDBMSs. 

In order to use  these 
pointers to perform the primary-foreign key joins more efficiently, we rewrite queries to replace
primary-foreign key equalities with RID equalities. 
Equality predicates in many columnar RDBMSs are primarily evaluated with hash-joins. 
To exploit the pointer-nature of predefined joins, we employ {\em sideways information passing} (sip) 
to speed up scans and indirectly other joins in query plans.
Specifically, the hash join operator keeps 
the RIDs from the build side in compact bitmaps and passes them to the relevant scans 
on the probe side to perform semi-joins. 
Because joins in RDBMSs are value-based, existing applications of sip pass 
information in probabilistic filters, often a bloom filter~\cite{neumann:scalable, patel:quickstep, kandula:dip}. 
This requires running hash functions both when creating the filter in the joins as well as
performing the semi-joins in scans. Since RIDs are dense integer-based IDs,
we directly pass a compact bitmap filter and avoid any hash computations. 

\vspace{2pt}
\noindent  {\bf $\bullet$ Indexing Sub-system:} 
A common way to represent many-to-many relationships between two sets of entities in relational databa\-ses
is to  have a table $F$ that contains two foreign keys on two other (not necessarily different) tables $P_1$ and $P_2$. 
For simplicity of terminology, we refer to such $F$ as a {\em relationship} table and $P_i$ as 
{\em entity} tables.
If the joins with both entity tables have been predefined to the system, users can additionally 
build an index on table $F$ 
on the two extended RID columns $RID_{p1}$ and $RID_{p2}$. This index 
is stored in adjacency list
format and serves two purposes. First, it is used to generate further information to pass when a query joins 
$P_1$, $F$, and $P_2$ and when a hash join operator builds a table of $P_1$ or $P_2$. 
Second, when a query refers to $F$ only to facilitate the join of tuples in $P_1$ and $P_2$, so contains 
no predicates on $F$ and projects out $F$'s columns, 
the index allows us to reduce the number of joins in query plans. 


We integrated our techniques into DuckDB~\cite{raasveldt:duckdb, raasveldt:duckdb-cidr}, a new columnar RDBMS that is 
actively being developed at Centrum Wiskunde \& Informatica~\cite{cwi}, and
call the extended system {\em GRainDB}. Unlike systems such as SQLGraph and IBM DB2 Graph, we  
modify the internals of the RDBMS to improve the performance on many-to-many joins. 
Unlike GR-Fusion and the envisioned GQ-Fast integration, our approach is
purely relational and does not require a separate graph-specific 
query processing codeline. As a result: (i) our approach directly leverages 
DuckDB's core components: the optimizer to generate efficient 
plans for the entire query, vector-based query processor, and bushy join plans;
and (ii) any database in the RDBMS can predefine a set of joins and build a RID index to 
improve performance.
We demonstrate that GRainDB improves the median query execution time 
of DuckDB by 3.6x on the relational JOB benchmark which contains many-to-many joins,
and by 22.5x on the LDBC SNB graph benchmark, making a columnar RDBMS competitive with 
the state-of-the-art GraphflowDB GDBMS~\cite{gupta:columnar}.
Because our approach is purely relational, 
GRainDB improves DuckDB even on some traditional 
relational analytics queries from TPC-H. In our detailed analysis, we show: (i) 
that our possibly bushy and sip- and hash-join-based plans can be more
efficient than left-deep 
index nested loop join plans on many queries, such
as those with selective predicates on tables that 
represent edges/relationships; and (ii) our use of sip makes the optimizer
of a system more robust because its semi-join computations can mitigate 
a poor join order selection of the optimizer.
Our code, queries, and data are avaliable here~\cite{graindb}.

%% file: related.tex
\vspace{-10pt}
\section{Related Work}
\label{sec:rw}
There are many native GDBMSs~\cite{neo4j, mhedhbi:sqs, tigergraph, dgraph, abulbasher:factorization, janusgraph} that employ
many read-optimized techniques,
such as specialized indices~\cite{mhedhbi:aplus}, factorization~\cite{abulbasher:factorization},
or worst-case optimal join algorithms~\cite{mhedhbi:optimizing, freitag:wco},
to be very efficient on analytical queries that contain large joins over many-to-many relationships
between entities.
However, two of the core techniques that appear in every GDBMS we are aware of are native 
graph storage in adjacency list indices and predefined pointer-based joins, where node IDs serve as pointers, i..e, positional
offsets, into these indices. Integration of these two core techniques into RDBMSs is the focus of this paper.
Below, we review prior work that leverage 
RDBMSs for supporting graph applications and the literature on sip and join indexes. 

GR-Fusion~\cite{hassan:grfusion, hassan:grfusion-edbt} is designed to perform graph querying natively inside an RDMBS. 
Users define graphs as views over tables, 
The topology of graph views are stored natively in an adjacency list index.
In contrast, the node and edge properties are stored as pointers to the underlying tables.
Users refer to the paths in a graph view as if they are a separate table 
using a new \texttt{Paths} construct in the FROM clause of SQL. 
Then,
part of the query that enumerates paths and their constraints are evaluated with special operators,
such as  VertexScan or PathScan, whose results are tuples that can be input to further relational operators.
Therefore this approach creates dual query processing pipelines inside 
the system. One advantage of this approach
is that 
the original relational operators remain unchanged because outputs from the graph pipeline
are regular tuples. 
However, GR-Fusion also has several shortcomings. First, PathScan enumerates only paths,
so some other patterns, such as stars, need to be evaluated by the vanilla relational query processor, 
so do not benefit from the native graph storage or fast join algorithms unless users manually decompose
these queries into paths.
Instead, our approach is purely relational and can improve equality joins on arbitrary queries,
including queries from traditional benchmarks. 
Second, paths are enumerated through DFS or BFS 
algorithms, which are akin to left-deep plans that use index nested loop join operators.
These plans can be suboptimal compared to bushy join plans,
which can be generated in GRainDB. This is further exacerbated if  
vertex and edge properties need to be scanned during DFS of BFS by following
pointers to the tables, which can lead to many random accesses. In contrast,
GRainDB uses adjacency list indices
to generate information to pass to scan operators (for semijoins) and in some cases
to reduce the number of join operators (See Section~\ref{subsec:join-merging}) but performs 
scans always sequentially. We intended to compare
our solution against GR-Fusion but the publicly available code has several errors 
and is not maintained.

GQ-Fast~\cite{lin:gq-fast} supports a restricted subset of SQL called ``relationship
queries'' which contain joins of tables that are similar to path queries, followed with aggregations. 
Similar to GRainDB and GR-Fusion, GQ-Fast stores relationship tables 
in CSR-like  indices. Unlike GRainDB and GR-Fusion, these indices also contain properties, i.e., non  ID columns,
of relationship tables and employ heavy-weight compression schemes. In addition, the system has
 a fully pipelined query processor that uses query compilation, which gives it performance advantages. 
However, similar to GR-Fusion, the joins are limited to paths and evaluated with 
left-deep index nested loop join operators that are equivalent to DFS traversals. In addition, unlike GRainDB  and 
GR-Fusion, GQ-Fast is implemented as a standalone system from scratch and does not integrate these
techniques into an underlying RDBMS to support more general queries, which is left as future work~\cite{lin:gq-fast}. However, even this envisioned integration is similar to GR-Fusion, where the 
GQ-Fast layer is a separate query processor whose outputs are given to the
query processor of the RDBMS. We intended to but could not compare against GQ-Fast because 
the system supports a very limited set of queries (e.g., none of the LDBC queries are supported) and 
the publicly available version is no longer maintained and has errors.

Another popular approach is to develop a translation layer between a graph data model and query language
to the relational model and SQL and leverage the underlying RDBMS without any modifications. Systems
such as  IBM DB2 Graph~\cite{tian:db2graph}, SQLGraph~\cite{sun:sqlgraph}, and SAP Hana's graph database extension~\cite{rudolf:hanagraph}
primarily provide a translation layer between the property graph data model and a query language or API, such 
as Gremlin~\cite{rodriguez:gremlin}, and convert the modeled graph into relational tables and queries into SQL.
This is a very attractive approach for commercial vendors because it is lightweight and it
requires almost no changes to the underlying RDBMS. 
Work in this space focus on optimizing the translation layer to minimize joins or how to utilize existing indexes of RDBMSs to speed
up processing. This approach is not performance focused and is limited by the 
underlying RDBMS's baseline performance. In contrast, our approach modifies the underlying RDBMS to improve its performance on 
some joins. Similar approaches have also been taken by several systems, such as Grail~\cite{fan:grail} and graph layers above the Vertica column store~\cite{jindal:vertica-graph} or the Aster system~\cite{simmen:aster-6}, that translate batch iterative graph computations, such as computing PageRank or finding connected components, into recursive SQL procedures.

SIP is a technique that is used in RDBMSs to avoid scanning large tables or 
indices or data from remote compute nodes~\cite{bancilhon:magic-sets, graefe:query, ives:sip, mumick:magic-sets, neumann:scalable, zhu:lip}. 
The use of sip closest to our work 
has been proposed by Neumann et al.~\cite{neumann:scalable} inside the 
RDF-3X system that manages RDF databases. 
This work has proposed using sip to avoid scans of large fractions of indices that store
 RDF triples. This work specifically targets 
queries with large joins but small outputs that contain sub-queries with non-selective filters. 
Evaluation of these  sub-queries in regular execution requires large index scans, but by passing information from
other sub-queries, the system can avoid scanning parts of the index. 
Zhu et al.~\cite{zhu:lip} have used sip in a similar fashion to avoid large table scans in 
in-memory star schema data warehouses when using left-deep query plans in queries. This paper 
has demonstrated that 
the difference between the best and worst performing left-deep plans shrink significantly when using
sip in contrast to without using sip, which makes the optimizer more robust. 
Our use of sip to integrate predefined
joins is similar to the use of sip in these works with several differences. In these systems and in RDBMSs in general,
joins are value-based so passing the values, which may be of arbitrary data types, 
requires compacting the keys 
in probabilistic data structures, specifically bloom filters.
This requires running hash functions both when creating the filter as well as
performing the semi-joins in scans. Since our pointer-based predefined joins are over 
dense integer-based ID, we can directly compact the keys in a deterministic bitmap 
filter and avoid any hash computations.
Similar to reference~\cite{zhu:lip}, we also demonstrate that using sip in graph or relational workloads with
large many-to-many joins makes the system more robust by analyzing GRainDB's plan space.

The analogue of adjacency list indices in our solution are the {\em RID indices} (Section~\ref{sec:rid-index}) that we use
to index tables that are part of predefined joins. Our RID indices can be seen as a form of 
join index. Valduriez originally introduced join indices~\cite{valduriez:join-index}
to index results of arbitrary join queries, e.g., consisting of equality or inequality predicates, 
and index for each RID of one table, the list of matching RIDs from one or more other tables.
Join indices are therefore simple materialized views. Similar to join indices, our RID indices 
store RID keys and list of RID values but are over base tables instead of results of join queries. 
Prior work on join indices have focused on their efficient use within RDBMSs, such as extending them to support
multi-table joins~\cite{oneil:join-index}, designing fast disk-based 
join algorithms~\cite{lei:stripe-join, li:jive-slam-join} or their application to top-k query processing~\cite{tsaparas:join-index-top-k}. In our context, we use RID indices primarily to generate information
to pass to scan operators and when a scan can be completely avoided, to reduce 
the number of joins in the query plan.

%% file: rid-materiallization.tex
\vspace{-5pt}
\section{RID Materialization}
\label{sec:rids}
We start by describing the changes at the physical data storage layer of the system. Users predefine their joins using a \texttt{PREDEFINE} \texttt{JOIN} 
command that we added to the SQL dialect in DuckDB. In this command users specify an equality join 
from a table $F(A_{f1}, ... A_{fk_f})$ to $P(A_{p1}, ..., A_{pk_p})$ on attributes 
$A_{ft_1}=A_{pz_1}$, ..., $A_{ft_{\ell}}=A_{pz_{\ell}}$, such that   $A_{ft_1}, ..., A_{ft_{\ell}}$ 
forms a foreign key to $P$.  
Upon executing this command, the system 
adds a new column \texttt{RID($A_{ft_1}, ..., A_{ft_{\ell}}$)} to $F$ that contains 
for each row $r_f \in F$, the RID of the row $r_p \in P$ to which $r_f$ has the foreign key.
This column is visible only to the system and not to users. RIDs in columnar RDBMSs serve as 
system-level pointers and can be directly used to compute the locations of rows in storage. So the \texttt{RID($A_{ft_1}, ..., A_{ft_{\ell}}$)}  column stores for each $r_f$ the pointer to the 
matching $r_p$, similar to how edges in GDBMSs point to their source or destination node records.
If $F$ contains foreign keys to multiple tables, multiple joins on $F$ can be predefined.
This is common for relationship tables that represents many-to-many joins, such as the 
\Follows table in the next example.

\begin{table}[!t]
    \begin{subtable}[b]{.5\linewidth}
      \centering
	\begin{tabular}{|c|c|}
	\multicolumn{2}{c}{{\bf Person}} \\
	\hline
	{\bf ID} & \bf{name}\\
	\hline
	101 & Mahinda \\
	202 & Karim  \\
	303 & Carmen \\ 
	404 & Zhang \\
	\hline
	\end{tabular}
     \caption{\Person table.}
    \end{subtable}%
    \begin{subtable}[b]{.5\linewidth}
      \centering
	\begin{tabular}{|c|c|c|}
	\multicolumn{3}{c}{{\bf Follows}} \\
	\hline
	\bf{ID1} & \bf{ID2} & \bf{year}\\
	\hline
	101 & 202 & 2021\\
	303 & 404 & 2019 \\
	101 & 303 & 2021 \\
	202 & 303 & 2020 \\
	101 & 404 & 2021 \\
	\hline
	\end{tabular}
        \caption{\Follows table.}
    \end{subtable} 
\captionsetup{justification=centering}
\vspace{-20pt}
\caption{Input tables for our running example.}
\vspace{-25pt}
\label{tab:tables}
\end{table}

\begin{example}
Table~\ref{tab:tables} shows a simple database with two tables, a \PersonNS(\texttt{ID}, \texttt{name}) table and a \FollowsNS(\texttt{ID1}, \texttt{ID2}, \texttt{year}) table, that will serve as our running example.  The \texttt{ID1}
and \texttt{ID2} columns in \Follows are both foreign keys to the \texttt{ID} column of \Person. 
Table~\ref{tab:extended-table} shows the extended \texttt{Follows} table (as \texttt{Follows'}) when a user predefines the 
\texttt{Person.ID} = \texttt{Follows.ID1} and \texttt{Person.ID} = \texttt{Follows.ID2} joins. 
The \texttt{Follows} table is extended with
\texttt{RID(ID1)} and \texttt{RID(ID2)} columns (abbreviated as \texttt{RID1} and \texttt{RID2}) that contain 
the RIDs of the rows in Person that match the values in the \texttt{ID1} and \texttt{ID2} columns, respectively. 
Both \Person and \Follows tables also have RID columns (abbreviated as \texttt{R}) that show 
the contiguous RIDs of the rows in these tables. These are shown in
gray to indicate that unlike \texttt{RID(ID1)}  and \texttt{RID(ID2)} columns, 
they are not materialized in storage.
\end{example}

\begin{table}[!t]
  \begin{subtable}[b]{.4\linewidth}
  \centering
  \begin{tabular}{|c|c|c|}
	\multicolumn{3}{c}{{\bf Person}} \\
	\hline
	{\color{gray} {\bf R}} & {\bf ID} & \bf{name}\\
	\hline
	{\color{gray} 0} & 101 & Mahinda \\
	{\color{gray} 1} & 202 & Karim  \\
	{\color{gray} 2} & 303 & Carmen \\ 
	{\color{gray} 3} & 404 & Zhang \\
	\hline
	\end{tabular}
  \caption{Extended \Person table.}
	\label{tab:extended-person-table}
  \end{subtable}%
  \begin{subtable}[b]{.5\linewidth}
      \centering
	\begin{tabular}{|c|c|c|c|c|c|}
	\multicolumn{5}{c}{{\bf Follows$'$}} \\
	\hline
	  {\color{gray} {\bf R}} & \bf{R1} & \bf{ID1} & \bf{R2} & \bf{ID2}  & \bf{year}\\
	\hline
	{\color{gray} 0} &\bf{0} & 101 & \bf{1} & 202 & 2021 \\
	{\color{gray} 1} &\bf{2} & 303 & \bf{3} & 404 & 2019 \\
	{\color{gray} 2} &\bf{0} & 101 & \bf{2} & 303 & 2021 \\
	{\color{gray} 3} &\bf{1} & 202 & \bf{2} & 303 & 2020 \\
	{\color{gray} 4} &\bf{0} & 101 & \bf{3} & 404 & 2021 \\
	\hline
	\end{tabular}
        \caption{Extended \Follows table.}
    \end{subtable} 
    \captionsetup{justification=centering}
    \vspace{-7pt}
    \caption{Extended tables. RID columns are abbreviated as \texttt{R} and are in 
gray to indicate that they are not materialized. The \texttt{RID(ID$_i$)} columns of \texttt{Follows'} 
are abbreviated as \texttt{R$i$}.}
\vspace{-25pt}
\label{tab:extended-table}
\end{table}

%% file: sip-join.tex
\vspace{-10pt}
\section{SJoin: SIP of RIDs}
\label{sec:sip-join}
Our implementation of predefined joins consists of two steps:

\noindent{\bf Step 1:  Rule-based query optimization.}
We use the system's default optimizer to generate a regular logical plan for the query. We
recursively traverse this plan and find each join operator that evaluates a predefined join 
from $F$ to $P$, e.g., the \PersonNS.\texttt{ID}=\FollowsNS.\texttt{ID1}.
In our implementation, these are \HashJoin operators because \linebreak DuckDB evaluates 
equality joins with \HashJoinNS. Upon finding these 
\HashJoinNS s, we perform one of two sets of actions:

\noindent {\em Case 1: $F$ is the build and $P$ is the probe side.} In this case we make the following 
changes to the operators in the plan tree: 
\begin{squishedlist}

\item \HashJoin is replaced with a new join operator we call \SJoin (explained momentarily in Step 2).

\item \ScanNS($F$) operator (on the build side sub-tree) is modified to (i) scan the materialized RID column of $F$; 
and (ii) if any of the original join attribute $F.A_{iz}$ is projected out later in the query, we remove the scan of $A_{iz}$ from the scanned columns of $F$.

\item \ScanNS($P$) operator (on the probe side) is replaced with a modified scan operator, which we refer to
as \ScanSJ, for {\bf scan} {\bf s}emi{\bf j}oin.

\end{squishedlist}
As we discuss momentarily below in Step 2, if we are in Case 1, we will perform sip to pass information
from $F$ to $P$ during evaluation.

\noindent {\em Case 2: $F$ is the probe and $P$ is the build side.} Now the changes are: 

\begin{squishedlist}

\item \HashJoin is now not replaced but we replace the join condition to be $P$.RID=$F.RID$($A_{it_1}, ..., A_{it_{\ell}}$). Note that because RIDs are integers and always form a single join attribute, this is more performant
if the original join predicate contains multiple columns or non-integer data types, e.g., strings.

\item \ScanNS($F$) operator (on the probe side sub-tree) is modified in exactly the same way as before.

\end{squishedlist}
If the plan is in Case 2, we cannot perform sip because we can pass information from $F$ to $P$ only if $F$ is on the build side. This is because we need to read the information to pass from $F$ to $P$ before $P$ is scanned. 
Alternatively, we can swap
the build and probe sides, but we chose not to overwrite the optimizer's choice here. This is because
as we next explain \SJoin is a modified hash join operator and the optimizer optimizes 
to put the smaller table on the build side to keep the constructed hash table small.

\noindent {\bf Step 2: Sideways information passing during query evaluation:}
If we are in Case 1, we then use sip during query evaluation from \SJoin operators to
the \ScanSJNS($P$) operators. \SJoin is a specialized hash join operator. \SJoin performs the join
on the replaced RID equality predicates instead of the original join columns in the query. 
In addition, \SJoin passes the 
materialized RID values from scanned $F$ tuples, which are pointers to $P$, to \ScanSJNS($P$) operators 
in its probe side. 
Similar to standard hash join, \SJoin first reads all of the tuples from its build side. These tuples contain
materialized RID values that are scanned from $F$ and point to the tuples in $P$. Using these RIDs,
\SJoin constructs two bitmask filters:

\begin{squishedlist}
\item {\em Zone bitmask:} For each zone of $P$, i.e., a block of tuples on disk, 
indicates whether the zone has any matching tuples joining with $F$. This bitmask contains 
1 bit for each zone and is constructed 
by taking the modulo of the RIDs with the zone size.

\item {\em Row bitmask:} Indicates whether each row $r_p$ of $P$ joins with an $F$ tuple.
This bitmask contains $|P|$ many bits and is constructed by directly setting the positions of the seen RIDs to 1.
\end{squishedlist}
Note that unlike existing applications of sip in DBMSs~\cite{graefe:query, ives:sip, neumann:scalable, kandula:dip}, 
the information we pass to scans do not need to be probabilistic filters, such as bloom filters.
This is because RID values are dense integers and their exact domain, which is 0 to the number of tuples in $P$,
is known by the system and can be compressed into a single bit. If $P$ is very large, 
the row bitmask can be large, in which case a system can resort to even smaller filters 
at a granularity level between zones and individual tuples.  
Once \SJoin receives all of the build side tuples, it passes both of these bitmasks 
to all of the \ScanSJNS($P$) operators in its probe side recursively. These filters are used to perform
the semijoin $P \ltimes F$ in the \ScanSJ operators as follows. Zone bitmask is 
used to skip over scanning zones of $P$ whose bits are 0. 
For zones with matching tuples, 
\ScanSJ operator scans the zone into vectors as regular scan operator and 
adds a new RID vector to the intermediate tuples 
that store the RIDs of the scanned tuples. This does not require any actual I/O
because RIDs of $P$ are virtual positional offsets of the tuples, which 
can directly be written into
 the RID vector. For example, if zones are of size 1024 and the second zone has been read, then this vector
contains values 1024 to 2047. Finally, to perform the semi-join, \ScanSJ attaches the row bitmask of this zone
as a selector vector to the intermediate tuples. This filters out the $P$ tuples without
matching $F$ tuples. 

\begin{figure*}[t!]
\vspace{-4pt}	
\centering
\captionsetup{justification=centering}
\begin{subfigure}[b]{0.35\linewidth}
\centering
  \includegraphics[width=0.95\linewidth,height=0.85\linewidth]{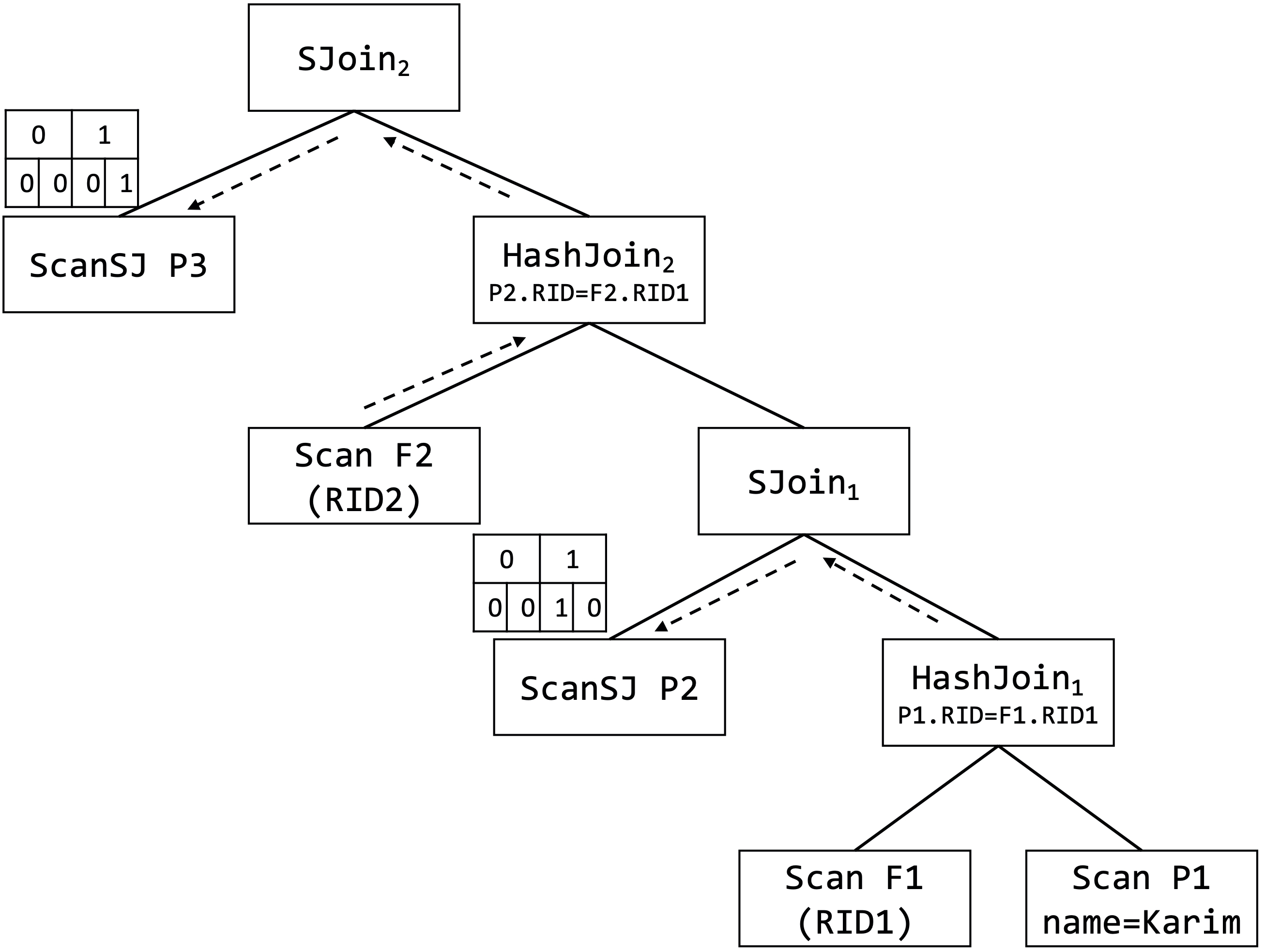}
\vspace{-5pt}
\caption{Plan without a RID index.}
\vspace{-10pt}
\label{fig:sj-no-index}
\end{subfigure}
\begin{subfigure}[b]{0.35\linewidth}
\centering
  \includegraphics[width=0.95\linewidth,height=0.85\linewidth]{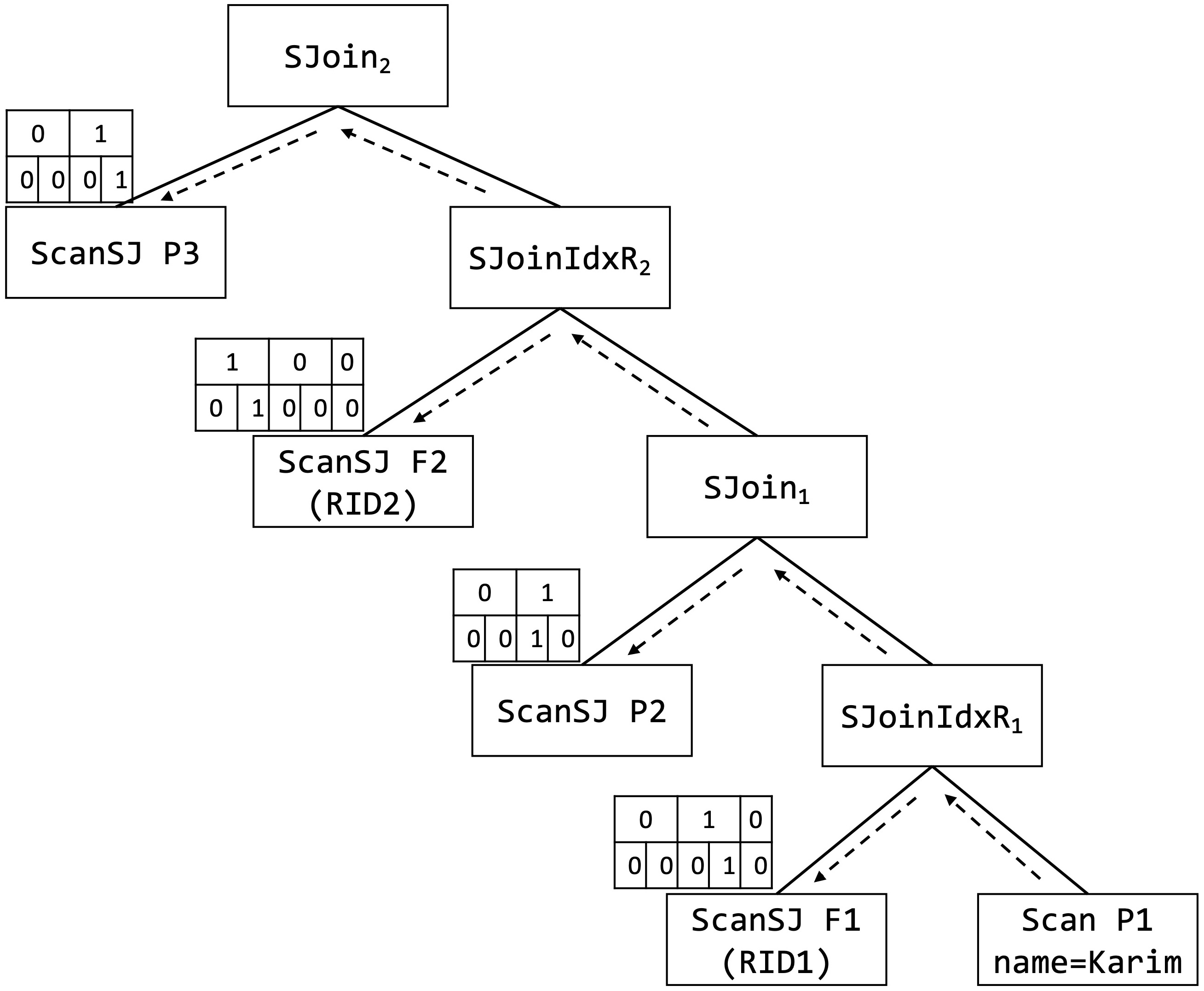}
\vspace{-5pt}
\caption{Plan with a RID index when columns of \Follows tables are in the final projection.}
\vspace{-10pt}
\label{fig:sjidxr}
\end{subfigure}
\begin{subfigure}[b]{0.29\linewidth}
\centering
  \includegraphics[width=0.95\linewidth]{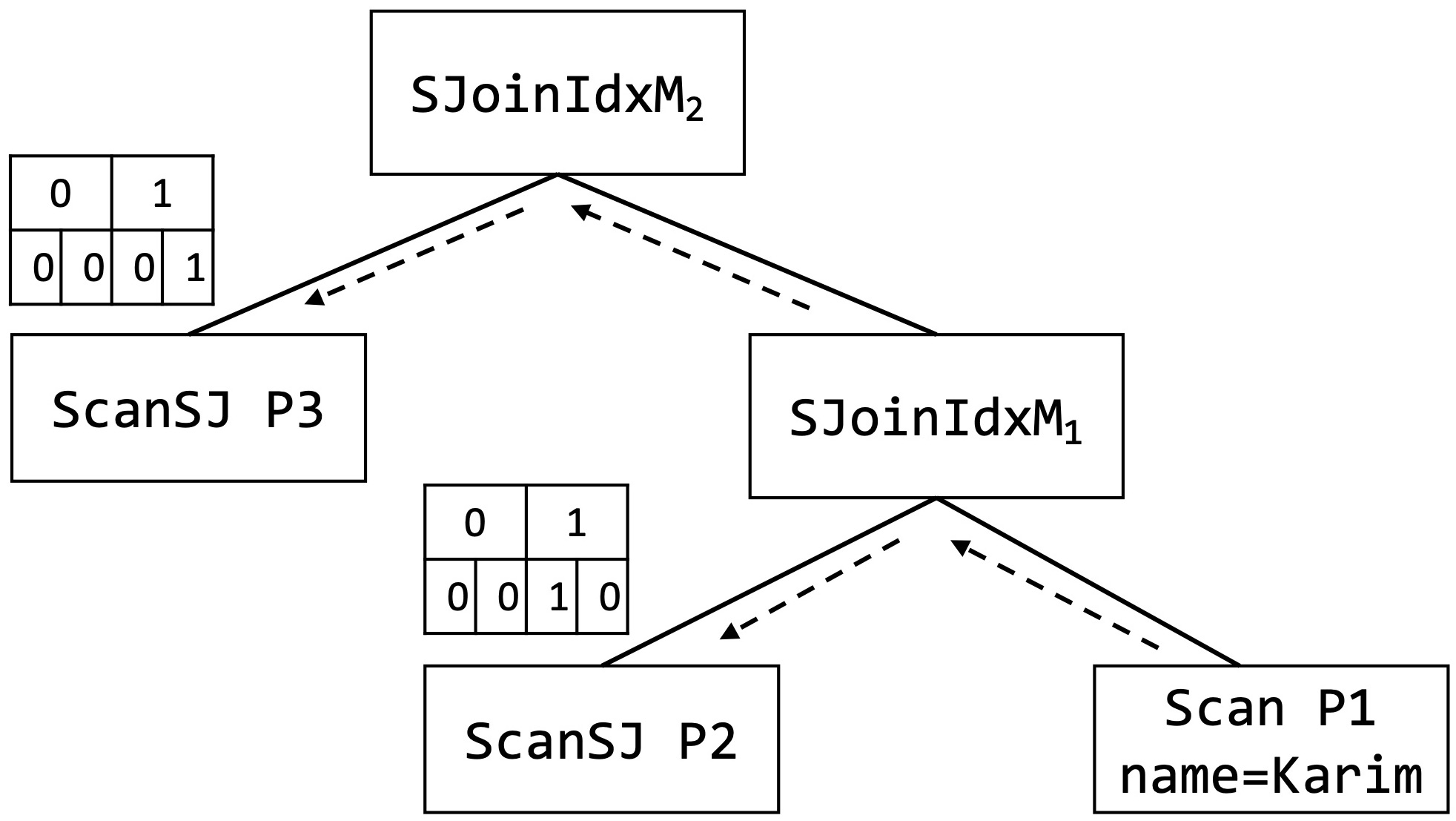}
\vspace{-5pt}
\caption{Plan with a RID index when columns of \Follows tables are projected out.}
\vspace{-10pt}
\label{fig:sjidxm}
\end{subfigure}
\caption{Example plans for our running example queries with different system configurations.}
\label{fig:sj-plans}
\end{figure*}

\vspace{-6pt}
\begin{example}
Consider a query that finds two hop friends of Karim: 
\begin{lstlisting}[numbers=none, mathescape=true, showstringspaces=false]
SELECT * 
FROM Pers. P1,Follows F1,Pers. P2,Follows F2,Pers. P3
WHERE P1.ID=F1.RID1 AND F1.RID2=P2.ID AND P2.ID=F2.RID1 
     AND F2.RID2=P3.ID AND P1.name = Karim
\end{lstlisting}
\vspace{-4pt}

Figure~\ref{fig:sj-no-index} shows an example plan for this query that has: (i) replaced two \texttt{HashJoins} with
\SJoin operators;  (ii) replaced two \Scan \Person table operators (for \texttt{P2} and \texttt{P3}) with \ScanSJ;
and (iii) modified the \Scan \Follows operators to read the materialized 
RID columns. 
\texttt{HashJoin$_1$} and \texttt{HashJoin$_2$} operators are not replaced with \SJoin because the \texttt{Scan}s of F1 and F2 are on their probe sides. 
Instead, we only modify their join predicates to be over RIDs.
The information passed from \SJoin operators are in the form of two bitmasks, which can be seen at the \texttt{ScanSJ P2} 
and \texttt{ScanSJ P3} operators. The top one is the zone bitmask and the bottom one the row bitmask.
The figure assumes zones of size 2. In our running example, \texttt{HashJoin$_1$}  joins the (1, 202, Karim) 
and (1, 202, P2.RID=2, 303, 2020) tuples, which is given to \texttt{SJoin$_1$}.  
Because the only matching \texttt{P2} in this tuple has RID 2, 
the row bitmask passed to \ScanSJ \texttt{P2} is [0, 0, 1, 0] and the zone bitmask is [0, 1]
because RID 2 is in the second zone. Therefore \ScanSJ \texttt{P2} only scans the second
zone and puts the [1, 0] selector vector to the two tuples in this zone (filtering out the tuple with RID 3). 
The output of \texttt{SJoin$_1$} is
(1, 202, Karim, 2, 303, Carmen, 2020) and the following \texttt{HashJoin$_2$} produces
(1, 202, Karim, 2, 303, Carmen, 2020, P3.RID=3, 404, 2019). This is given to \texttt{SJoin$_2$} (during build), which passes the [0, 1] zone bitmask
and the [0,0,0,1]  row bitmask to \ScanSJ \texttt{P3}. The final output is 
(1, 202, Karim, 2, 303, Carmen, 2020, 3, 404, Zhang, 2019).
\end{example}



%% file: rid-index.tex
\vspace{-10pt}
\section{RID Index and Its Applications}
\label{sec:rid-index}
Next, we describe  two applications of indexing the RID values in table $F$ that contains materialized RIDs.
We call this index the {\em RID index}. Section~\ref{subsec:reverse-sj} describes our RID index and 
our first application, which is performing reverse semijoin of $F \ltimes P$ through sip. 
Section~\ref{subsec:join-merging} considers the case when $F$ is a relationship table,
so contains two predefined joins and describes an optimization that merges two consecutive joins 
in a plan to avoid the scan of $F$ completely.

\subsection{Reverse Semijoins}
\label{subsec:reverse-sj}
In our approach of evaluating predefined joins so far,  we can pass RID values only from $F$ to scans of $P$ 
and not vice versa, so we can only perform $P \ltimes F$ through sip. 
In many settings, $F$ is a much larger table than
$P$, and the ability to perform $F \ltimes P$ is very beneficial.
For example, in LDBC benchmark with scale 30, \texttt{Knows} table is 41x larger than \texttt{Person}. 
However, given a row $r_p \in P$, we cannot directly
find from the RID value of $r_p$ the RIDs of rows $r_{f1}, ..., r_{fp} \in F$ that join with $r_p$, 
as this list is not materialized in 
$P$. In order to perform this reverse semijoin, we need an index on $F$ 
that for each $r_p$ returns this list. 
We call this index the {\em RID index}. In our implementation, 
users can construct RID index on any table $F$ on which at least one join has been predefined (say to a table $P$).
Therefore, $F$ already has a materialized $\texttt{RID($A_{it_1}, ..., A_{it_{\ell}}$)}$ column and its own virtual RID
column. The RID index stores for each value in the $\texttt{RID($A_{it_1}, ..., A_{it_{\ell}}$)}$ 
column the RIDs of $r_{f1}, ..., r_{fp} \in F$ that join
with $r_p$. 
RID index is the analogue of adjacency list indices in GDBMSs and similar to many GDBMSs
we store them in memory using a compressed sparse row data structure~\cite{bonifati:graphs-book}.

Recall that in absence of a RID index, we 
could not replace the join operators in the system's original plan
if $F$ was in the probe side of the join (Case 2 in Section~\ref{sec:sip-join}). When there is a RID index, we replace
such join operators with a modified \SJoin operator we call \SJoinIdxR and all of the \ScanNS($F$) operators 
on the probe side with \ScanSJNS($F$) operators. The \texttt{Idx} suffix is for using 
the RID {\bf i}n{\bf d}e{\bf x} and $R$ suffix is for {\bf r}everse. Similar to \SJoin, \SJoinIdxR builds a hash table, now of tuples from $P$ and
constructs the bitmasks for sip as follows: For each tuple $r_p$ from the build side, \SJoinIdxR consults the 
RID index on $F$ to find the RIDs of the $F$ tuples 
that join with $r_p$ and sets the bits corresponding to these RIDs. Then, similar to \SJoin, these bitmasks 
are passed to the  \ScanSJNS($F$) operators, which perform $F \ltimes P$ semijoins. 

\begin{figure}[!t]
\centering
\captionsetup{justification=centering}
  \includegraphics[width=0.6\linewidth,height=0.24\linewidth]{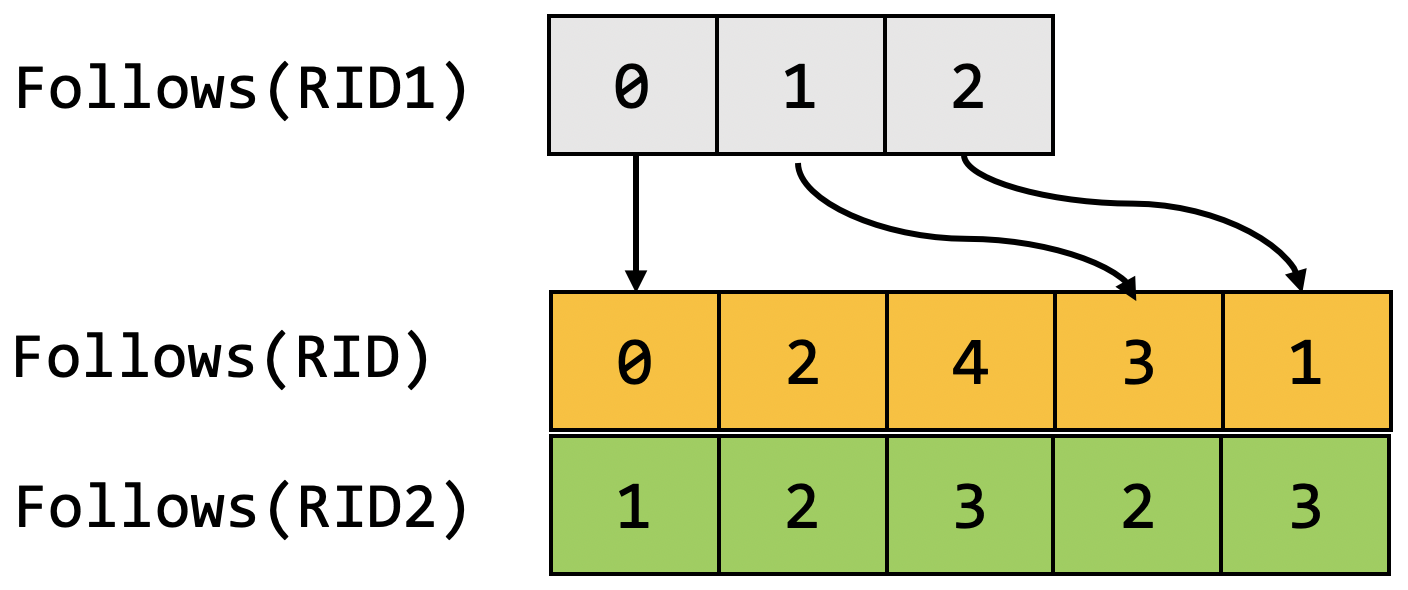}
\vspace{-10pt}
\caption{An example RID index on Follows in the CSR format. 
In graph terms, for each RID of a source entity \texttt{Person} (\texttt{RID1})
stores a set of edges and destination entities (\texttt{Follows} RID, \texttt{Person} RID) pairs,
where destination \texttt{Person} RIDs are stored in \texttt{Follows(RID2)} column.  
}
\vspace{-10pt}
\label{fig:rid-index}
\end{figure}

\begin{example}
Figure~\ref{fig:rid-index} shows the RID index that indexes the (RID1, RID) columns of the \Follows table, such
that for each RID of a row $r_p$ from the \Person table, we have a list of RIDs of matching \Follows tuples.
Ignore the \texttt{Follows(RID2)} values in the figure for now. Figure~\ref{fig:sjidxr}
shows the plan we now generate in presence of this RID index. The two \HashJoin operators from the plan
in Figure~\ref{fig:sj-no-index} are replaced with \SJoinIdxR operators and the previous \Scan operators of the
\Follows table are replaced with \ScanSJ operators. The figure also shows 
bitmasks that the new \ScanSJ operators take. For example, the \ScanSJ $F1$ operator takes a tuple bitmask with only
the index 3 set to 1 and zone bitmask with only index 2 set to 1. This is because the RID 
of the (1, 202, Karim) tuple is 1 and 1's list of matching RIDs contains only the
RID 3 of \Follows, because 202 joins with (3, 1, 202, 2, 303, 2020) (see Table~\ref{tab:extended-table}). 
This can also be seen from the RID index for Person.RID=1 in Figure~\ref{fig:rid-index}. 
\end{example}

\subsection{Extended RID Index and Join Merging}
\label{subsec:join-merging}

Many-to-many joins between two tables $P_1$ and $P_2$ that represent two (possibly same) 
sets of entities are often facilitated through a third relationship table $F$. 
In these settings, many queries use the table $F$ to join $P_1$ and $P_2$. This is,
for example, the case in our running example, where each \Follows table is joined with two \Person tables.
Therefore, it can be beneficial to predefine two joins on $F$. In this case,
each row $r_f$ of $F$ would contain the virtual RID of $F$ and two materialized RIDs, one for row $r_{p_1} \in P_1$
and the other $r_{p_2} \in P_2$ that $r_f$ joins with.
Consider building a RID index from the RIDs of $P_1$
to lists of RIDs of $F$ tuples. So for each RID of $P_1$, say $i_1$, we store a list $L_{i_1}=\{r_{f_1}, ..., r_{f_k}\}$ 
of RIDs of $F$ rows that have $i_1$ in their materialized RID column for $P_1$. We can also extend 
$L_i$ to store the RIDs of $P_2$ tuples along with the RIDs of F as follows: $\{(r_{f_1}, r_{p2_1}), ..., (r_{f_k}, r_{p2_k})\}$. This is similar to how GDBMSs store both the 
edge IDs and neighbor node IDs in their adjacency lists.
Analogous to {\em forward} and {\em backward} adjacency list indices in GDBMSs,
one can similarly build a second RID index
that now stores for each RID of $P_2$ a list of RIDs of joining $F$ and $P_1$ tuples. 
Figure~\ref{fig:rid-index} is an example ``forward'' extended RID index for the \texttt{Follows} table, 
that stores for each ``source'' \texttt{Person} tuple $r_p$, the list of RIDs of the joining \texttt{Follows} tuples, 
shown as \texttt{Follows(RID)} values, as well as the RIDs of the 
``destination'' \texttt{Person} tuples that these \texttt{Follows} tuples point to,
shown as \texttt{Follows(RID2)} values.

Consider a query that performs a join of $P_1$ $\bowtie$ $F$ $\bowtie$ $P_2$ with the predefined conditions,
but $F$ is only used to facilitate the join, so: (i) there are no filters, group by and 
aggregations, or others joins on $F$; and (ii) the final projection does 
not contain any columns of $F$. Then we can use an extended RID index to directly join the $P_1$ tuples with $P_2$ tuples, 
without ever scanning  the $F$ table and joining it with $P_1$ or $P_2$. 
We call this the {\em join merging optimization}. Specifically,
in our query optimization step, we look for two consecutive join operators $J_1$, evaluating the predefined $P_1 \bowtie F$, and $J_1$'s parent $J_2$, evaluating $P_2 \bowtie F$ such that conditions (i) and (ii) above
are satsified. 
Note that if the query satisfies condition (i), i.e., $F$ is not involved in any other joins,
the only operator on the probe side of $J_1$ must be the scan of $F$.
We replace $J_1$ and $J_2$ with a new \SJoinIdxM operator $S$, where M
stands for {\bf m}erged. $S$ takes as its build side $J_1$'s build side and as its probe side $J_2$'s probe side,
and we drop the scan of $F$, i.e., the probe side of $J_1$.
During evaluation for each $P_1$ tuple $r_p$, $S$ looks for the RIDs of joining $P_2$ tuples 
directly from the RID index, and passes these RIDs as bitmasks to the \ScanSJNS($P_2$) operators
on its probe side, without ever scanning $F$.  The join with $F$ happens implicitly while accessing the RID
index to read the RIDs of $P_2$ tuples.
 
Note that if a query needs to scan $F$, the extended index is not directly useful as
the scanned $F$ rows already materialize the $P_2$ RIDs 
which can directly be used to construct the bitmasks for sip.

\begin{example}
Figure~\ref{fig:sjidxm} shows our plan in the presence of an extended RID index from RID1 to RID2 columns of 
\Follows. Observe that compared to the plan in Figure~\ref{fig:sjidxr}, we have merged \SJoinIdxRNS$_1$ and \SJoinNS$_1$
into a new \SJoinIdxMNS$_1$ operator and \SJoinIdxRNS$_2$ and \SJoinNS$_2$ into a new \SJoinIdxMNS$_2$ operator.
\end{example}

%% file: implementation.tex
\section{Implementation Considerations}
We next elaborate on two system components, optimizer and update handling,
under our proposed integration of predefined joins. In our proposed solution,
we have chosen to use the default join optimizer of DuckDB to generate
an initial plan $P_{d}$ and then replace some of the hash joins in a rule-based approach
with our \texttt{S-Join} variants to obtain $P_{d}^{*}$.
Even if $P_{d}$ is the best default 
join order of DuckDB, in principle,
modifying another plan $P$ with predefined joins can outperform $P_{d}^{*}$.
Therefore, one can extend our integration to develop a sip-aware 
optimizer to generate such plans.  This opportunity arises 
for example on a query $T_1 \bowtie .... \bowtie T_k$, where 
 assume table $T_k$ contains a very selective predicate and is very small after the predicate. 
 Suppose further that the database has RID indices
 to pass information from $T_k$ to each $T_i$. 
Without sip, a left deep plan that first joins the small $T_{k}$ with $T_{k-1}$ and 
iteratively with $T_{k-2}$, $T_{k-3}$ until $T_1$   
would be more efficient than a left-deep plan with the reverse join order, i.e., one that first joins large tables 
and finally the smallest $T_k$.
However, under sip, the reverse plan can be more efficient because it can pass information
from $T_k$ to each $T_i$.
 In Section~\ref{sec:evaluation}, we present a plan spectrum study on a subset of the JOB benchmark
queries to study the effects of predefined joins  on the plan space of DuckDB
and how much room for improvement we found for adopting a sip-aware optimizer. 
 %
%
Although we found several queries for which we could obtain non-negligible improvements, broadly we found 
$P_{d}^{*}$ and $P_{s}^{*}$ competitive, which is why we chose not to modify the optimizer of the system
within the scope of this work.

Second, although we do not focus on updates within the scope of this work, 
our integration requires further considerations for handling updates. 
Insertions of a tuple $t$ to a table $F$ that has a predefined join to $P$ requires
finding the RID of the tuple in $P$ that $t$ refers to and inserting it in the  
system-visible \texttt{RID($A_{it_1}, ..., A_{it_{\ell}}$)} column $F$. If there is a 
RID index on $F$, possibly an extended one, we need to further update the index. 
Deletions also require additional handling. 
Observe that because we use RIDs as pointers, we materialize them in system-visible RID columns or a RID index.
Therefore once  a tuple $t$ is assigned a RID, it needs to remain fixed. Suppose a tuple $t$ with RID $k$ 
is deleted. In addition to removing any references 
to $k$ in system-visible RID columns or RID indices, the system also needs to keep track of the 
gap in $k$ and assign it to the next inserted table. That is the system cannot shift tuples with
RIDs $k+1$, which would change a large number of RIDs and require updating the references to these RIDs.  
Reusing gaps or IDs is common practice both in RDBMSs and GDMBSs, e.g., 
MySQL reuses
gaps left by deleted tuples to use for new insertions~\cite{mysql:delete}, and Neo4j keeps deleted nodeIDs 
in a separate
file and reassigns them to later inserted nodes~\cite{ neo4j:delete}.
DuckDB, on which we base our work, currently leaves these gaps as they are there 
is an open issue in the system for handling them as the system is further 
developed~\cite{duckdb:delete}.

%% file: evaluation.tex
\section{Evaluation}
\label{sec:evaluation}
We next evaluate our proposed predefined join support that is implemented inside DuckDB.
We call this version of DuckDB as GRainDB. 
Our goal is to demonstrate and validate several behaviors of our implementation. 
First, we aim to demonstrate that predefined joins provide fast join capabilities 
on several relational and graph workloads, improving the performance 
of vanilla DuckDB significantly as well as being competitive with a state-of-the-art specialized GDBMSs 
on many queries. We also demonstrate that our approach does not incur major overheads on workloads that are
not amenable to benefiting from predefinition of joins.
Second, we aim to perform an ablation study to demonstrate that each of our optimizations that facilitated different
levels of integration has additional benefits. Third, we aim to compare the performance characteristics 
of our approach against index nested loop join-based implementations that are
prevalent in GDBMSs and prior approaches that integrate predefined joins into RDBMSs~\cite{hassan:grfusion-edbt, lin:gq-fast}. 
Fourth, we demonstrate
that sip makes DuckDB's optimizer more robust by analyzing the plan spectrum of DuckDB and GRainDB on a suite 
of queries.

\subsection{Setup}
\label{subsec:evaluation-setup}

\noindent {\bf Baseline Systems:}  
We compare GRainDB against vanilla DuckDB and GraphflowDB, a state-of-art academic graph database system, as
demonstrated in several prior publications~\cite{kankanamge:graphflow, mhedhbi:optimizing, mhedhbi:aplus, gupta:columnar}. We use the most performant version of GraphflowDB from
reference~\cite{gupta:columnar}.
We also performed preliminary experiments with
Neo4j's community edition, but as with several prior work~\cite{kankanamge:graphflow, mhedhbi:aplus, gupta:columnar} did not find it competitive with 
GraphflowDB (or GRainDB on many queries) and omit these experiments. 
We emphasize that our goal in this paper is not to argue that an RDBMS can be more efficient, even after integrating
these approaches, than an efficient GDBMS because specialized systems should be expected to be more performant on the 
workloads they optimize for. However, in our evaluations we will demonstrate that our approach can be 
competitive with a state-of-the-art GDBMS on many queries from graph workloads.

We also intended to compare against GQ-Fast and GR-Fusion.
GQ-Fast and GR-Fusion are both academic prototype systems.
However, the publicly available versions of both systems have errors on sample queries and are out of maintenance
and we failed to setup these systems on our desired benchmarks. One of our goals in the GQ-Fast and GR-Fusion
comparisons were to show that the pure left-deep and index nested loop join-based plans used by these
approaches can be suboptimal to bushy and hash join-based 
plans of GRainDB. 
Instead, we will perform this comparison against similar plans from Neo4j and GraphflowDB. The GraphflowDB
 version we use~\cite{gupta:columnar} also only supports such plans.

\noindent {\bf Benchmarks:}
We expect predefined joins to provide performance improvements on queries 
with the following properties:
\begin{squishedenumerate}
\item {\em Existence of predefined joins:} As a necessary condition, the query must contain 
at least one predefined join, so that we can replace a value-based hash join operator 
with our \texttt{S-Join} variants. 
\item {\em Existence of selective predicates on $F$ and/or $P$:}  This is critical because when $F$ (or $P$) has a 
selective predicate, the semi-joins used by sip more effectively reduces the scan of
$P$ (or $F$) and probes in the following hash join.
\item {\em Existence of one/many-to-many joins:} We also expect to see 
performance improvements when queries contain one/many-to-many joins for two reasons.
First, predefined joins primarily improves join performance (as opposed to say aggregations),
and the join performance is often an important runtime factor in queries with one/many-to-many joins, which
are challenging as they lead to growing intermediate results.
Second, the reverse semijoins and join-merging optimizations
can benefit primarily queries with 
one/many-to-many joins. This is because these optimizations require a RID index, which is
generally built on tables that represent one/many-to-many 
relationships between two other tables that represent entities, such as \texttt{Follows} in our running example.
\end{squishedenumerate}

In light of these, we used one relational and one graph benchmark that 
contain queries that satisfy these properties and
for a more complete evaluation, a second relational workload that does not.
\begin{squishedlist}
\item Join order benchmark (JOB) on the IMDB dataset~\cite{leis2015good}, which contains 
more than 2.5 M movie titles produced by 235K different companies with over 4 M actors.
When using GRainDB, we predefine every one-to-many primary foreign key relationship in the database and
for tables that represent many-to-many relationships, such as \texttt{movie-companies}, we build a RID index.
\item LDBC Social Network Benchmark~\cite{angles:ldbc-snb} (SNB) benchmark at scale factor 10 and 30, which is a commonly
used graph benchmark that models a social networking application with users, forums, and posts. 
In relational format, LDBC10 dataset contains 8 entity (i.e, node) and 10 relationship (i.e., edge) tables,
with a total number of 36.5M and 123.6M tuples, respectively. LDBC30 contains 106.8M entity and 385.2M relationship tuples. 
We use SNB primarily to compare against GraphflowDB (and Neo4j, which was not competitive).
GraphflowDB is an academic prototype system that does not implement several language features, such as recursive queries. 
Therefore, we slightly modified the benchmark and refer to it as {\em SNB-M}, for {\bf m}odified.
We removed queries involving shortest paths and decomposed queries with variable-length joins into multiple 
queries, each of which has a fixed path join (we denote each version with a suffix -$\ell$, where $\ell$ denotes the length).
SNB-M  contains variants of 18 out of 21 queries from the original SNB interactive simple (IS) 
and interactive complex (IC) benchmarks. 
\iflong
Our full SNB-M queries are listed in Appendix~\ref{app:snb-m}.
\else
In the longer version of this paper~\cite{graindb:long-version}, we list our full SNB-M queries.
\fi
SNB is generated in both relational and property graph formats. We use the relational format in DuckDB and GRainDB.
For every edge type in the graph format of SNB, e.g., \texttt{Knows} edges,
we build a RID index
over the corresponding table in GRainDB.
\item TPC-H benchmark at scale factor 10. We include TPC-H
to perform a sanity check that making the primary-foreign key joins on such traditional workloads 
does not hurt performance. We do not expect GRainDB to 
provide meaningful improvements on TPC-H as it does not contain selective many-to-many joins.
We predefined every one-to-many primary foreign key relationships in GRainDB, such as \texttt{customer} and \texttt{orders}. 
Although we did not expect performance speedups, we still found several queries on which we obtained non-negligible runtime improvements.
\end{squishedlist}
In our detailed evaluation in Section~\ref{subsec:evaluation-detailed}, we also use modifications of some of the join queries 
from JOB and SNB-M to create microbenchmarks.

\noindent\textbf{System Configurations and Hardware:}
We set DuckDB to the in-memory mode. GraphflowDB is already an in memory system. 
DuckDB is still in early stage and does not integrate full cardinality estimation. We observed
that this limits its ability to choose good plans on many instances, 
especially in queries with large joins and selective predicates.
To isolate the influence of join order selection, we injected true cardinalities into the system. 
\iflong
In Appendix~\ref{app:duckdb-optimized}, we present a demonstrative experiment on the JOB benchmark that 
this uniformly improves the performance of both DuckDB and GRainDB.
\else
In the longer version of this paper~\cite{graindb:long-version}, we present a demonstrative experiment on the JOB benchmark that 
this uniformly improves the performance of both DuckDB and GRainDB.
\fi
The GraphflowDB version we use does not
contain an optimizer, so does not need to estimate cardinalities. We manually picked the systems' best join order, which
for many queries was obvious. For example, 
many of the queries in our benchmarks are path queries that have a highly selective predicate on the left-most node, 
in which case we picked the plan that evaluates the join from left to right.

All experiments were conducted on a machine with two Intel E5-2670 @2.6GHz CPUs and 256 GB of RAM, consisting of 16 physical cores and 32 logical cores. Because GraphflowDB runs only in serial mode, we set DuckDB to run in serial mode as well. All reported times are averages of five successive runs after a warm-up running. Our measurements reflect the end-to-end query evaluation time, and a timeout of 10 minutes is imposed on each running.

\vspace{-10pt}
\subsection{End-To-End Benchmarks}
\label{subsec:evaluation-end2end}
\begin{figure}[!tbp]
	\centering
	\captionsetup{justification=centering}
  	\includegraphics[width=1\linewidth]{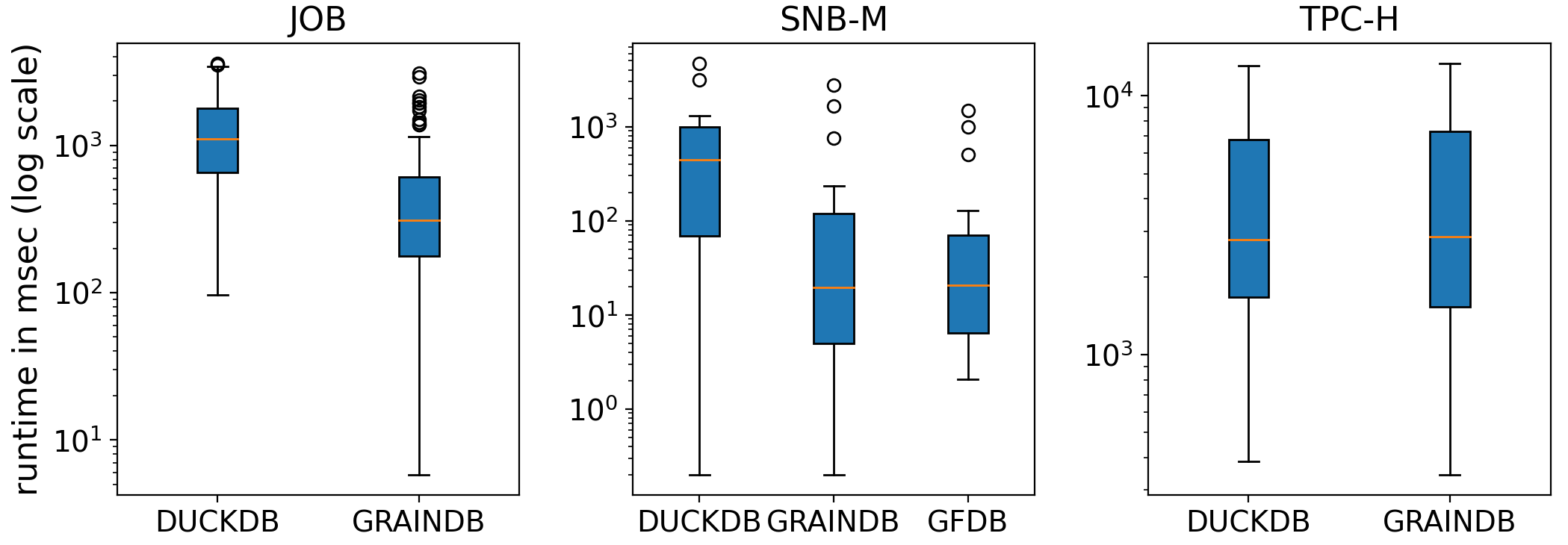}
\vspace{-10pt}
\caption{Runtimes (in ms) of DuckDB and GRainDB on JOB, SNB-M and TPC-H, and GraphflowDB on SNB-M.}
\label{fig:end2end}
\vspace{-20pt}
\end{figure}

We first present end-to-end evaluations on JOB, SNB-M, and TPC-H. We compare DuckDB and GRainDB on
JOB and TPC-H, and DuckDB, GRainDB, and GraphflowDB on SNB-M. 
We expect to see large performance improvements of GRainDB over DuckDB on JOB and SNB-M because
queries in JOB and SNB-M, with a few exceptions, satisfy the three properties we reviewed in Section~\ref{subsec:evaluation-setup}. 
In contrast, we do not expect broad improvements in TPC-H, but expect minor overheads as well.
For reference, Figure~\ref{fig:end2end} presents box plot
charts that show the performances of these systems on these workloads. Each boxplot shows the distribution of 
the runtimes of the queries in the workloads, specifying the 5th, 25th, 50th, 75th, and 95th percentiles of the
distribution with marks.


\subsubsection{JOB: Relational Workload with Selective Many-to-Many Joins}
\label{subsubsec:evaluation-job}
The box plots of DuckDB and GRainDB on JOB are shown in Figure~\ref{fig:end2end}.
As we expect, we observe GRainDB outperforms DuckDB by large margins. 
JOB contains 113 queries. Table~\ref{tab:end2end-job-percentiles} lists detailed percentiles 
for query execution times of DuckDB and GRainDB of these queries. We see consistent large runtime 
reductions for each percentile. For example, the 25th percentile, median, and 75th percentile query execution times reduce
respectively  from 652.4ms to 176.4ms (3.7x), from 1110ms to 309ms (3.6x), and from 1797ms to 614.2ms (2.9x). 
For reference, Table~\ref{tab:end2end-job} presents the execution times of a subset of the queries in JOB. Specifically, JOB queries contain
between 2 to 6 variants and we present the first variant of each query in the table. 
\iflong
The full table can be found in Appendix~\ref{app:job-tpch-full}. 
\else
The full table can be found in the longer version of this paper~\cite{graindb:long-version}.
\fi
Importantly, we see consistent runtime improvements on all queries, with a few exceptions. 
Table~\ref{tab:end2end-job} also presents the reduction on the 
amount of scanned tuples for each query in DuckDB and GRainDB. 
Although runtime reductions depend on many factors besides the reductions in scanned tuples,
such as the actual outputs from the joins or how complex the predicate expressions are, 
this is still a good proxy for explaining when sip 
and predefined joins improve performance. For example, we observe that the queries in which we observe the largest 
improvement factors, such as Q6a, Q21a, Q27a and Q32a, also have large reductions in scanned tuples by 348.9x, 182.2x, 185.4x, and 53.8x.
Similarly queries with negligible improvements, such as Q5a and Q20a also have respectively no or small (1.3x)
reductions in scanned tuples.

\begin{table}[!t]
	\setlength{\tabcolsep}{3.0pt}
	\def\arraystretch{1.0}%
	\centering
	\hspace{1.7cm}{}
	\begin{tabular}{ c|c|c|c|c|c|c|c }
		& \textbf{Min} & \textbf{5th} & \textbf{25th} & \textbf{50th} & \textbf{75th} & \textbf{95th} & \textbf{Max} \\
		\hline
				
		{\texttt{DuckDB}} & 96.0 & 203.6 & 652.4 & 1110.0 & 1797.0 & 2939.9 & 3584.6 \\ 
		
		{\texttt{GRainDB}}& 5.8 & 27.4 & 176.4 & 309.0 & 614.2 & 1878.5 & 3104.4 \\
	\end{tabular}\vspace{0.1 cm}
	\captionsetup{justification=centering}
	\caption{Detailed percentiles of runtimes (in ms) for DuckDB and GRainDB on JOB.}
	\label{tab:end2end-job-percentiles}
	\vspace{-30px}
\end{table}

\begin{table*}
	\bgroup
	\setlength{\tabcolsep}{3.0pt}
	\def\arraystretch{1.0}%
	\centering
	\hspace{0.5cm}
	\begin{tabular}{ |c|c|c|c|c|c|c|c|c|c|c|c|c|c|c|c|c|c| }
		\hline
		& \textbf{1a} & \textbf{2a} & \textbf{3a} & \textbf{4a} & \textbf{5a} & \textbf{6a} & \textbf{7a} & \textbf{8a} & \textbf{9a} & \textbf{10a} & \textbf{11a} & \textbf{12a} & \textbf{13a} & \textbf{14a} & \textbf{15a} & \textbf{16a} & \textbf{17a} \\ 
		\hline
				
		{\texttt{DuckDB}} & 234.2 & 207 & 1491.4 & 216 & 96 & 885.4 & 879.8 & 1307.4 & 1933.2 & 1404.2 & 264.2 & 684 & 981.2 & 1644.2 & 987.8 & 652.4 & 1164.8 \\ 
		\hline	

		\multirow{2}{*}{\texttt{GRainDB}}& 34.2 & 154.0 & 328.0 & 114.2 & 116.4 & 57.0 & 203.4 & 349.8 & 554.8 & 614.2 & 44.4 & 242.4 & 336.8 & 202.2 & 283.4 & 177.6 & 486.0 \\ 
		& \textbf{6.8x} & \textbf{1.3x} & \textbf{4.5x} & \textbf{1.9x} & \textbf{0.8x} & \textbf{15.5x} & \textbf{4.3x} & \textbf{3.7x} & \textbf{3.5x} & \textbf{2.3x} & \textbf{6.0x} & \textbf{2.8x} & \textbf{2.9x} & \textbf{8.1x} & \textbf{3.5x} & \textbf{3.7x} & \textbf{2.4x} \\ 
		\hline

		{\texttt{Scan Reduction}} & 751.2x & 23.9x & 97.2x & 58.7x & 1x & 348.9x & 47.0x & 9.2x & 6.0x & 13.0x & 73.7x & 44.8x & 42.4x & 142.1x & 7.4x & 14.4x & 10.7x \\ 
		\hline

	\end{tabular}\vspace{0.2 cm}
\newline

	\begin{tabular}{ |c|c|c|c|c|c|c|c|c|c|c|c|c|c|c|c|c| }
		\hline
		 & \textbf{18a} & \textbf{19a} & \textbf{20a} & \textbf{21a} & \textbf{22a} & \textbf{23a} & \textbf{24a} & \textbf{25a} & \textbf{26a} & \textbf{27a} & \textbf{28a} & \textbf{29a} & \textbf{30a} & \textbf{31a} & \textbf{32a} & \textbf{33a} \\ 
		\hline
				
		{\texttt{DuckDB}} & 1797.0 & 2632.4 & 1118.4 & 1629.2 & 1471.8 & 866.6 & 2554.8 & 2318.8 & 1074.6 & 761.4 & 2068.4 & 2742.6 & 2198.0 & 2523.8 & 126.0 & 336.2 \\ 
		\hline
				
		\multirow{2}{*}{\texttt{GRainDB}} & 612.6 & 491.6 & 1071.6 & 54.2 & 864.0 & 296.8 & 788.8 & 1376.6 & 733.4 & 44.4 & 240.2 & 266.6 & 673.0 & 612.0 & 8.2 & 149.4 \\ 
		& \textbf{2.9x} & \textbf{5.4x} & \textbf{1.0x} & \textbf{30.1x} & \textbf{1.7x} & \textbf{2.9x} & \textbf{3.2x} & \textbf{1.7x} & \textbf{1.5x} & \textbf{17.1x} & \textbf{8.6x} & \textbf{10.3x} & \textbf{3.3x} & \textbf{4.1x} & \textbf{15.4x} & \textbf{2.3x} \\ 
		\hline

		{\texttt{Scan Reduction}} & 13.8x & 7.8x & 1.3x & 182.2x & 7.6x & 8.2x & 8.7x & 32.2x & 15.7x & 185.4x & 156.9x & 10.8x & 282.2x & 165.5x & 53.8x & 2.6x \\ 
		\hline

	\end{tabular}

	\egroup
		\captionsetup{justification=centering}
	\caption{Runtimes (in ms) of \texttt{DuckDB} and \texttt{GRainDB} on each query in JOB.}
	\label{tab:end2end-job}
	\vspace{-20px}
\end{table*}


\begin{table}
	\setlength{\tabcolsep}{3.0pt}
	\def\arraystretch{1.0}%
	\centering
	\hspace{1.7cm}{}
	\begin{tabular}{ c|c|c|c|c|c|c|c }
		& \textbf{Min} & \textbf{5th} & \textbf{25th} & \textbf{50th} & \textbf{75th} & \textbf{95th} & \textbf{Max} \\
		\hline
				
		{\texttt{DuckDB}} & 0.2 & 1.5 & 68.4 & 441.8 & 989.0 & 2762.5 & 4647.0 \\ 
		
		{\texttt{GRainDB}}& 0.2 & 0.7 & 5.0 & 19.6 & 119.4 & 1482.4 & 2768.0 \\
		{\texttt{GraphflowDB}}& 2.1 & 2.5 & 6.4 & 20.8 & 70.3 & 888.2 & 1473.6 \\
	\end{tabular}\vspace{0.1 cm}
	\captionsetup{justification=centering}
	\caption{Detailed percentiles of runtimes (in ms) for DuckDB, GRainDB, and GraphflowDB on SNB-M.}
	\label{tab:end2end-snb-percentiles}
	\vspace{-20px}
\end{table}

\subsubsection{SNB-M: Graph Workload with Selective Many-to-Many Joins}
\label{subsubsec:evaluation-snb}
The box plots of DuckDB, GRainDB, and GraphflowDB on SNB-M are shown in Figure~\ref{fig:end2end}.
Table~\ref{tab:end2end-snb-percentiles} also lists detailed percentiles for query execution times of the systems. 
We see that GraphflowDB outperforms DuckDB by large margins on SNB-M.
Specifically for the 25th percentile, median, and 75th percentile query execution times, GraphflowDB outperforms
DuckDB respectively by 10.7x (68.4ms vs 6.4ms), 22.5x  (441.8ms vs 20.8ms), and 14.1x (989.0ms vs 70.3ms).
However, 
by predefining the joins in SNB-M, GRainDB closes this performance gap significantly, making
DuckDB competitive with GraphflowDB on majority of the queries. 
Specifically for the 25th percentile, median, and 75th percentile query execution times, GRainDB and GraphflowDB 
compare as follows:  5.0ms vs 6.4ms (0.78x), 19.6ms vs 20.8ms (0.94x), and 119.4ms vs 70.3ms (1.7x).
This shows that
our implementation of predefined joins can make a columnar RDMBS competitive with a performant 
GDBMS on a workload that GDBMSs are optimized for.


\begin{table*}[t!]
	\bgroup
	\setlength{\tabcolsep}{3.0pt}
	\def\arraystretch{1.0}%
	\centering
	\hspace{0.5cm}
	\begin{tabular}{ |c|c|c|c|c|c|c|c|c|c|c|c|c|c|c| }
		\hline
		& \textbf{IS1} & \textbf{IS2} & \textbf{IS3} & \textbf{IS4} & \textbf{IS5} & \textbf{IS6} & \textbf{IS7} & \textbf{IC1-1} & \textbf{IC1-2} & \textbf{IC1-3} & \textbf{IC2} & \textbf{IC3-1} & \textbf{IC3-2} \\ 
		\hline
				
		{\texttt{DuckDB}} & 0.8 & 524.8 & 36.6 & 0.2 & 4.5 & 148.0 & 989.0 & 38.0 & 72.0 & 110.5 & 926.0 & 1177.8 & 4647.0 \\ 
		\hline	

		\multirow{2}{*}{\texttt{GRainDB}}& 1.2 & 19.6 & 3.4 & 0.2 & 0.6 & 5.0 & 11.0 & 4.0 & 6.4 & 38.2 & 134.8 & 119.4 & 1665.0 \\ 
		& \textbf{0.7x} & \textbf{26.8x} & \textbf{10.8x} & \textbf{1.0x} & \textbf{7.5x} & \textbf{29.6x} & \textbf{90.0x} & \textbf{9.5x} & \textbf{11.2x} & \textbf{2.9x} & \textbf{6.9x} & \textbf{9.9x} & \textbf{2.8x} \\ 
		\hline
	

		\multirow{2}{*}{\texttt{GraphflowDB}} & 6.8 & 3.0 & 2.5 & 42.7 & 82.7 & 66.4 & 72.0 & 2.1 & 6.4 & 70.3 & 47.8 & 11.5 & 505.4 \\ 
		& \textbf{0.1x} & \textbf{175.8x} & \textbf{14.5x} & \textbf{0.005x} & \textbf{0.05x} & \textbf{2.2x} & \textbf{13.7x} & \textbf{18.3x} & \textbf{11.3x} & \textbf{1.6x} & \textbf{19.4x} & \textbf{102.6x} & \textbf{9.2x} \\ 
		\hline

	\end{tabular}\vspace{0.2 cm}
\newline

	\begin{tabular}{ |c|c|c|c|c|c|c|c|c|c|c|c|c| }
		\hline
		 & \textbf{IC4} & \textbf{IC5-1} & \textbf{IC5-2} & \textbf{IC6-1} & \textbf{IC6-2} & \textbf{IC7} & \textbf{IC8} & \textbf{IC9-1} & \textbf{IC9-2} & \textbf{IC11-1} & \textbf{IC11-2} & \textbf{IC12} \\ 
		\hline
				
		{\texttt{DuckDB}} & 402.0 & 636.0 & 3125.0 & 244.6 & 471.2 & 1186.8 & 1017.0 & 441.8 & 1312.6 & 35.8 & 68.4 & 788.4 \\ 
		\hline
			
		\multirow{2}{*}{\texttt{GRainDB}}& 54.0 & 174.0 & 2768.0 & 13.0 & 22.0 & 33.2 & 14.0 & 113.6 & 752.0 & 2.8 & 9.0 & 234.8 \\ 
		& \textbf{7.4x} & \textbf{3.7x} & \textbf{1.1x} & \textbf{18.8x} & \textbf{21.4x} & \textbf{35.7x} & \textbf{72.6x} & \textbf{3.9x} & \textbf{1.8x} & \textbf{12.8x} & \textbf{7.6x} & \textbf{3.4x} \\ 
		\hline


		\multirow{2}{*}{\texttt{GraphflowDB}}& 12.3 & 20.8 & 984.0 & 8.1 & 127.2 & 2.8 & 2.8 & 55.6 & 1473.7 & 2.6 & 14.4 & 28.8 \\ 
		& \textbf{32.6x} & \textbf{30.6x} & \textbf{3.2x} & \textbf{30.2x} & \textbf{3.7x} & \textbf{426.6x} & \textbf{359.5x} & \textbf{7.9x} & \textbf{0.9x} & \textbf{13.8x} & \textbf{4.8x} & \textbf{27.3x} \\ 
		\hline

	\end{tabular}

	\egroup
		\captionsetup{justification=centering}
	\caption{Runtimes (in ms) of \texttt{DuckDB}, \texttt{GRainDB}, and \texttt{GraphflowDB} on each query in SNB-M.}
	\label{tab:end2end-snb}
	\vspace{-25px}
\end{table*}

Table~\ref{tab:end2end-snb} shows the detailed execution times of each query for
all systems. 
We see that GRainDB outperforms DuckDB on almost all queries by up to 90x, except for IS1 and IS4, 
which are two small queries executed within 2ms. Similarly, GraphflowDB  outperforms DuckDB in most queries by up to 426.6x.
We observe that there are also 9 queries in which GRainDB outperforms GraphflowDB by large margins. 
We analyzed each of these queries to study GRainDB's performance advantages over GraphflowDB. First are 
 IS01 and IS04-IS07, which are point lookup queries over large base tables with inexpensive joins. Here,
GraphflowDB resorts to sequential scans of these tables/nodes while GRainDB (and DuckDB) use a primary key index.
This is not an inherent limitation of GraphflowDB plans and can be remedied if GraphflowDB also supports primary key indexes.
On the remaining 4 queries, GRainDB plans have two separate advantages.
\begin{squishedlist}
\item {\em Bushy vs Left-deep Plans:} IC1-3, IC6-2, IC11-2 are three queries in which GRainDB outperforms
DuckDB and uses a bushy plan. We describe IC6-2 as an example.
IC6-2 is a complex query with 8 joins in SQL. In graph version, this is a 5-path
query with selective predicates on both ends of the path. 
GraphflowDB does not implement bushy plans, so this query is implemented with a left-deep plan.
This is less performant than the bushy plan that GRainDB uses that breaks the path into two parts. 
\iflong
Figure~\ref{fig:plan-gfdb} and~\ref{fig:plan-graindb} in Appendix~\ref{app:ic6-2-plans} show the plans from both systems.
\else
In the longer version of this paper~\cite{graindb:long-version}, we show the plans from both systems.
\fi
This is an example of when the left-deep plan-based 
approaches to evaluate such path queries, which are used in systems like GR-Fusion and GQ-Fast, can be suboptimal
to bushy plans.
\item {\em Hash Join vs Index Nested Loop Joins  and Scanning Edges Before vs After Joins:} 
The IC9-2 query 
is a smaller query with  4 joins in SQL. This is a 3-path query that also has filters on both ends. 
Now both systems use left-deep plans. 
The majority of the time in this query is spent in the very last join, which requires joining 7681 tuples from a \texttt{Person}
table with a \texttt{Comments} table. 2.7M of these tuples successfully join with the 7681 tuples and 2.4M of these also
pass a filter on the \texttt{Comments} table. In graph terms,
7681 \texttt{Person} nodes have 2.7M outgoing \texttt{Comment} edges (so an average degree of 351). 
As every GDBMS we are aware of, GraphflowDB follows these steps: (i) {\em joining nodes with edges}:
looks up the edges of each of 7681 keys in a large adjacency list index that point to \texttt{Comment}s.
This effectively performs 7681 random lookups into a hash table of size 26.5M, 
and then generates 2.7M intermediate tuples. (ii) {\em property scan and edge filtering}: 
reads the necessary properties of the 2.7M \texttt{Comment}s and runs the filter predicates on these edges. In contrast, 
GRainDB, as typical of columnar RDBMSs for evaluating equality joins, follows these steps: (i) 
{\em hash table build:} creating a hash table of size 7681; (ii) {\em edge scanning and filtering: }
sequentially scanning a large \texttt{Comments} table with 26.5M tuples and running a predicate on them which returns
2.4M tuples; (iii) {\em joining nodes with edges:} and finally doing 2.4M lookups into this very small hash table and performing the join.
Now the joins happen after a sequential scan and filter of the ``edge'' table, leveraging 
columnar RDBMS techniques for
highly optimized for sequential scans and filters of large columns. In addition, in the final join, now the lookups are into a very small hash-table 
instead of a large adjacency list index. This is more performant than performing the joins by lookups into a large index and
non-sequentially scanning and filtering the joined edges.
It is interesting to note that no GDBMS we are aware of (nor GRFusion or GQFast's approaches to
perform predefined joins) generates plans that can sequentially scan and filter all of the edges
and then join these edges with their source nodes. GDBMSs typically joins nodes with their edges by performing lookups using the nodes as keys. 
For example, Neo4j's plan on the same query is the same as GraphflowDBs. 
We will present a more controlled experiment to demonstrate this difference in Section~\ref{subsubsec:evaluation-selectivity}.
\end{squishedlist}

Finally, this experiment gives us a point of comparison for the memory consumption of our RID indices. 
We profiled the memory consumption of GRainDB's RID indices and GraphflowDB's 
adjacency list indices. For each RID index we have on SNB-M, there is a corresponding adjacency list in GraphflowDB. 
In total, GRainDB's RID indices take 5.9GB while GraphflowDB's indices take
2.8GB. This is expected because GraphflowDB implements several compression 
techniques~\cite{gupta:columnar}, such as compressing trailing 0s in IDs, which in SNB-M reduces 8 byte IDs to 4 bytes, and compressing empty/null adjacency lists.
Instead, we store each RID in 8 bytes. As acknowledged in reference~\cite{gupta:columnar}, the compression techniques
in GraphflowDB are modifications of widely adopted  techniques from columnar RDBMSs 
and can be integrated into our index implementation to close this gap.

\subsubsection{TPC-H: Traditional OLAP Workloads}
\label{subsubsec:evaluation-tpch}
For completeness of our work and to verify that predefined joins have small overheads in a workload 
that does not contain many queries with selective many-to-many joins, we
also compared the performances of DuckDB and GRainDB on TPC-H. The box plots of DuckDB and GRainDB are shown in Figure~\ref{fig:end2end}. 
\iflong
Table~\ref{tab:tpch-full} in the Appendix~\ref{app:job-tpch-full} also gives the detailed
execution time of each query.
\else
The longer version of this paper~\cite{graindb:long-version} shows detailed execution time of each query.
\fi
As expected we do not see significant speedups or
slowdowns on this benchmark. GRainDB replaces value-based hash joins with predefined joins
in 13 of the 22 queries in TPC-H. The median runtime improvement out of these queries is 1.1x, with the maximum slow-down
and speedup of 0.8x (so 1.2x slowdown) and 2.6x, respectively. Interestingly, even on a benchmark of 
traditional analytical queries, we found two queries with one/many-to-many joins on which replacing value-based joins
with predefined joins lead to visible speedups (2.6x for Q2 and 1.8x for Q3) and no queries visibly slowed down,
indicating the low performance overheads of our implementation when queries are not suitable to
benefitting from predefined joins.

\subsection{Detailed Evaluation}
\label{subsec:evaluation-detailed}
We next provide a more detailed evaluation consisting of (i) an ablation study to verify that each of our optimizations on top of 
DuckDB leads to additional performance benefits; (ii) a controlled experiment 
comparing the performances of index nested loop join-based plans (adopted in GDBMSs and 
systems such as GR-Fusion and GQ-Fast)
and our hash-join-based plans when joining records from relationship tables with entity tables under varying selectivities; 
and (iii) an analysis of the effects of our sip-based predefined joins in the plan space of DuckDB on a suite of queries.

\subsubsection{Ablation Study.}
\label{subsubsec:evaluation-ablation}
\begin{figure}[!t]
\centering
	\captionsetup{justification=centering}
  \includegraphics[width=0.65\linewidth,height=0.4\linewidth]{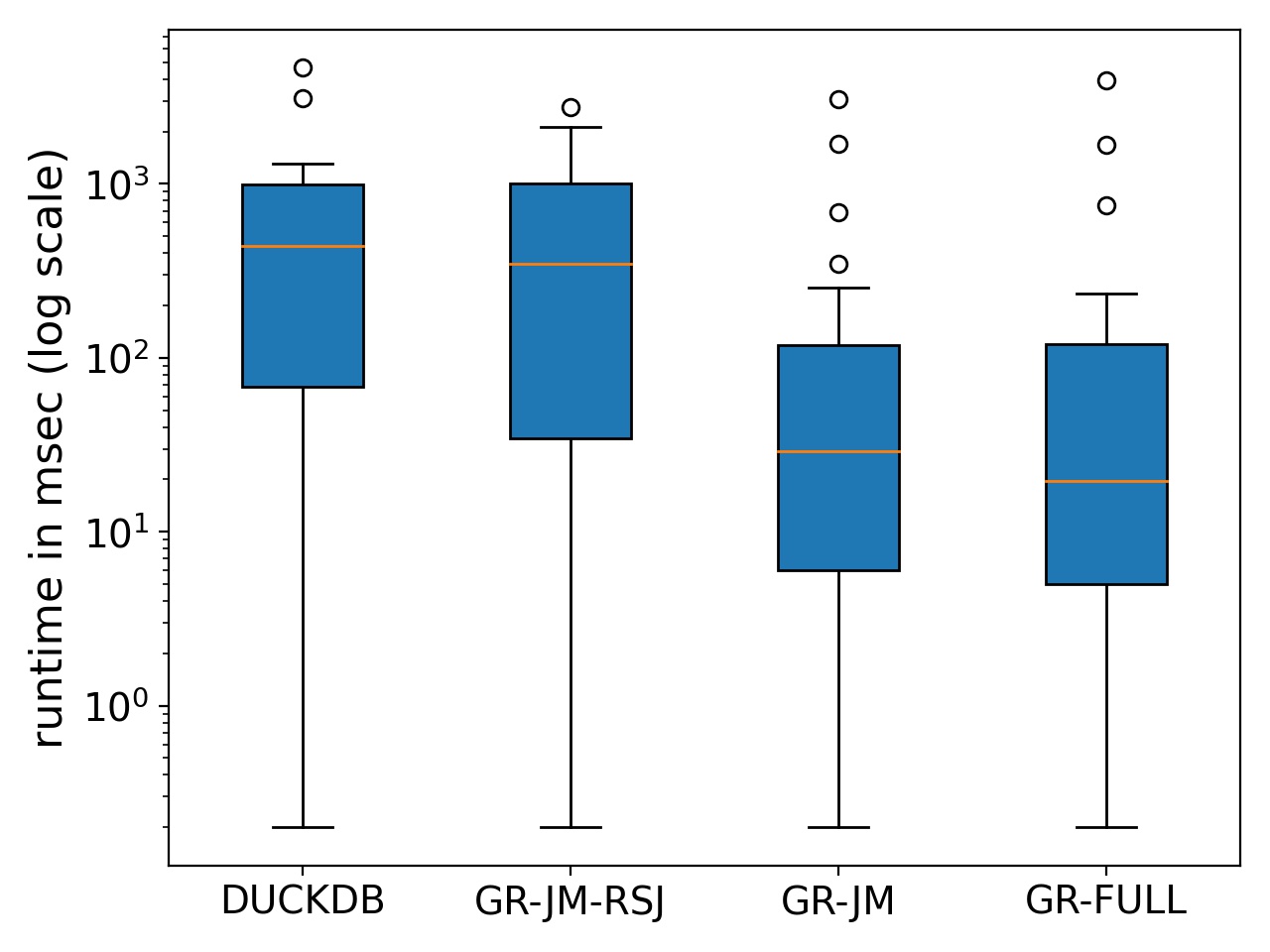}
\vspace{-10pt}
\caption{Ablation tests for different optimizations in GRainDB. DuckDB implements no optimizations. 
GRainDB-JM-RSJ only implements RID materialization.
GRainDB-JM in addition implements reverse semijoins. GRainDB in addition implements
join merging.}
\vspace{-10pt}
\label{fig:ablation-test}
\end{figure}

We performed an ablation study, to show the positive performance benefits of each of the optimizations we integrated
into DuckDB: (i) RID materialization (Section~\ref{sec:rids}); (ii) reverse semijoins (Section~\ref{subsec:reverse-sj}); 
and (iii) extended RID index and join merging (Section~\ref{subsec:join-merging}). Note that our optimizations are not independent of each other,
e.g., without RID materialization we cannot perform either reverse semijoins or join merging. 
We therefore turned them off in a specific order and in growing sets. 
We first turned off extended RID index and join merging (-JM), then, we turned off
reverse semijoins (-JM-RSJ), and finally we turned off all optimizations, which gives us vanilla DuckDB.
Then we ran each version of the system on the SNB-M benchmark. Figure~\ref{fig:ablation-test} shows the box plot
charts of each version of the system.  GR-FULL in the figure is the configuration with all optimizations on. 
We see that each optimization has a positive effect on performance, which can be
seen by inspecting the median and 25 percentile lines, which consistently shift down as we add more optimizations.
We see most impact from the reverse semijoin optimization, which is expected as it allows passing information from
smaller entity tables ($P$ in our notation) to much larger relationship tables 
($F$ in our notation). 
\iflong
For reference, we show the runtime numbers of each query on each system configuration 
in Appendix~\ref{app:snb-m-ablation}.
\else
In the longer version of this paper~\cite{graindb:long-version}, we show the runtime numbers of each query on each system configuration.
\fi
We observe queries where RID materialization leads up to 29.6x additional improvement (IS6),
reverse semijoins up to 40.9x (IC7) additional improvements, and join merging up to 7.3x (IC2) additional improvements.

\begin{figure}[!t]
	\centering
	 \captionsetup{justification=centering}
	\begin{subfigure}[t]{0.22\textwidth}
		\centering
  		\includegraphics[width=1\linewidth]{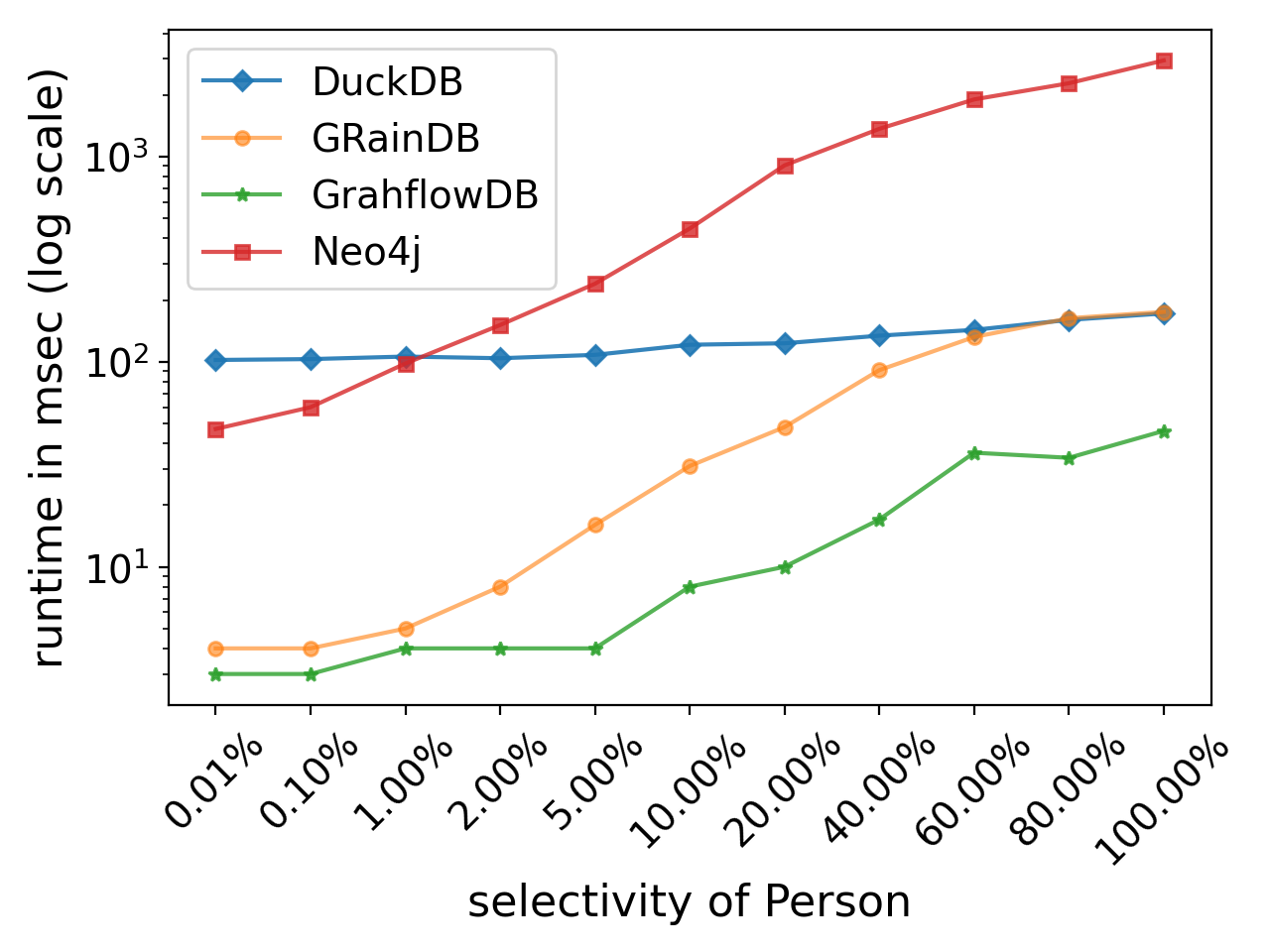}
  		\caption{MICRO-P.}
  		\label{fig:detailed-selectivity-person}
  	\end{subfigure}
  	\begin{subfigure}[t]{0.22\textwidth}
		\centering
  		\includegraphics[width=1\linewidth]{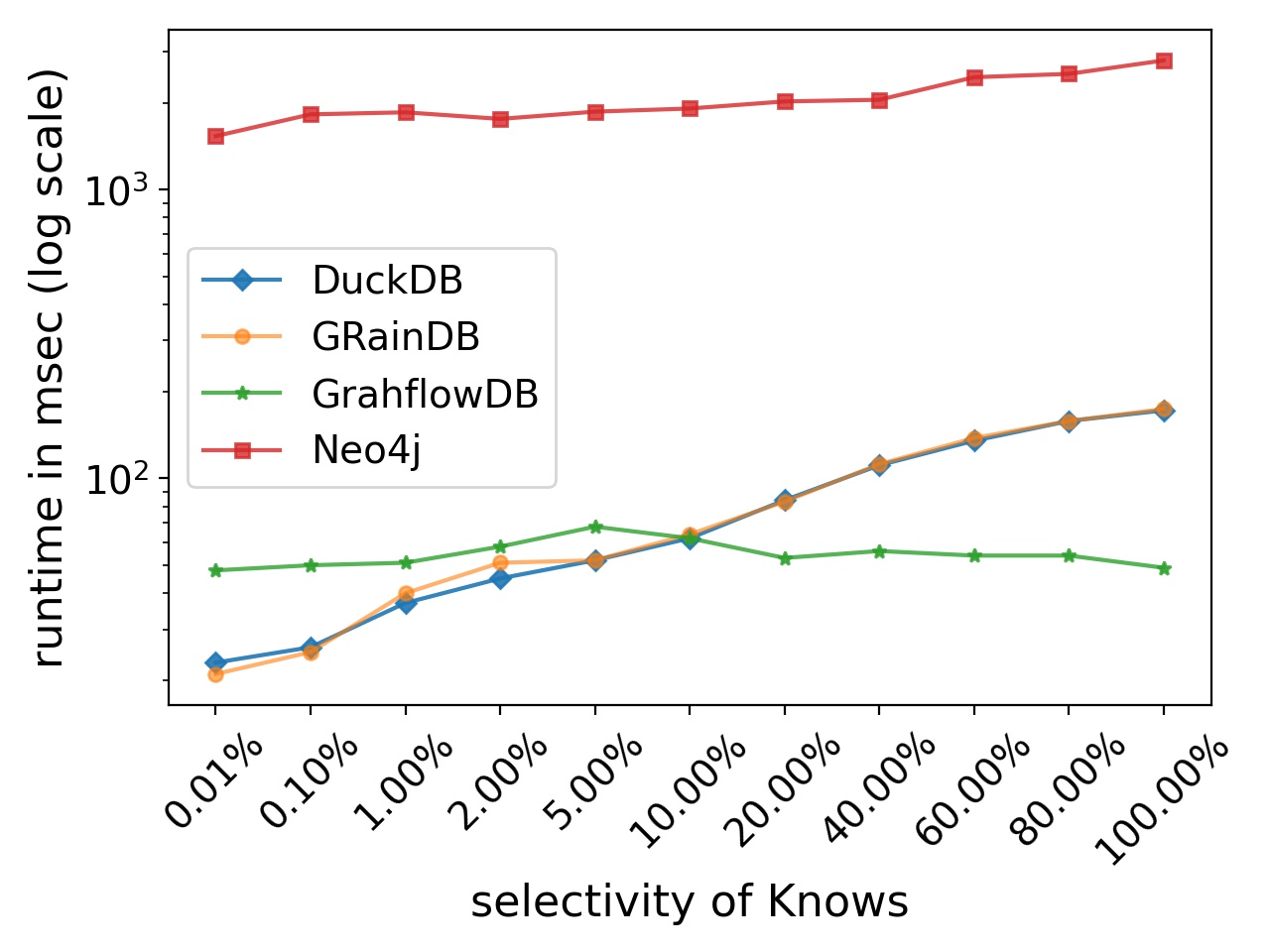}
  		\caption{MICRO-K.}
  		\label{fig:detailed-selectivity-knows}
  	\end{subfigure}
\vspace{-10pt}
\caption{Runtimes (in ms) of DuckDB, GRainDB, GraphflowDB and Neo4j. The left and right figures show, respectively, 
the times on 
MICRO-P and MICRO-K queries, which vary the selectivity of \texttt{Person} and \texttt{Knows}, respectively.}
\label{fig:selectivity}
\vspace{-10pt}
\end{figure}
 
\subsubsection{Performance of Predefined Joins Under Varying Entity vs Relationship Table Selectivity}
\label{subsubsec:evaluation-selectivity}
We next do a controlled experiment to demonstrate the behavior of our sip- and hash-join based implementation
of predefined joins under various selectivities on the $P$ and $F$ tables. 
Our goals are twofold: (i) to show the cases when sip yields performance improvements; and (ii) to 
demonstrate the different performance behaviors of index nested loop joins, which GDBMSs use, vs hash joins, which
many RDBMSs, including DuckDB, use for equality joins.  We take the LDBC30 dataset and the 1-hop 
\lstinline{$(p_1$$:$$Person)$$\xrightarrow{e:Knows}$$(p_2$$:$$Person)$} query, where the \texttt{Person}
and \texttt{Knows} tables have 18.4K and 7.5M tuples, respectively. This is representative of the general
case when there are many more relationship/edge tuples than node/entity tuples in databases. 
We then run two sets of micro-benchmark
queries: (1) MICRO-P: we fix a predicate with 99.9\% selectivity on the \texttt{creationDate} 
property of \texttt{Knows} and vary the selectivity of a predicate on the \texttt{id} property of \texttt{Person} 
between 0.01\% to 100\%. 
(2) MICRO-K: we now fix a predicate with 99.9\% selectivity on the \texttt{id} 
property of \texttt{Person} and vary the \texttt{creationDate} property of \texttt{Knows} between 0.01\% to 100\%.
We run each set of queries on DuckDB, GRainDB, GraphflowDB and Neo4j. Our goal in including Neo4j in these
experiments, which was omitted in our baselines, 
is to show that the two GDBMSs behave very similarly albeit in different performance levels.
Before we discuss the results, we note that in both sets of queries GraphflowDB and Neo4j's
executions are always as follows: (i) scan the \texttt{Person} nodes and their \texttt{id} property and run the filter on \texttt{id};
(ii) join these nodes with their \texttt{Knows} edges by index nested loop join using the \texttt{Knows} adjacency list index;
(iii) read the \texttt{creationDate} property of the joined edged and run a filter. This execution is standard in every GDBMS we are 
aware of, where joins are always from nodes to edges and  we will momentarily show that this is in 
fact too rigid and can be suboptimal. This is also the execution in systems such as GR-Fusion and GQ-Fast. 

Figure~\ref{fig:detailed-selectivity-person} shows the results for 
MICRO-P. First, we note that on all MICRO-P queries, DuckDB makes \texttt{Person} the build side 
as it is already much smaller than \texttt{Knows} and gets even smaller as we decrease the selectivity on the predicate on \texttt{Person}.
Therefore, in GRainDB, as we decrease the selectivity, we can pass selective information to \texttt{Knows} table 
and decrease the amount of scanned \texttt{Knows} tuples and hash table probes. 
Therefore we see GRainDB to outperform
DuckDB significantly and close the gap with GraphflowDB's performance at these lower selectivities.
 Second, observe that both GDBMSs have consistent upward curves, indicating that their
runtimes decrease as selectivity on the \texttt{Person} nodes decreases. This happens because the amount of
join work that GDBMSs perform decreases proportionately as fewer \texttt{Person} nodes pass the filter. We cannot observe this desirable
behavior with DuckDB because although its cost of hash table build decreases, its probe cost, which is 
the dominant cost here, does not. In fact, although broadly Neo4j is not competitive with other systems, it can
still outperform DuckDB at lower selectivities on MICRO-P, because of its performance gains from decreasing selectivity on \texttt{Person}. 
Unlike DuckDB, GRainDB however also behaves similarly to GDBMSs and obtains this
desirable behavior because it can also decrease the amount of probes through sip.
Finally, we note that it is expected in this microbenchmark that GraphflowDB is the most performant system 
at all selectivity levels because it can decrease its probes with decreasing selectivity and by default has several
advantages over DuckDB and GRainDB, such as not incurring the cost of building hash tables or accessing
data without going through a buffer manager because it is an in-memory system (which DuckDB does).

We next analyze the results of MICRO-K, shown in Figure~\ref{fig:detailed-selectivity-knows}. First observe that now
the GDBMSs do not react as positively to the decreasing selectivity on \texttt{Knows}. This is because now selectivity on
\texttt{Person} is fixed, so the amount of probes GDBMSs perform is fixed. So both Neo4j and GraphflowDB curves
are relatively straight (similar to the DuckDB curve in Figure~\ref{fig:detailed-selectivity-person}). 
Now note that DuckDB has a downward curve. This is because at all selectivity levels except 0.01\% and 0.01\%, 
DuckDB chooses \texttt{Person} as the build side (recall that the predicate on \texttt{Person} is not selective), so
decreasing the selectivity proportionately decreases the probe amount. 
DuckDB can even outperform GraphflowDB when the selectivity 
is low enough. Note also that as expected GRainDB does not improve the
performance of DuckDB now because although it passes information from \texttt{Person} to \texttt{Knows}, since \texttt{Person}
does not have a selective predicate (it is fixed at 99.9\%), this information is not useful. However, we also do not observe visible overheads. 
We see minor benefits at the lowest two 
selectivity levels, when DuckDB starts to choose \texttt{Knows} as the build side, and can pass selective information
to scans of \texttt{Person}. Although we do not observe major improvements, this shows the flexibility of join processing 
in RDBMS, where there is no notion of node vs edge tables and for hash joins, systems can make any table the probe
or build side. In contrast, in every GDBMS we are aware of, 
first node records are scanned and then the adjacency list indices are probed with the IDs of these nodes 
to perform the join (and not vice versa). As the MICRO-K benchmark demonstrates
this can sometimes prevent them from benefiting from selective predicates on the edge records.


\begin{figure}[ht!]
	\centering
	\captionsetup{justification=centering}
    \begin{subfigure}[b]{0.20\textwidth}
		\centering
		\includegraphics[scale=0.25]{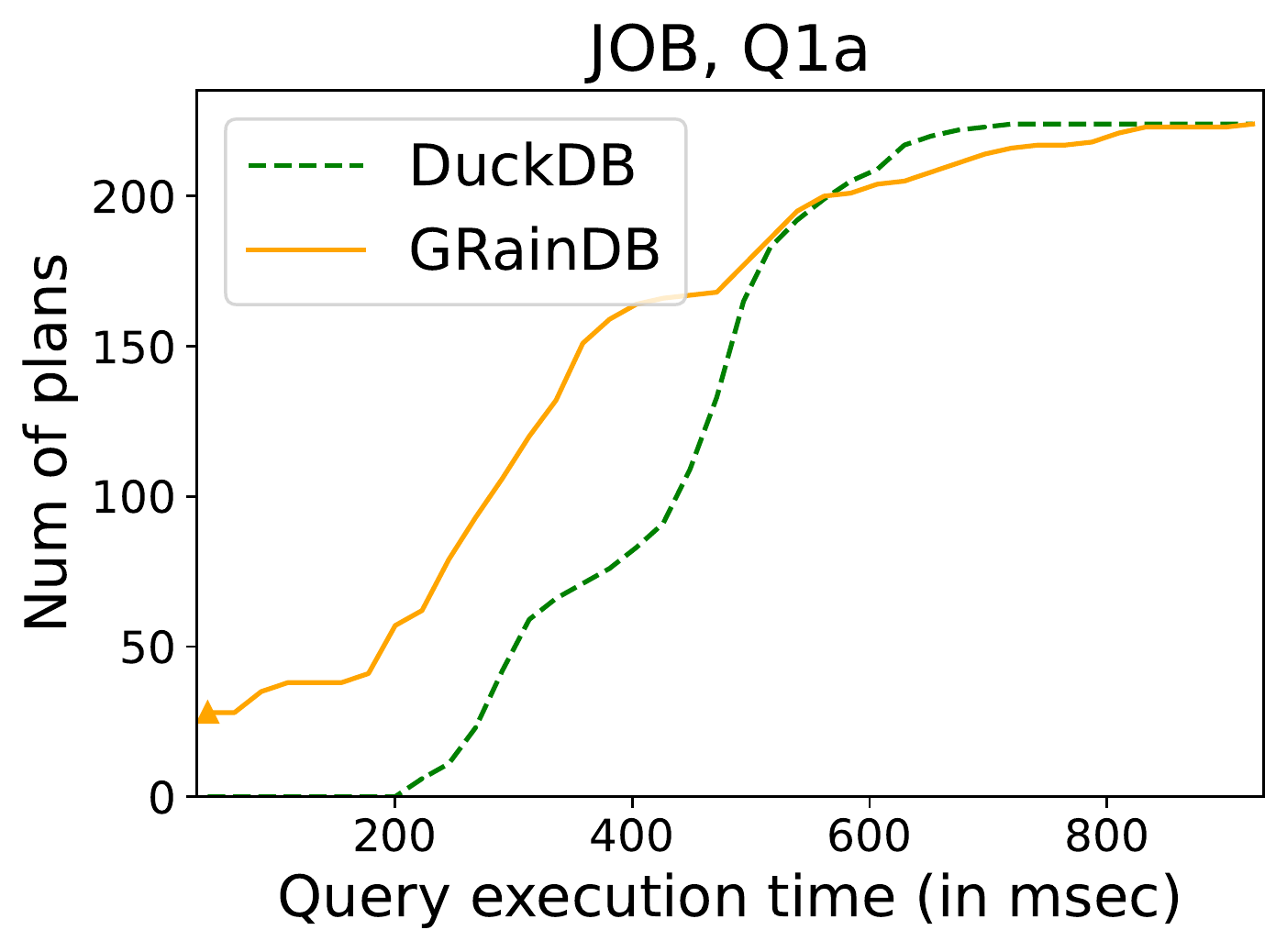}
		\vspace{-2.5mm}
		\label{fig:spectrum-job-q1}
	\end{subfigure}
	\begin{subfigure}[b]{0.20\textwidth}
		\centering
		\includegraphics[scale=0.25]{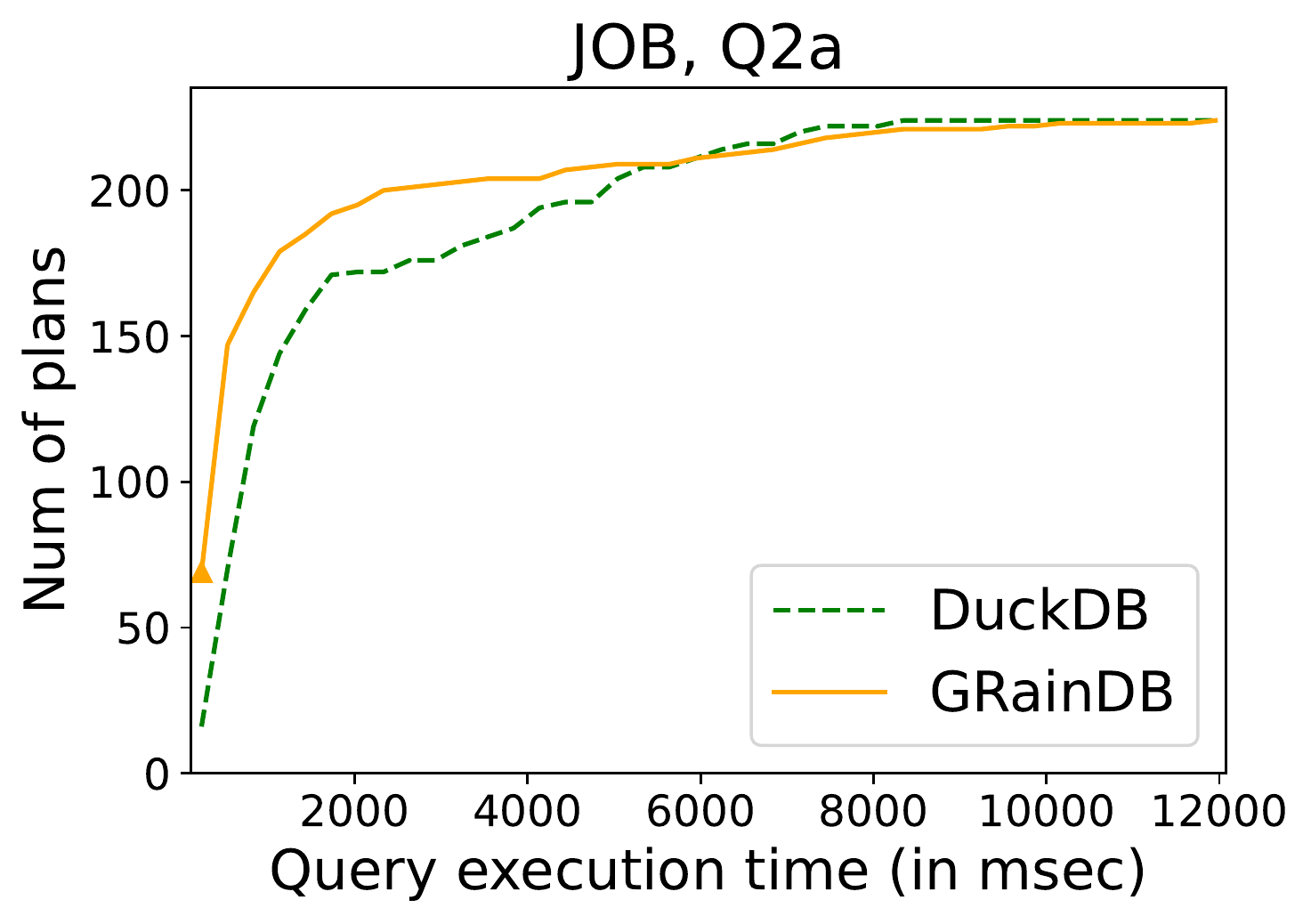}
		\vspace{-2.5mm}
		\label{fig:spectrum-job-q2}
	\end{subfigure}
	\begin{subfigure}[b]{0.20\textwidth}
		\centering
		\includegraphics[scale=0.251]{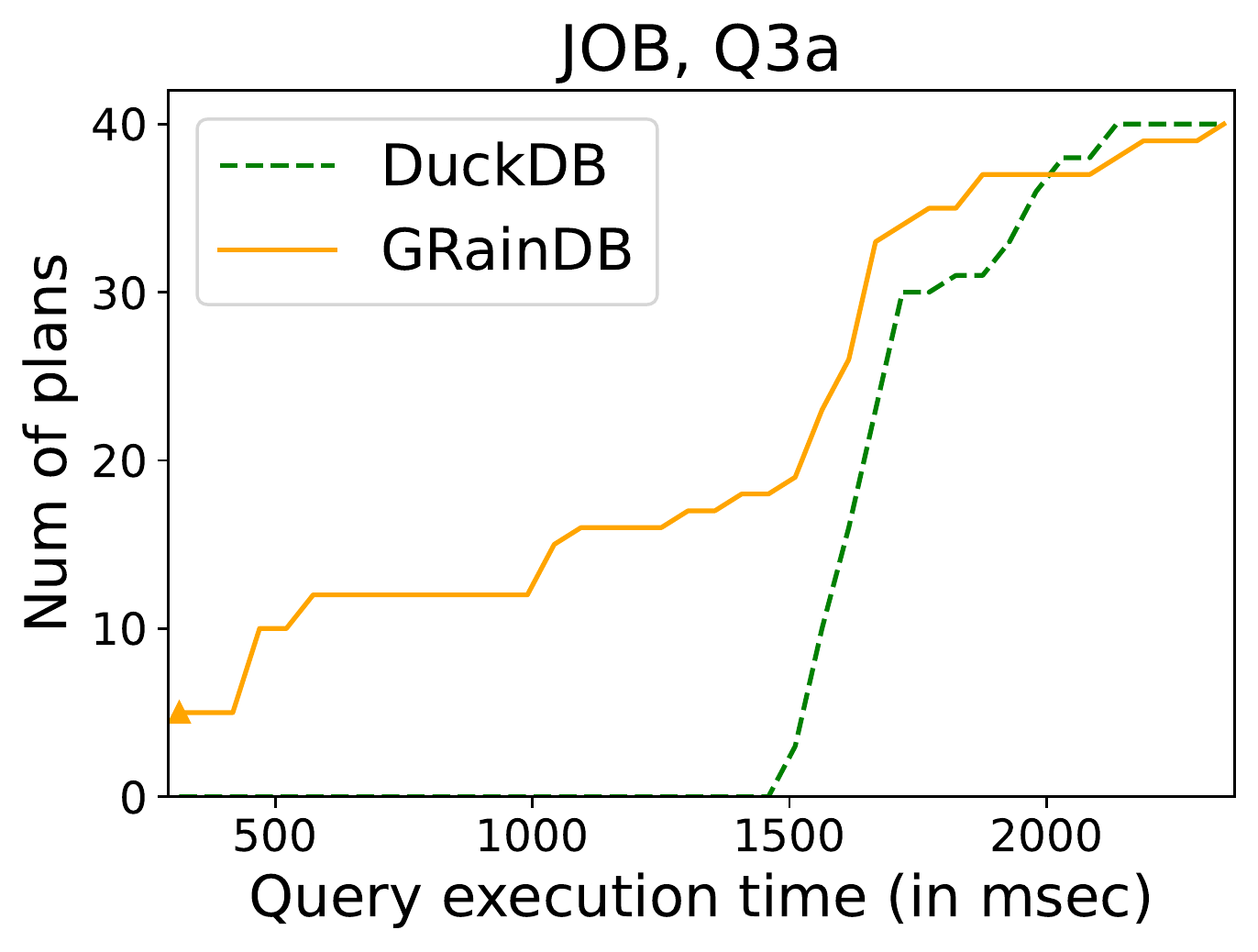}
		\vspace{-2.5mm}
		\label{fig:spectrum-job-q3}
	\end{subfigure}
	\begin{subfigure}[b]{0.20\textwidth}
		\centering
		\includegraphics[scale=0.25]{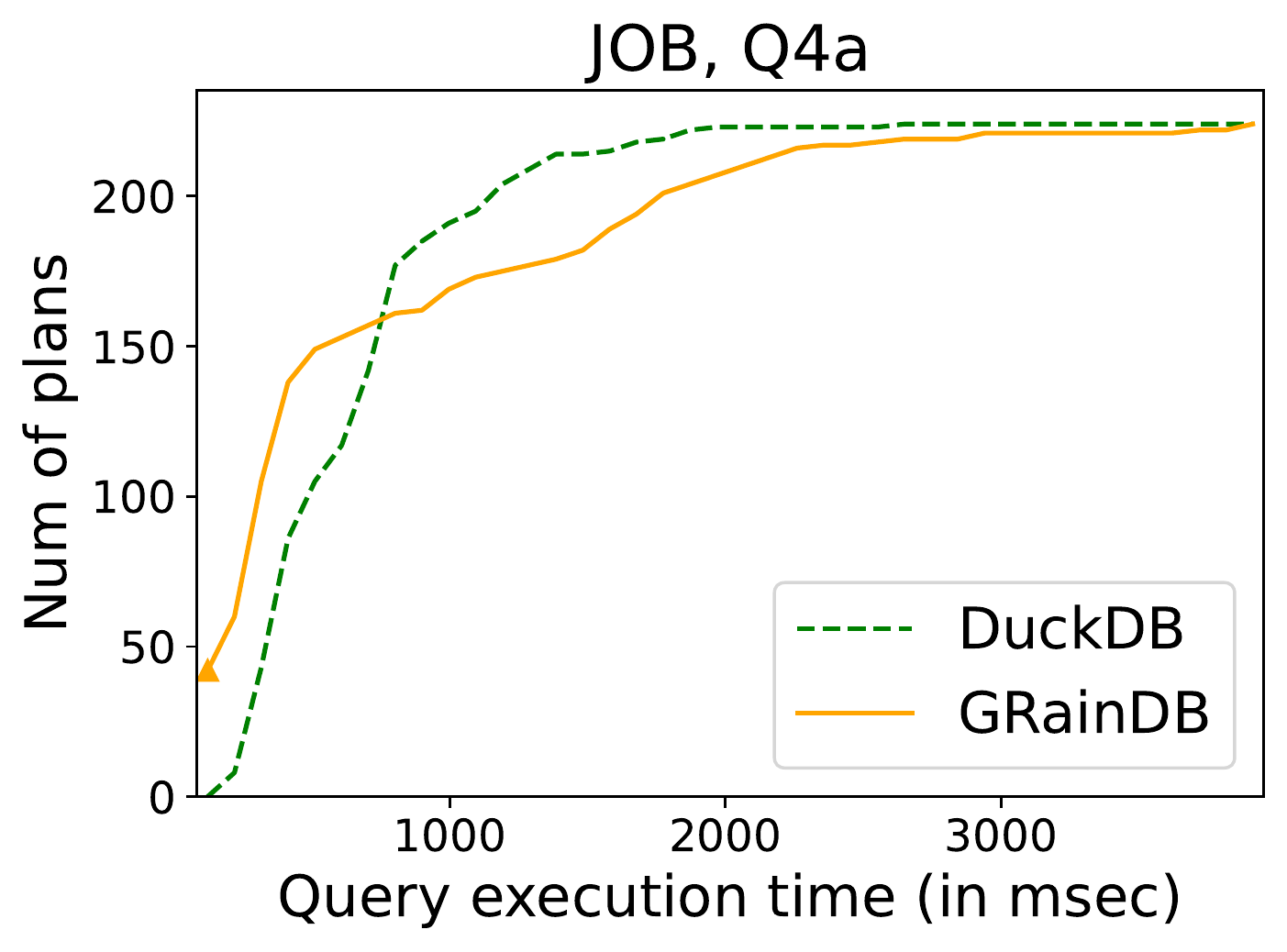}
		\vspace{-2.5mm}
		\label{fig:spectrum-job-q4}
	\end{subfigure}
	\begin{subfigure}[b]{0.20\textwidth}
		\centering
		\includegraphics[scale=0.25]{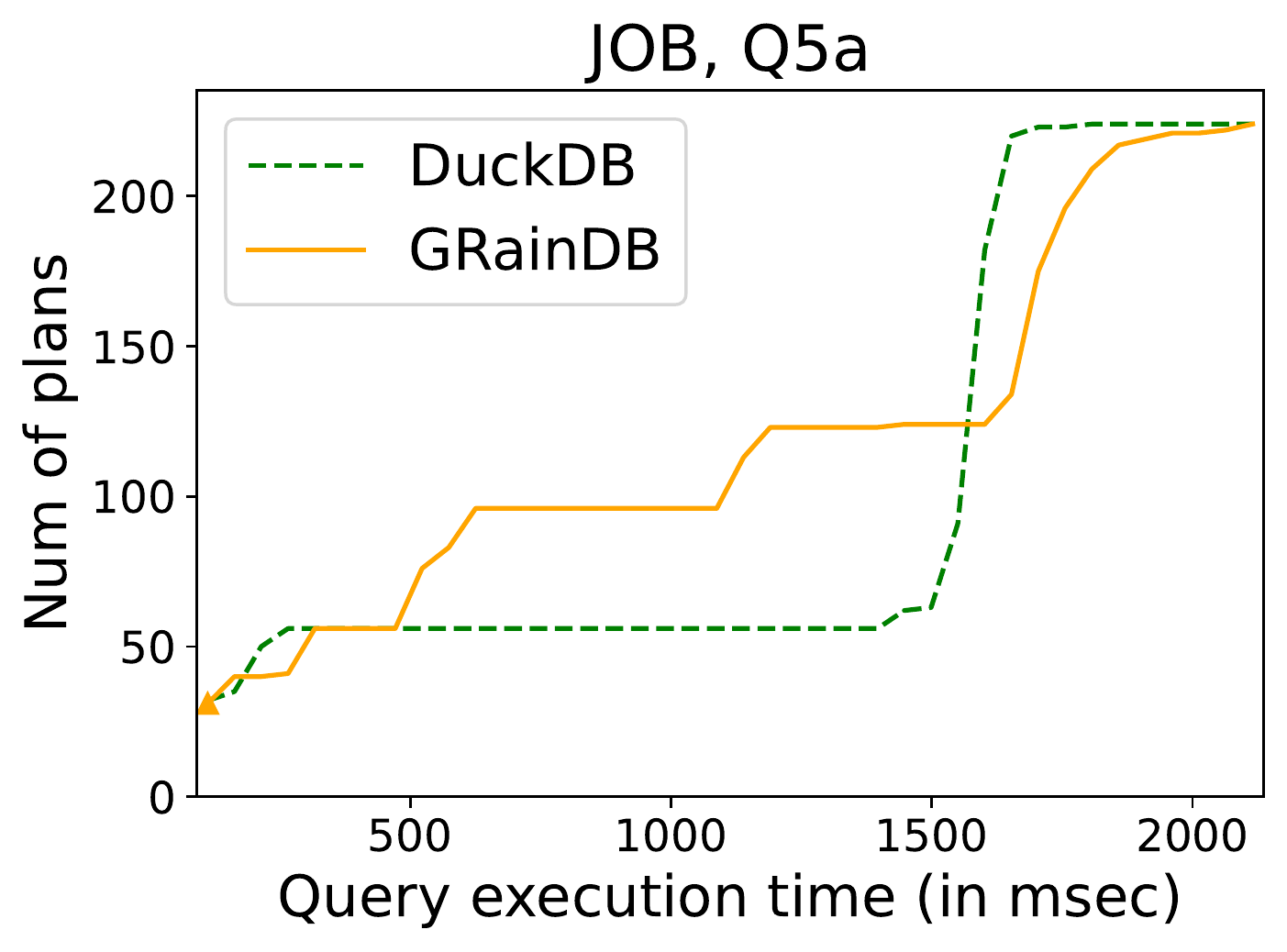}
		\vspace{-2.5mm}
		\label{fig:spectrum-job-q5}
	\end{subfigure}
	\begin{subfigure}[b]{0.20\textwidth}
		\centering
		\includegraphics[scale=0.25]{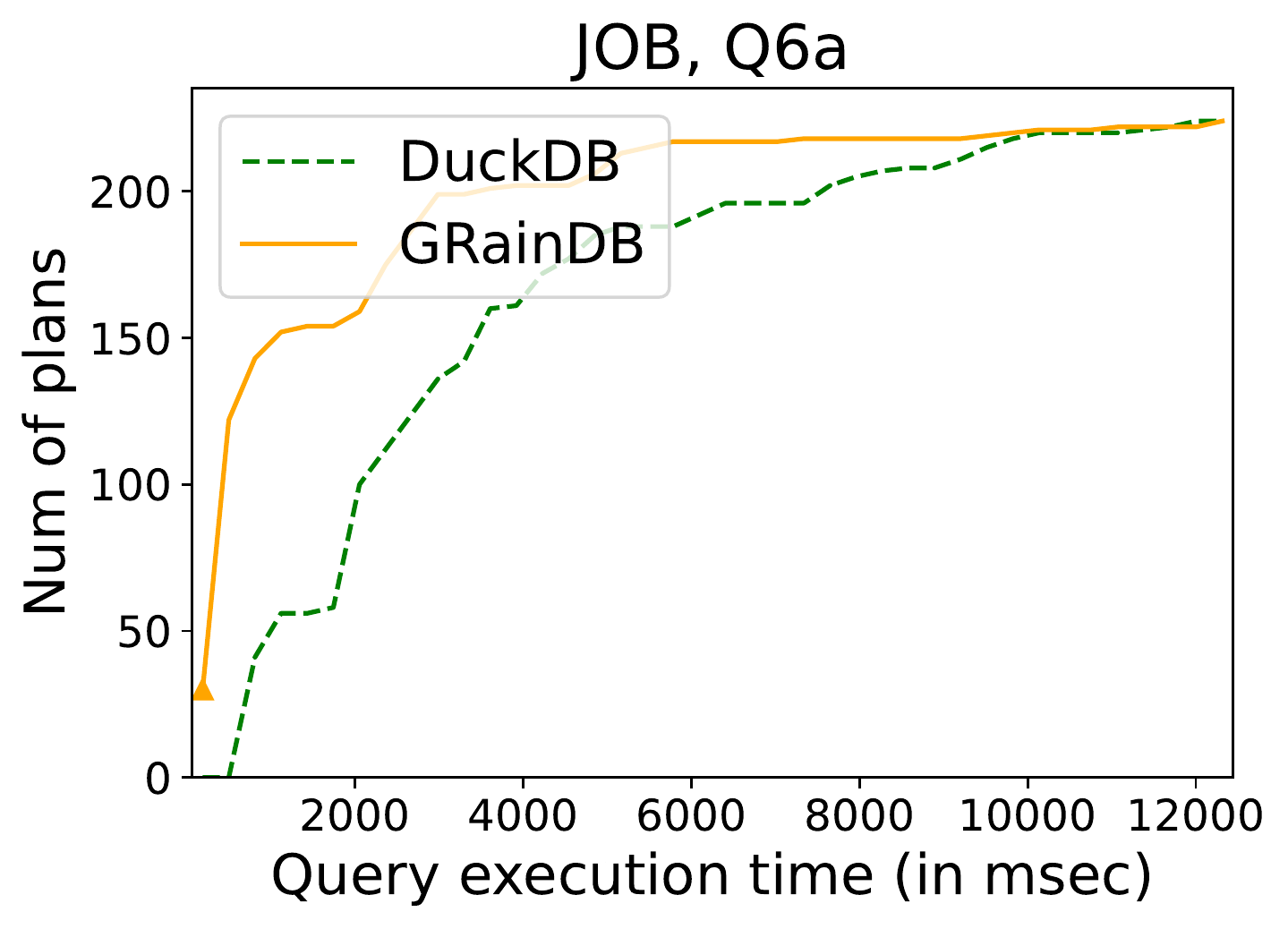}
		\vspace{-2.5mm}
		\label{fig:spectrum-job-q6}
	\end{subfigure}
	\vspace{-12pt}
	\caption{Cumulative distributions of the number of DuckDB and GRainDB plans (y-axis) that have runtimes below different thresholds (x-axis).}
    \vspace{-22pt}
	\label{fig:plan-spectrum}
\end{figure}

\vspace{-10pt}
\subsubsection{Effectiveness of Predefined Joins in the Plan Space and Room for Improvement for a Sip-aware Optimizer.}
\label{subsubsec:evaluation-join-order}
Our next set of experiments analyze the effects predefined joins have on the plan space of DuckDB. 
Prior work has observed that sip-based~\cite{zhu:lip} query processing makes systems broadly more 
robust to join order selection by decreasing the performance differences between different join orders
and more importantly by increasing the number of join orders that perform competitive with the best default order. 
To demonstrate that our proposed solution also has similar effects on the plan space, 
we picked the first six query groups in JOB, and  for the first two variants of each query (so a total of 12 queries) 
performed a plan spectrum 
analysis as follows. We take each plan $P$ for each $Q$, corresponding to one join order, and execute 
both the default version of $P$ (call it $P_d$) and the version where we apply our optimizations to change
value-based joins with sip-based predefined joins (call it $P_d^*$). 
We then plotted two cumulative distribution lines for 
$Q$, one for the set of $P_d$ and one for $P_d^*$ plans, which show the number of plans (on y-axis) for different 
runtime value cutoffs, e.g., 100ms, 200ms, ..., 1000ms (on x-axis).  Figure~\ref{fig:plan-spectrum} shows the 
distributions we obtained for the first variants of these queries. 
\iflong
The remaining charts are shown in Appendix~\ref{app:cumulative-distribution}.
\else
The remaining 6 charts, which are similar to the ones in Figure~\ref{fig:plan-spectrum}, can be found in the longer version of our paper~\cite{graindb:long-version}.
\fi
Dashed and straight lines are the distributions of $P_d$ and $P_d^*$ plans, respectively.
We observe on left sides of the curves, which summarize the best performing plans, 
the line showing  $P_d^*$ is consistently above the line for $P_d$. 
This shows that by predefining joins, we obtain larger sets of good plans. In many of these queries 
we also observe runtimes that were not achievable by any default plan. For example, on Q1a, while there are 
60 plans with a runtime of $\le 200$ms under predefined joins, there is no such plan 
with default value-based joins. In addition, there are now a set of 35 plans with runtime of $\le 100$ms.
Therefore not only are the best plans under predefined joins more performant,
the optimizer is more robust to making mistakes when picking 
a plan, as there is a larger set of good performing plans. 
We also observe that on the right ends of many curves, 
which plot the set of worst-performing plans, the curve for $P_d^*$ plans is now below the curve for $P_d$ 
plans. This is also expected because we expect there to be some plans that do not benefit from 
predefined joins and instead incur minor overheads that \texttt{S-Join} 
operators incur, e.g., to prepare bitmasks.

\begin{table}[!t]
	\setlength{\tabcolsep}{1.3pt}
	\centering
	\small
	\begin{tabular}{ c|c|c|c|c|c|c|c|c|c|c|c|c }
		& \textbf{Q1a} & \textbf{Q1b} & \textbf{Q2a} & \textbf{Q2b} & \textbf{Q3a} & \textbf{Q3b} & \textbf{Q4a} & \textbf{Q4b} & \textbf{Q5a} & \textbf{Q5b} & \textbf{Q6a} & \textbf{Q6b} \\
		\hline

		{\texttt{$P_{Duck}^*$}}& 34 & 3 & 154 & 143 & 328 & 502 & 114 & 73 & 116 & 146 & 57 & 97 \\
		{\texttt{$P_{opt}^*$}}& 31 & 3 & 77 & 67 & 287 & 135 & 72 & 47 & 110 & 112 & 32 & 82 \\
	\end{tabular}
	\vspace{0.1cm}
	\captionsetup{justification=centering}
	\caption{Runtimes (in ms) of $P_{Duck}^*$ and $P_{opt}^*$ on JOB queries.}
	\label{tab:detailed-plan-space}
	\vspace{-30pt}
\end{table}

Let us call the $P_d^*$ of a plan $P_d$ the {\em predefined version} of $P_d$.
Recall that for a query $Q$, GRainDB's plan, $P_{Duck}^*$ is the predefined version of the plan $P_{Duck}$ 
that DuckDB's default optimizer picks for $Q$. Next, on these 12 queries, we analyzed the potential room for 
improvement on our rule-based approach if a system implements a sip-aware optimizer. We do a thought process and
assume that an oracle sip-aware optimizer could pick the best GRainDB plan $P_{opt}^*$, i.e., the best performing
$P_d^*$, and compare it against $P_{Duck}^*$. 
Table~\ref{tab:detailed-plan-space} shows this comparison.
Although we did not find large rooms of improvement on most of these queries, 
 we still found several queries, Q2a, Q2b, and Q3b with >2x improvements. The largest 
 improvement is on Q3b, from 502ms to 135ms (3.7x). 
Q3b is a join query with 4 tables, and contains a selective predicate on a table called \texttt{keyword} that returns 
only 30 of the 134K tuples in this table. 
$P_{opt}^*$ is a left-deep plan where the last-join has \texttt{keyword} on its build side. Normally 
putting this table as the last join on a left-deep plan is not efficient because joining the smaller
tables first and creating smaller intermediate results is more efficient. However, under sip, 
this is a good plan because this can lead
to iterative information passing to reduce the amount of scans in other tables. Instead $P_{Duck}^*$ uses a bushy plan. 
In principle, a sip-aware optimizer that can accurately estimate the selectivities of our zone and bit mask filters 
can further improve our performance by generating such plans that are normally not efficient but become efficient
due to sip. However, our analysis indicates that our rule-based approach also generates
competitive plans on almost all of these queries.

%% file: conclusion.tex
\vspace{-7pt}
\section{Conclusions}
We described a novel approach to integrate predefined and pointer-based joins, which
are prevalent in GDBMSs, into columnar
RDBMSs. Our approach is based on materializing and optionally indexing RIDs similar to
how edges are indexed in adjacency lists.
In contrast to native GDBMSs and prior implementations of predefined joins in RDBMSs~\cite{hassan:grfusion-edbt,
hassan:grfusion, lin:gq-fast, lin:gq-fast-demo} that use such indices in index nested loop joins,
we use them primarily to generate semi-join filters that are passed from hash join operators to scans.
This ensures sequential scans when results of joins need to access other properties of joined tuples
for further processing, such as running predicates. 
Unlike prior approaches that propose a graph-specific optimizer and query processor that produce
left-deep join plans, our approach directly leverages the default optimizer of the system to generate an
arbitrary and possibly bushy plan, and transforms this plan by replacing some join and scan operators. 
We also described an optimization that can use an extended RID index
to avoid scans of a relationship table entirely in some settings.
We demonstrated the practicality of our approach by implementing it in the DuckDB system~\cite{raasveldt:duckdb, raasveldt:duckdb-cidr} 
and demonstrated its performance benefits on both relational as well as graph 
workloads that contain queries with large selective many-to-many
joins that can benefit from predefinition, such as the LDBC SNB benchmark on which our approach
makes DuckDB competitive with the state-of-the-art GraphflowDB GDBMS. 

%% file: appendix.tex
\begin{appendix}

\section{SNB-M Queries}
\label{app:snb-m}
For completeness, we include all modified LDBC SNB queries used in our evaluation.

\begin{query}
\noindent IS1
{\em 
\begin{lstlisting}[numbers=none, showstringspaces=false, belowskip=0pt ]
SELECT p.firstname, p.lastname, p.birthday, p.locationip,
 p.browserused, pl.placeid, p.gender, p.creationdate
FROM person p, place pl
WHERE person.id=933 AND person.placeid=place.placeid;
\end{lstlisting}
}
\end{query}

\begin{query}
\noindent IS2
{\em 
\begin{lstlisting}[numbers=none, showstringspaces=false, belowskip=0pt ]
SELECT m1.id, m1.creationdate, m2.id, p2.personid,
 p2.firstname, p2.lastname
FROM person p1, comment m1, post m2, person p2
WHERE m2.creatorid=p2.personid AND m1.replyof_post=m2.id
 AND m1.creatorid=p1.personid AND p1.id=933;
\end{lstlisting}
}
\end{query}

\begin{query}
\noindent IS3
{\em 
\begin{lstlisting}[numbers=none, showstringspaces=false, belowskip=0pt ]
SELECT p2.personid, p2.firstname, p2.lastname, k.creationdate
FROM knows k, person p1, person p2
WHERE p1.id=933 AND p1.personid=k.person1id
 AND k.person2id=p2.personid;
\end{lstlisting}
}
\end{query}

\begin{query}
\noindent IS4
{\em 
\begin{lstlisting}[numbers=none, showstringspaces=false, belowskip=0pt ]
SELECT c.content, c.creationdate
FROM comment c
WHERE id=4947802324993;
\end{lstlisting}
}
\end{query}

\begin{query}
\noindent IS5
{\em 
\begin{lstlisting}[numbers=none, showstringspaces=false, belowskip=0pt ]
SELECT p.personid, p.firstname, p.lastname
FROM comment c, person p
WHERE c.id=4947802324993 AND c.creatorid=p.personid;
\end{lstlisting}
}
\end{query}

\begin{query}
\noindent IS6
{\em 
\begin{lstlisting}[numbers=none, showstringspaces=false, belowskip=0pt ]
SELECT f.forumid, f.title, p.personid, p.firstname, p.lastname
FROM comment m1, post m2, person p, forum f
WHERE m1.id=4947802324993 AND m1.replyof_post=m2.id
  and m2.forumid=f.forumid AND f.moderatorid=p.personid;
\end{lstlisting}
}
\end{query}

\begin{query}
\noindent IS7
{\em 
\begin{lstlisting}[numbers=none, showstringspaces=false, belowskip=0pt ]
SELECT m2.id, m2.content, m2.creationdate, p.personid,
 p.firstname, p.lastname
FROM comment m1, comment m2, person p
WHERE m1.id=8246337208329 AND m2.replyof_comment=m1.id
  AND m2.creatorid=p.personid;
\end{lstlisting}
}
\end{query}

\begin{query}
\noindent IC1-1
{\em 
\begin{lstlisting}[numbers=none, showstringspaces=false, belowskip=0pt ]
SELECT p2.id, p2.lastname, p2.birthday, p2.creationdate,
 p2.gender, p2.browserused, p2.locationip, pl.name
FROM person p1, knows k, person p2, place pl
WHERE p1.personid=k.person1id AND k.person2id=p2.personid
 AND p2.placeid=pl.placeid AND p1.id=933 AND
 p2.firstname='Rahul';
\end{lstlisting}
}
\end{query}

\begin{query}
\noindent IC1-2
{\em 
\begin{lstlisting}[numbers=none, showstringspaces=false, belowskip=0pt ]
SELECT p2.id, p2.lastname, p2.birthday, p2.creationdate,
 p2.gender, p2.browserused, p2.locationip, pl.name
FROM person p1, knows k1, knows k2, person p2, place pl 
WHERE p1.id=933 and p2.firstname='Rahul'
 AND p1.personid=k1.person1id AND k1.person2id=k2.person1id 
 AND k2.person2id=p2.personid AND p2.placeid=pl.placeid;
\end{lstlisting}
}
\end{query}

\begin{query}
\noindent IC1-3
{\em 
\begin{lstlisting}[numbers=none, showstringspaces=false, belowskip=0pt ]
SELECT p2.id, p2.lastname, p2.birthday, p2.creationdate, 
 p2.gender, p2.browserused, p2.locationip, pl.name
FROM person p1, knows k1, knows k2, knows k3,
 person p2, place pl
WHERE p1.id=933 and p2.firstname='Rahul'
 AND p1.personid=k1.person1id AND k1.person2id=k2.person1id 
 AND k2.person2id=k3.person1id AND k3.person2id=p2.personid
 AND p2.placeid=pl.placeid;
\end{lstlisting}
}
\end{query}

\begin{query}
\noindent IC2
{\em 
\begin{lstlisting}[numbers=none, showstringspaces=false, belowskip=0pt ]
SELECT p2.id, p2.firstname, p2.lastname, c.id,
 c.content, c.creationdate
FROM person p1, knows k, person p2, comment c 
WHERE p2.personid=c.creatorid AND c_creationdate<1338552000
 AND k.person2id=p2.personid AND p1.personid=k.person1id
 AND p1.id=933;
\end{lstlisting}
}
\end{query}

\begin{query}
\noindent IC3-1
{\em 
\begin{lstlisting}[numbers=none, showstringspaces=false, belowskip=0pt ]
SELECT p2.id, p2.firstname, p2.lastname
FROM person p1, knows k1, person p2, comment m1, 
 place pl1, comment m2, place pl2
WHERE m1.creationdate>=1313591219 AND m1.creationdate<1513591219
 AND m2.creationdate>=1313591219 AND m2.creationdate<1513591219
 AND p1.personid=k1.person1id AND k1.person2id=p2.personid
 AND m2.creatorid=p2.personid AND m1.locationid=pl1.placeid
 AND m1.creatorid=p2.personid AND m2.locationid=pl2.placeid
 AND p1.id=933 AND pl1.name='India' AND pl2.name='China';
\end{lstlisting}
}
\end{query}

\begin{query}
\noindent IC3-2
{\em 
\begin{lstlisting}[numbers=none, showstringspaces=false, belowskip=0pt ]
SELECT p2.id, p2.firstname, p2.lastname
FROM person p1, knows k1, knows k2, person p2, comment m1, 
 place pl1, comment m2, place pl2
WHERE m2.creationdate>=1313591219 AND m2.creationdate<1513591219 
 AND m1.creationdate>=1313591219 AND m1.creationdate<1513591219
 AND p1.id=933 AND pl1.name='India' AND pl2.name='China'
 AND p1.personid=k1.person1id AND k2.person1id=k1.person2id
 AND k2.person2id=p2.personid AND m2.creatorid=p2.personid
 AND m1.locationid=pl1.placeid AND m1.creatorid=p2.personid
 AND m2.locationid=pl2.placeid;
\end{lstlisting}
}
\end{query}

\begin{query}
\noindent IC4
{\em 
\begin{lstlisting}[numbers=none, showstringspaces=false, belowskip=0pt ]
SELECT t_name
FROM knows k1, person p1, knows k2, person p2, post ps,
 post_tag mt, tag t
WHERE mt.tagid=t.tagid AND ps.id=mt.messageid AND
 p2.personid=ps.creatorid AND k2.person2id=p2.personid 
 AND p1.personid=k2.person1id AND p1.personid=k1.person1id 
 AND p1.id=933 AND ps.creationdate>=1313591219 
 AND ps.creationdate<1513591219;
\end{lstlisting}
}
\end{query}

\begin{query}
\noindent IC5-1
{\em 
\begin{lstlisting}[numbers=none, showstringspaces=false, belowskip=0pt ]
SELECT f.title
FROM person p1, knows k1, person p2, forum_person fp, forum f,
 post m
WHERE f.forumid=m.forumid AND fp.forumid=f.forumid 
 AND p2.personid=fp.personid AND k1.person2id=p2.personid 
 AND p1.personid=k1.person1id AND p1.id=933
 AND fp.joindate>=1353819600;
\end{lstlisting}
}
\end{query}

\begin{query}
\noindent IC5-2
{\em 
\begin{lstlisting}[numbers=none, showstringspaces=false, belowskip=0pt ]
SELECT f.f_title
FROM person p1, knows k1, knows k2, person p2, forum_person fp,
 forum f, post m
WHERE f.forumid=m.forumid AND fp.forumid=f.forumid 
 AND p2.personid=fp.personid AND k2.person2id=p2.personid 
 AND k1.person2id=k2.person1id
 AND p1.personid=k1.person1id AND p1.id=933
 AND fp.joindate>=1353819600;
\end{lstlisting}
}
\end{query}

\begin{query}
\noindent IC6-1
{\em 
\begin{lstlisting}[numbers=none, showstringspaces=false, belowskip=0pt ]
SELECT t2.t_name
FROM person p1, knows k1, person p2, post m, post_tag mt1,
 tag t1, post_tag mt2, tag t2
WHERE mt1.tagid=t1.tagid AND m.id=mt1.messageid AND
 mt2.tagid=t2.tagid AND m.id=mt2.messageid AND 
 m.creatorid=p2.personid AND k1.person2id=p2.personid
 AND p1.personid=k1.person1id AND p1.id=933 AND 
 t1.t_name='Rumi' AND t2.t_name!='Rumi';
\end{lstlisting}
}
\end{query}

\begin{query}
\noindent IC6-2
{\em 
\begin{lstlisting}[numbers=none, showstringspaces=false, belowskip=0pt ]
SELECT t2.t_name
FROM person p1, knows k1, knows k2, person p2, post m,
 post_tag mt1, tag t1, post_tag mt2, tag t2
WHERE mt1.tagid=t1.tagid AND m.id=mt1.messageid
 AND mt2.tagid=t2.tagid AND m.id=mt2.messageid
 AND m.creatorid=p2.personid AND k2.person2id=p2.personid
 AND k1.person2id=k2.person1id AND p1.personid=k1.person1id
 AND p1.id=933 AND t1.t_name='Rumi' AND t2.t_name!='Rumi';
\end{lstlisting}
}
\end{query}

\begin{query}
\noindent IC7
{\em 
\begin{lstlisting}[numbers=none, showstringspaces=false, belowskip=0pt ]
SELECT p2.personid, p2.firstname, p2.lastname,
 l.creationdate, c.content
FROM person p1, comment c, likes_comment l, person p2
WHERE p2.personid=l.personid AND c.id=l.messageid
 AND c.creatorid=p1.personid AND p1.id=933;
\end{lstlisting}
}
\end{query}

\begin{query}
\noindent IC8
{\em 
\begin{lstlisting}[numbers=none, showstringspaces=false, belowskip=0pt ]
SELECT c.creatorid, p2.firstname, p2.lastname, c.creationdate,
 c.id, c.content
FROM person p1, post ps, comment c, person p2
WHERE c.creatorid=p2.personid AND c.replyof_post=ps.id AND
 p1.personid=ps.creatorid AND p1.personid=933;
\end{lstlisting}
}
\end{query}

\begin{query}
\noindent IC9-1
{\em 
\begin{lstlisting}[numbers=none, showstringspaces=false, belowskip=0pt ]
SELECT p2.firstname, p2.lastname, c.creationdate
FROM person p1, knows k1, person p2, comment c
WHERE p2.personid=c.creatorid AND k1.person2id=p2.personid
 AND p1.personid=k1.person1id AND p1.id=933
 AND c.creationdate<1342840042;
\end{lstlisting}
}
\end{query}

\begin{query}
\noindent IC9-2
{\em 
\begin{lstlisting}[numbers=none, showstringspaces=false, belowskip=0pt ]
SELECT p2.firstname, p2.lastname, c.creationdate
FROM person p1, knows k1, knows k2, person p2, comment c
WHERE p2.personid=c.creatorid AND k2.person2id=p2.personid
 AND k1.person2id=k2.person1id AND p1.personid=k1.person1id 
 AND p1.id=933 AND c.creationdate<1342840042;
\end{lstlisting}
}
\end{query}


\begin{query}
\noindent IC11-1
{\em 
\begin{lstlisting}[numbers=none, showstringspaces=false, belowskip=0pt ]
SELECT p2.id, p2.firstname, p2.lastname, o.name, pc.workfrom
FROM person p1, knows k1, person p2, person_company pc,
 organisation o, place pl
WHERE o.placeid=pl.placeid AND pc.organisationid=o.organisationid
 AND p2.personid=pc.personid AND k1.person2id=p2.personid
 AND p1.personid=k1.person1id AND p1.id=933
 AND pc.workfrom<2016 AND pl.name='China';
\end{lstlisting}
}
\end{query}

\begin{query}
\noindent IC11-2
{\em 
\begin{lstlisting}[numbers=none, showstringspaces=false, belowskip=0pt ]
SELECT p2.id, p2.firstname, p2.lastname, o.name, pc.workfrom
FROM person p1, knows k1, knows k2, person p2,
 person_company pc, organisation o, place pl
WHERE p1.id=933 AND pc.workfrom<2016 AND pl.name='China'
 AND p2.personid=pc.personid AND k1.person2id=k2.person1id
 AND p1.personid=k1.person1id AND k2.person2id=p2.personid
 AND o.placeid=pl.placeid AND pc.organisationid=o.organisationid;
\end{lstlisting}
}
\end{query}

\begin{query}
\noindent IC12
{\em 
\begin{lstlisting}[numbers=none, showstringspaces=false, belowskip=0pt ]
SELECT f.personid, friend.p_firstname, friend.p_lastname 
FROM person p1, knows k, person f, comment c, post ps,
 post_tag pt, tag t, tagclass tc1, tagclass tc2
WHERE tc1.subclassoftagclassid=tc2.tagclassid
 AND t.tagclassid=tc1.tagclassid AND mt.tagid=tag.tagid
 AND c.replyof_post=ps.id AND c.creatorid=f.personid
 AND ps.id=mt.messageid AND k.person2id=f.personid AND p1.id=933
 AND p1.personid=k.person1id AND tc2.tc_name='Person';
\end{lstlisting}
}
\end{query}

\section{DuckDB and GRainDB With Default and Optimized Join Orders}
\label{app:duckdb-optimized}
\begin{figure}[!t]
  \centering
  \captionsetup{justification=centering}
  \includegraphics[width=\linewidth]{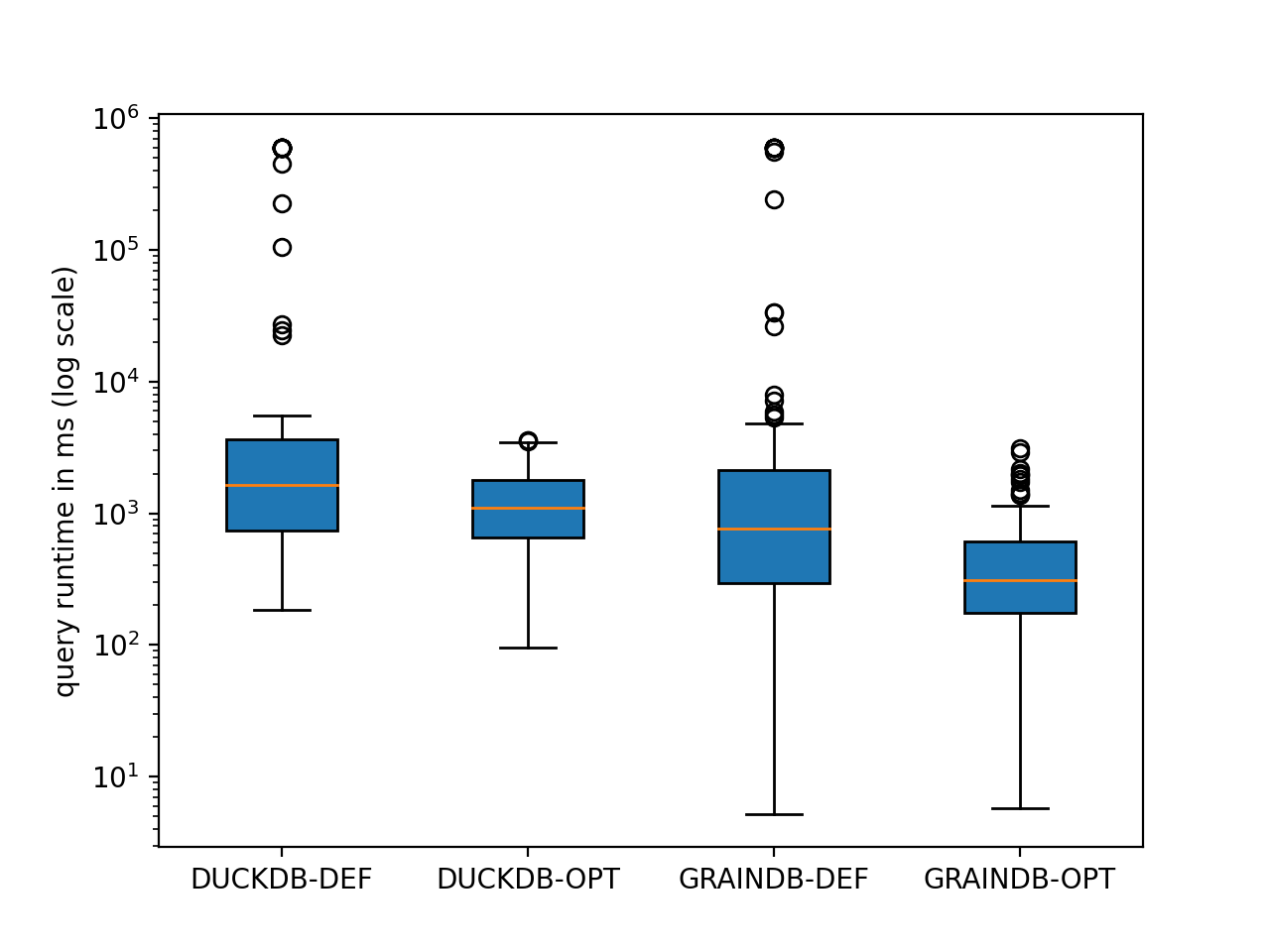}
  \vspace{-10pt}
  \caption{The query runtime (ms) of DuckDB and GRainDB on JOB with the system-default (DUCKDB-DEF, GRAINDB-DEF) and optimized (DUCKDB-OPT, GRAINDB-OPT) join orders. The boxplot shows the 5th, 25th, 50th, 75th, and 95th percentiles.}
  \vspace{-10pt}
  \label{fig:optimized-plans}
\end{figure}

To isolate the influence of join order selection, instead of using DuckDB's default join orders, we injected true cardinalities into the system to generate optimized join orders.
Figure~\ref{fig:optimized-plans} shows the boxplots of running DuckDB on JOB with default and optimized join orders.
We see with optimized join orders, the system reduces outlied runtimes largely, and improves the query performance in general.
For 113 queries in JOB, DuckDB performs timeouts on 15 of them under default join orders, while 0 under optimized join orders.
And the 25th percentile, median, and 75th percentile query execution times reduce respectively from 741.0ms to 652.4ms (1.4x), from 1656.4ms to 1110ms (1.5x), and from 3624.6 to 1797.0ms (2.0x). 
Recall that in GRainDB, we directly apply the same join order as DuckDB.
And the performance improvement due to optimized join orders applies on GRainDB uniformly.
The number of timeout queries in GRainDB reduces from 9 to 0 after replacing default join orders with optimized ones.
And the 25th percentile, median, and 75th percentile query execution times reduce respectively from 2125.0ms to 614.2ms (3.5x), from 770.0ms to 309.0ms (2.5x), and from 295.6 to 176.4ms (1.7x).
Moreover, noticeably in either default or optimized join orders, GRainDB shows better performance over DuckDB.

\section{Query Execution Time of DuckDB and GRainDB on JOB}
\label{app:job-tpch-full}
In Section~\ref{subsubsec:evaluation-job}, we presented box plots and detailed percentiles for query execution times of DuckDB and GRainDB on all JOB queries, and also execution times of the first variant of each query. In Table~\ref{tab:job-full}, we show detailed query execution time of DuckDB and GRainDB on all 113 queries on JOB.

\section{Bushy and Left-deep Plans for SNB-M IC6-2}
\label{app:ic6-2-plans}
\begin{figure}
\centering
    \includegraphics[scale=0.08]{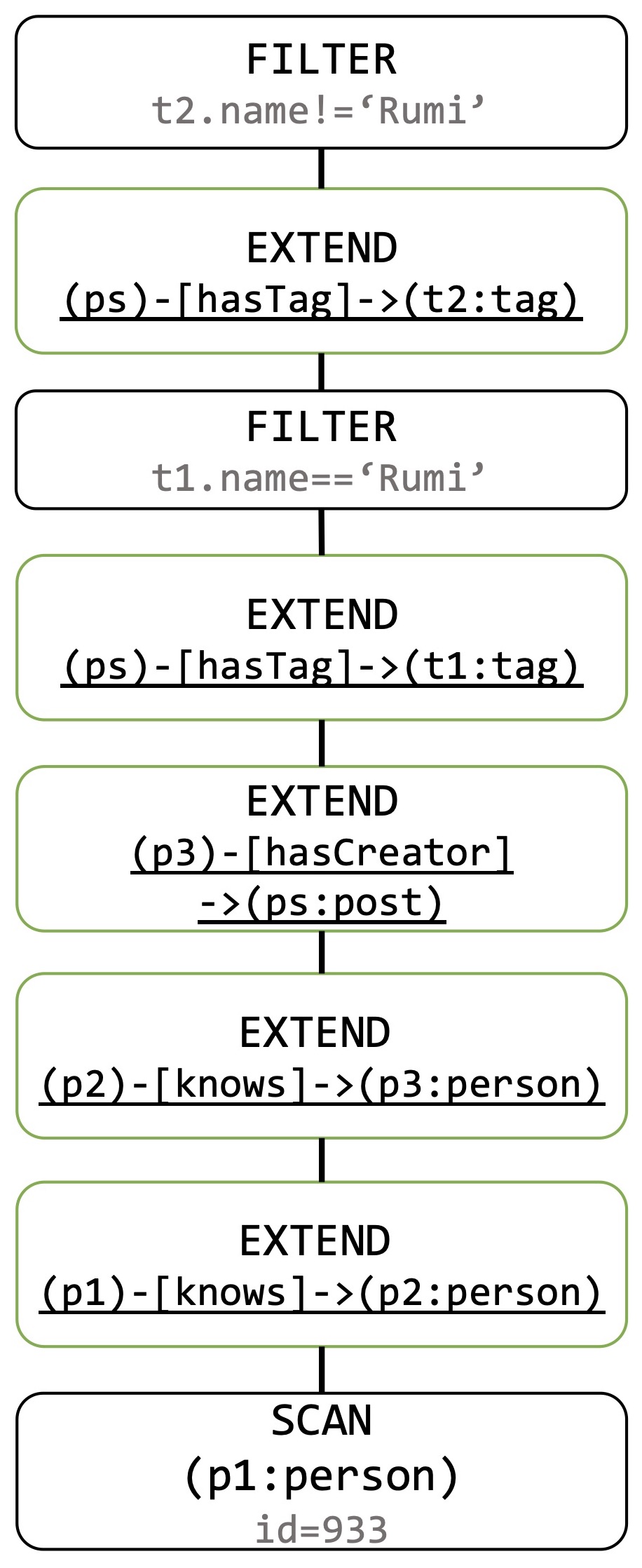}
    \caption{The Left-deep Plan in GraphflowDB. The final Projection is omitted.}
    \vspace{12pt}
    \label{fig:plan-gfdb}
\vspace{-10pt}
\end{figure}

\begin{figure}
\centering
    \includegraphics[scale=0.08]{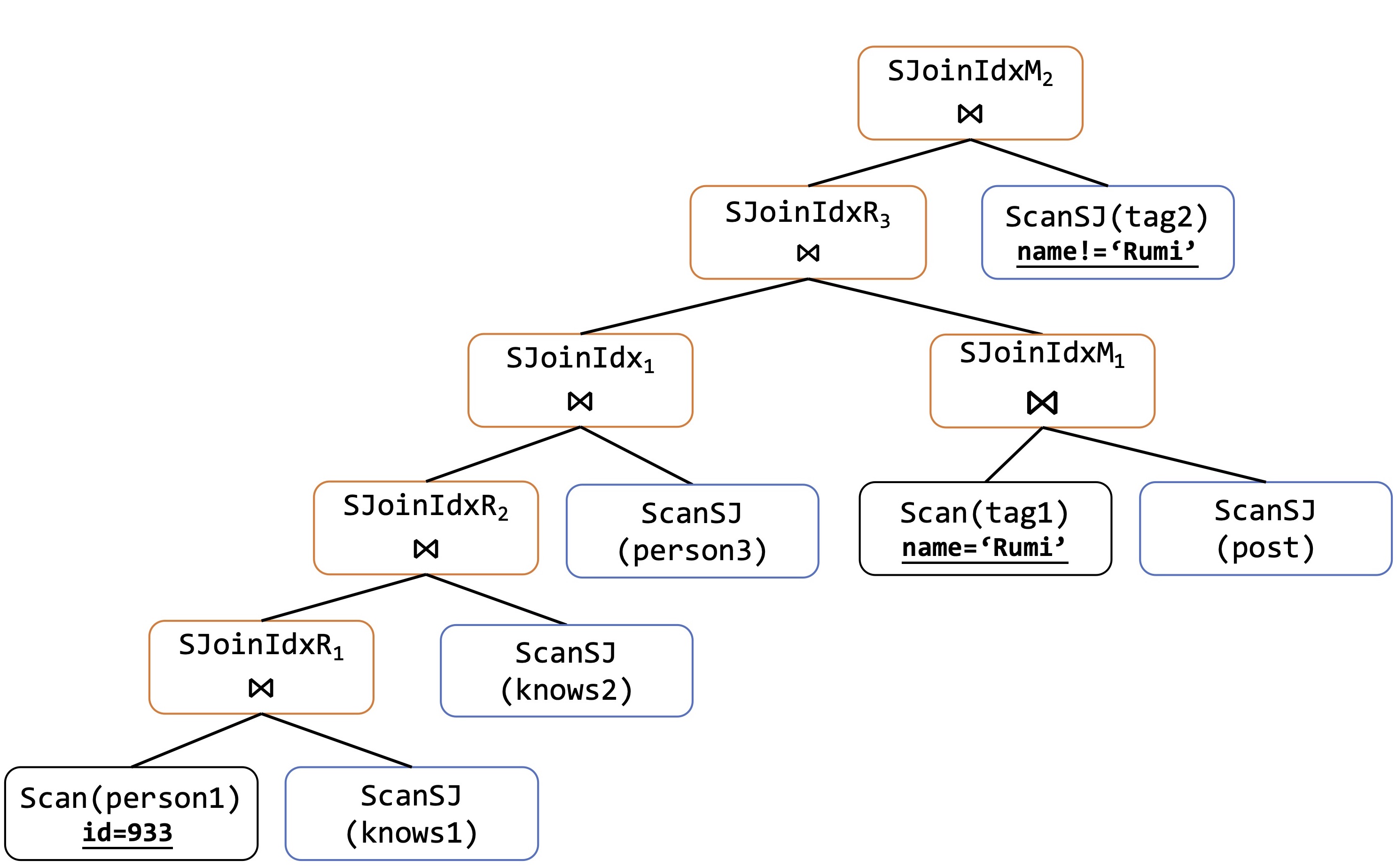}
    \caption{The Bushy Plan in GRainDB. The final Projection is omitted.}
    \vspace{12pt}
    \label{fig:plan-graindb}
\vspace{-10pt}
\end{figure}

Figure~\ref{fig:plan-gfdb} presents the left-deep plan in GraphflowDB. The \texttt{EXTEND} operator is the join operator in GraphflowDB, which takes as input k tuples and extends each tuple $t$ to one or more matches from adjacency list indexes. 
The join algorithm of \texttt{EXTEND} is essentially index nested-loop join.
In this left-deep plan, GraphflowDB starts with scanning and filtering $Person$ on $id=933$, and extends to $p2$ and $p3$ through $knows$ relationship, then extends to $post$ created by $p3$, and further extends to all tags of $post$ ($tag1$ and $tag2$) and apply filters on tag names.
Figure~\ref{fig:plan-graindb} demonstrates the bushy plan in GRainDB, in which hash joins are replaced by SJoin variants, and $SJoinIdxM_{1}$ and $SJoinIdxM_{2}$ merges two consecutive joins \texttt{tag JOIN post_tag} and \texttt{post JOIN post_tag} into one, respectively.
Compared to the left-deep plan, the bushy plan takes advantage of both selective filters $person1.id=933$ and $tag1.name='Rumi'$ to reduce intermediate result size.

\section{Query Execution Time of DuckDB and GRainDB on TPC-H}
\label{app:job-tpch-full}
In Section~\ref{subsubsec:evaluation-tpch} we presented box plots of DuckDB and GRainDB on TPC-H, and showed no significant speedups or slowdowns on this benchmark. In Table~\ref{tab:tpch-full}, we list detailed query execution time of DuckDB and GRainDB on all 22 queries on TPC-H. 
In all queries, the biggest slowdown is from 6863.8ms to 8217.8ms (0.8x) on Q7.
Though not expected, we still found performance speedups on Q2 and Q3.
On Q2, we reduce runtime from 1324.0ms to 519.0s (2.6x), and on Q3, from 2757.0ms to 1515.0ms.

\vspace{-5pt}
\section{Query Execution Time for Ablation Tests on SNB-M}
\label{app:snb-m-ablation}
Section\ref{subsubsec:evaluation-ablation} showed the settings and box plots of our ablation study. In Table~\ref{tab:snb-m-ablation}, we present detailed query execution time for each query in SNB-M under different system configurations, including GR-FULL, GR-JM, GR-JM-RSJ, and DuckDB (all optimizations turned off).

\section{Plan Spectrum}
\label{app:cumulative-distribution}
\begin{figure}[ht!]
  \centering
  \captionsetup{justification=centering}
    \begin{subfigure}[b]{0.20\textwidth}
    \centering
    \includegraphics[scale=0.25]{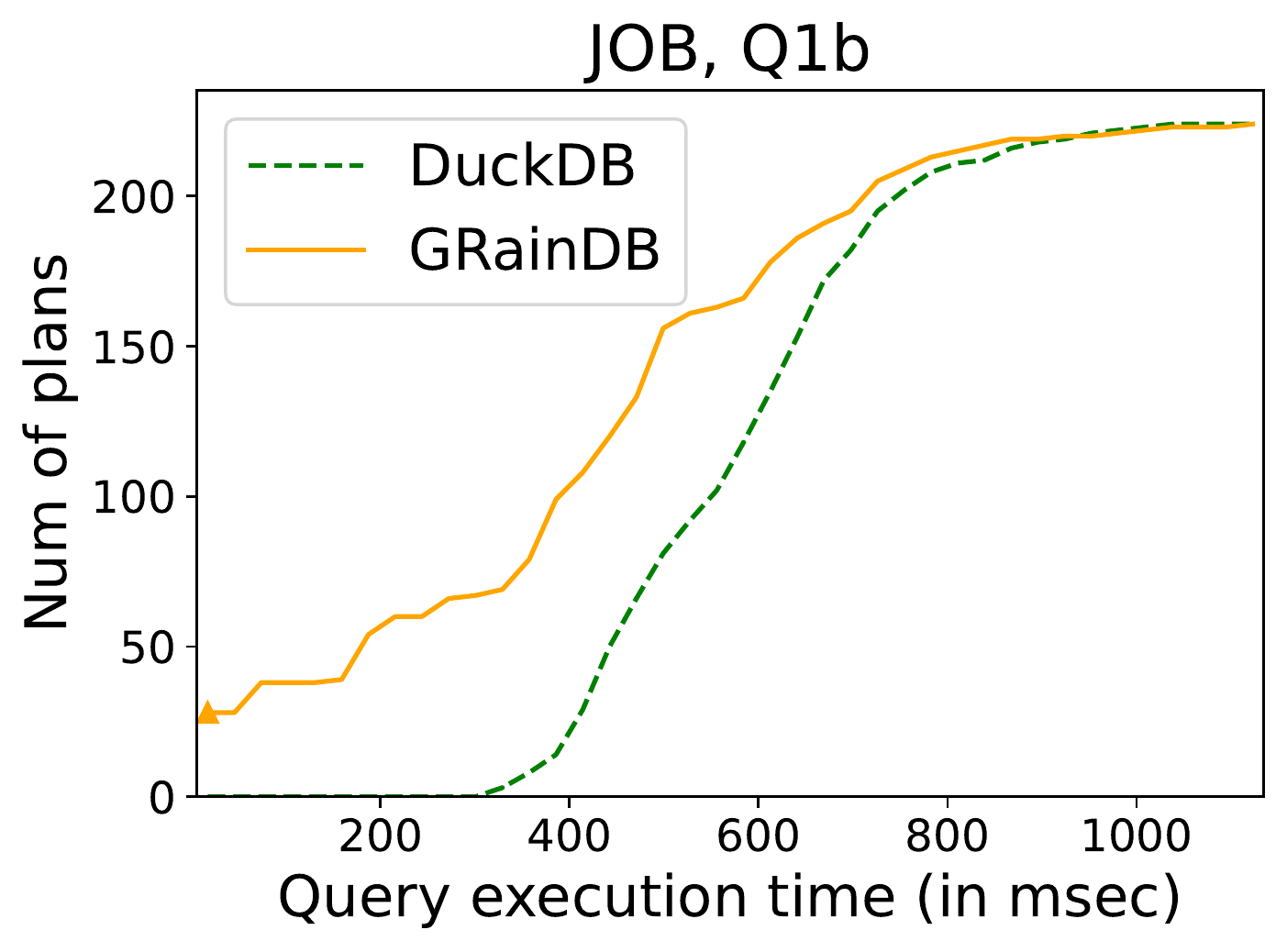}
    \vspace{-2.5mm}
    \label{fig:spectrum-job-q1b}
  \end{subfigure}
  \begin{subfigure}[b]{0.20\textwidth}
    \centering
    \includegraphics[scale=0.25]{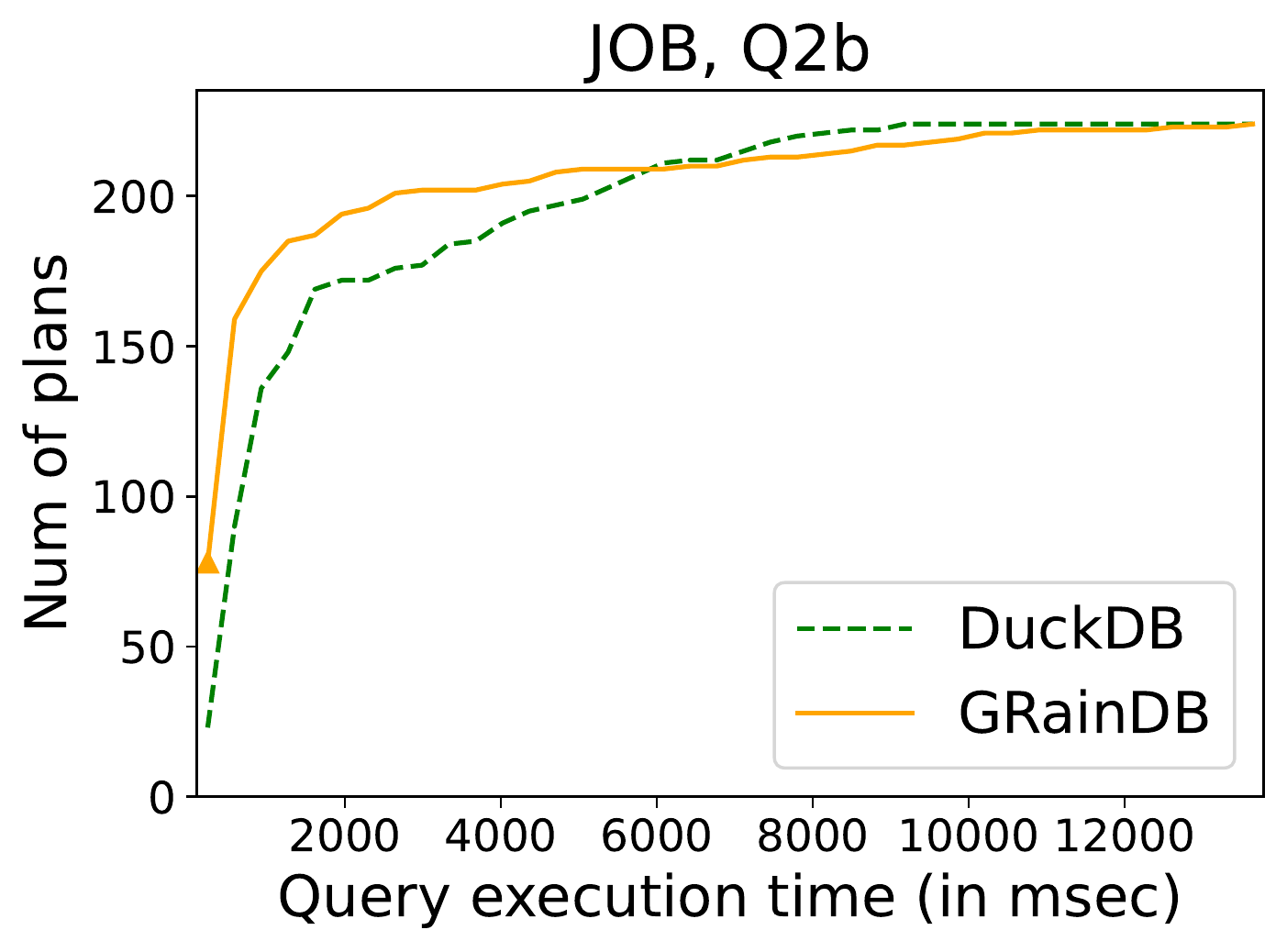}
    \vspace{-2.5mm}
    \label{fig:spectrum-job-q2b}
  \end{subfigure}
  \begin{subfigure}[b]{0.20\textwidth}
    \centering
    \includegraphics[scale=0.251]{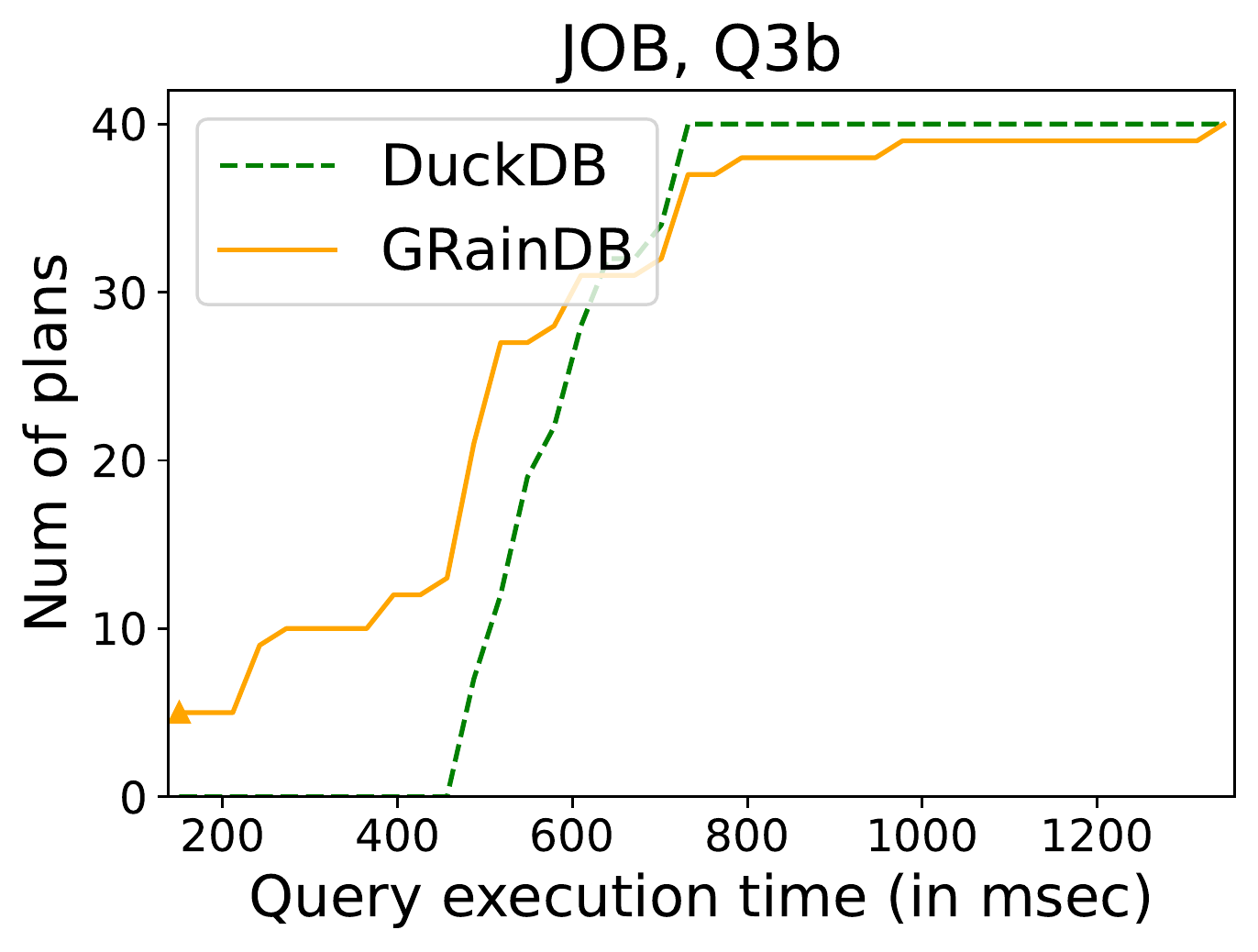}
    \vspace{-2.5mm}
    \label{fig:spectrum-job-q3b}
  \end{subfigure}
  \begin{subfigure}[b]{0.20\textwidth}
    \centering
    \includegraphics[scale=0.25]{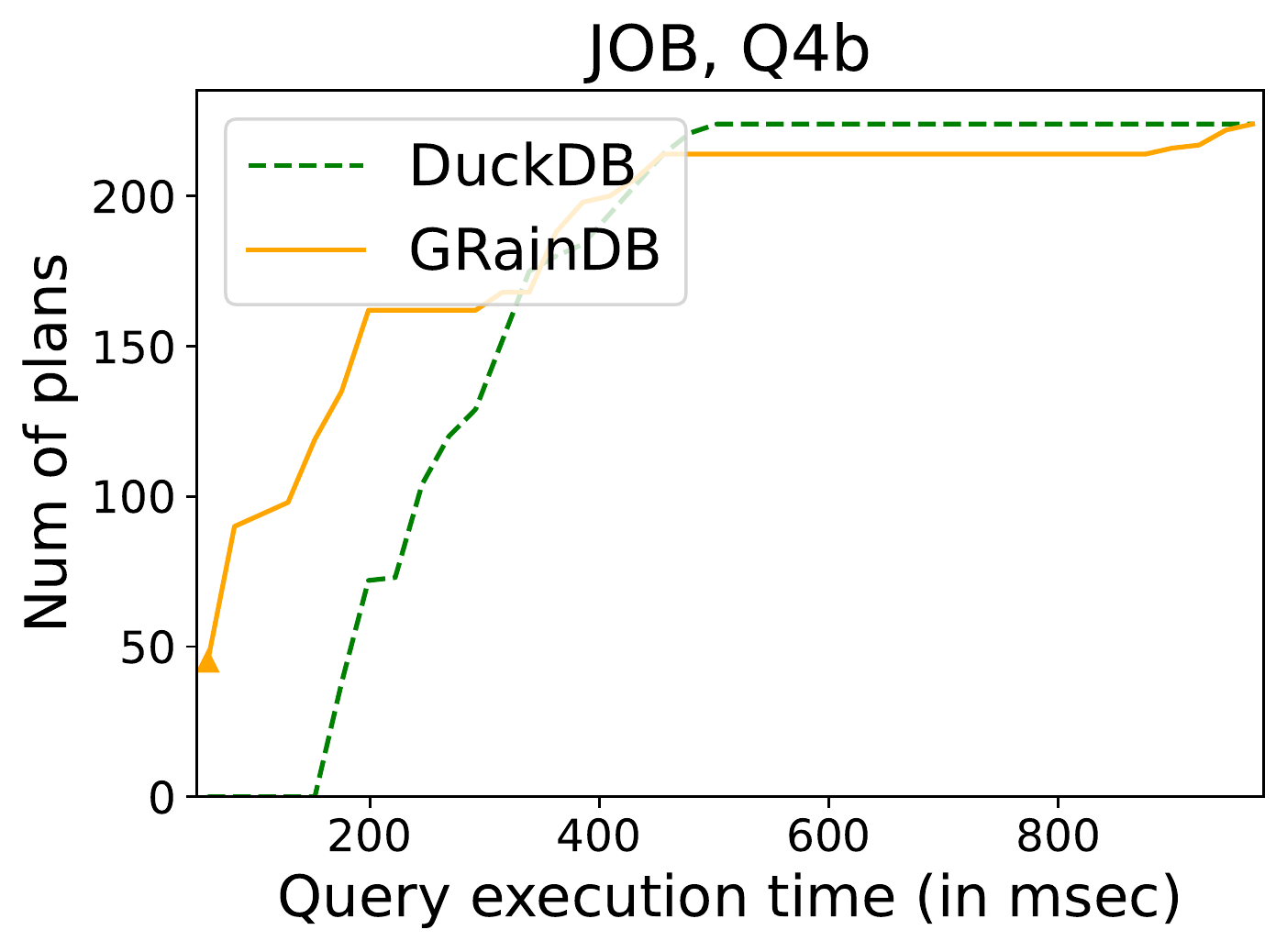}
    \vspace{-2.5mm}
    \label{fig:spectrum-job-q4b}
  \end{subfigure}
  \begin{subfigure}[b]{0.20\textwidth}
    \centering
    \includegraphics[scale=0.25]{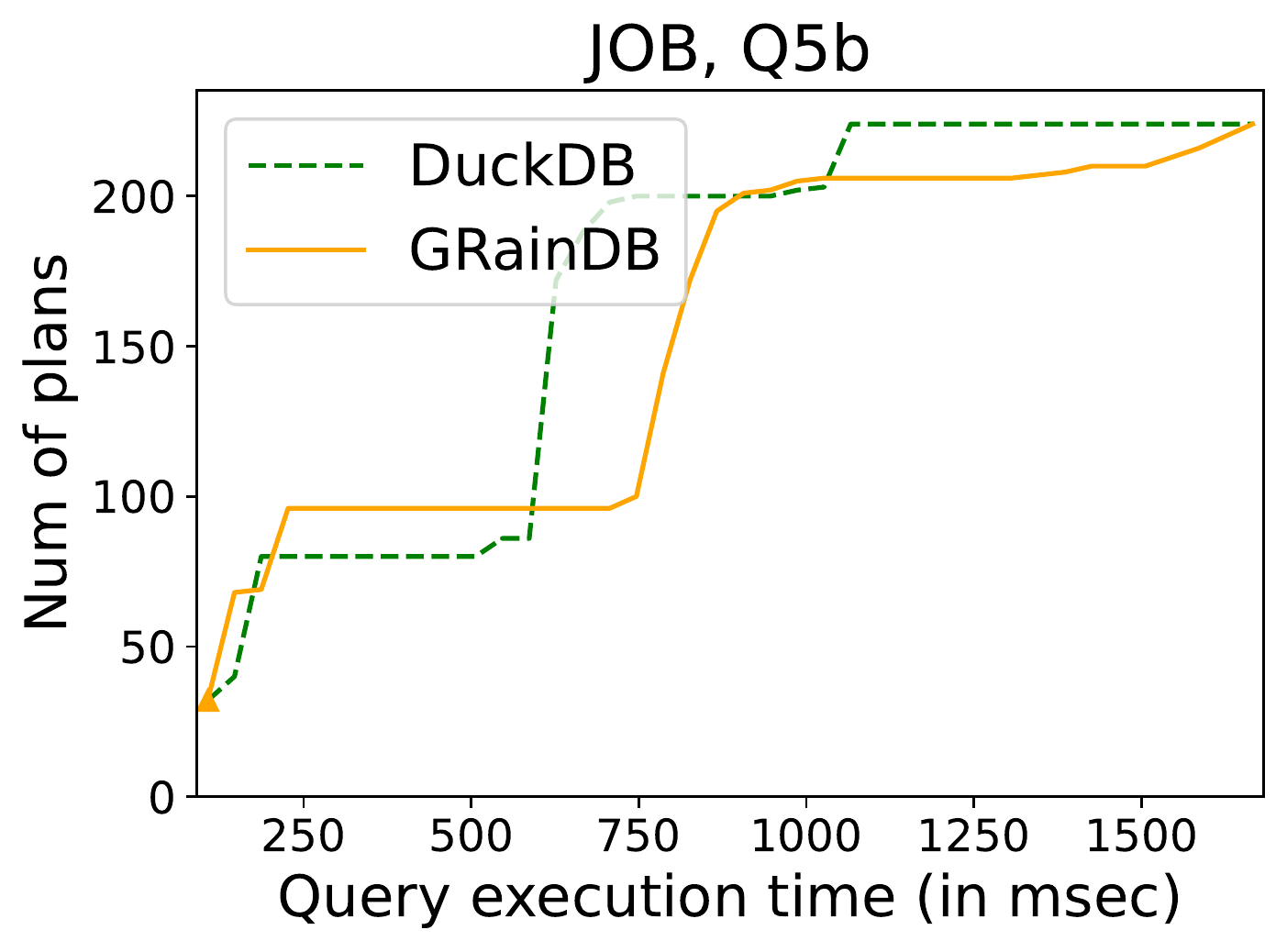}
    \vspace{-2.5mm}
    \label{fig:spectrum-job-q5b}
  \end{subfigure}
  \begin{subfigure}[b]{0.20\textwidth}
    \centering
    \includegraphics[scale=0.25]{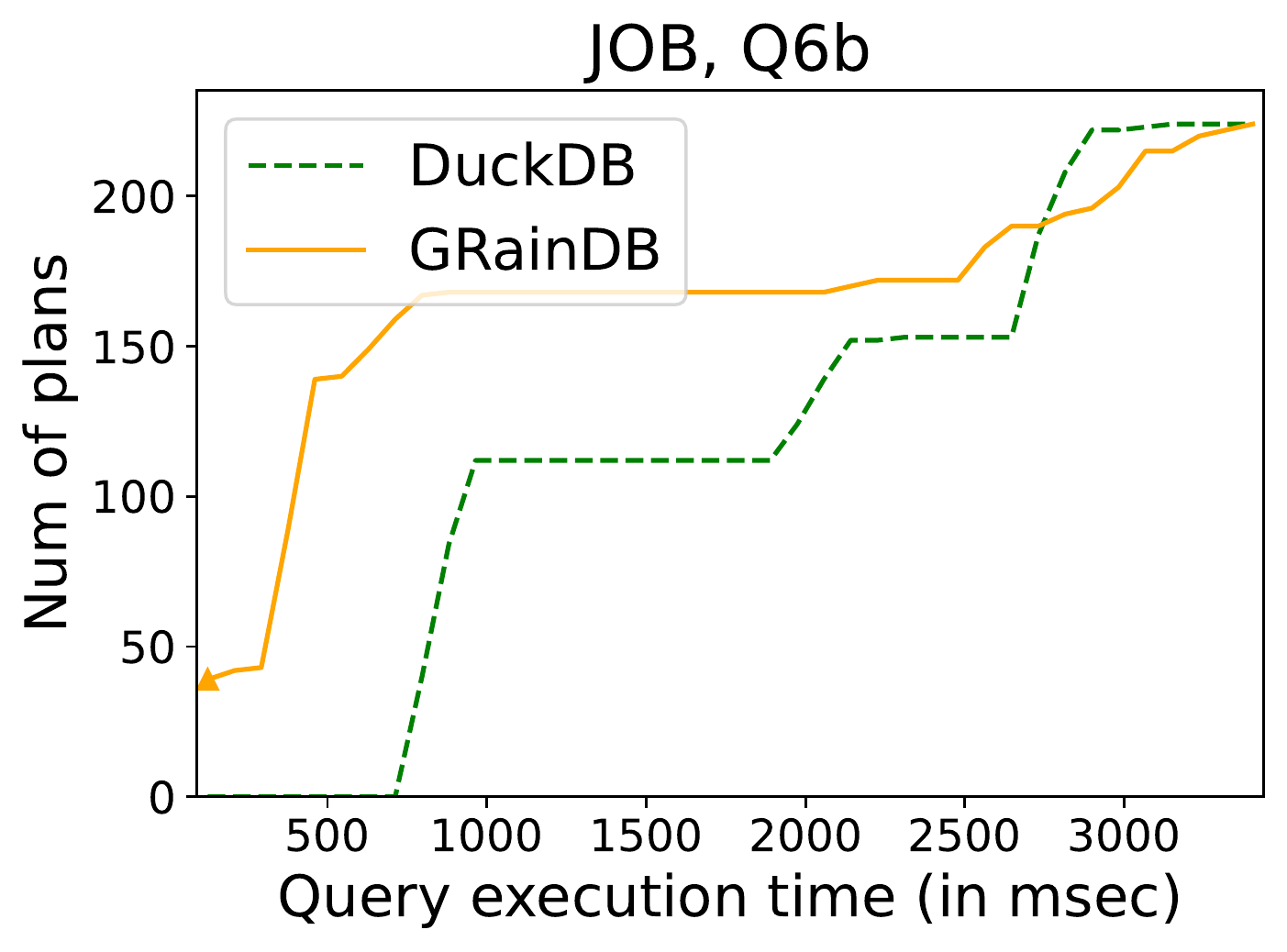}
    \vspace{-2.5mm}
    \label{fig:spectrum-job-q6b}
  \end{subfigure}
  \vspace{-12pt}
  \caption{Cumulative distributions of the number of DuckDB and GRainDB plans (y-axis) that have runtimes below different thresholds (x-axis) on JOB Q1b-Q6b.}
    \vspace{-22pt}
  \label{fig:job-spectrum}
\end{figure}

Figure~\ref{fig:job-spectrum} shows the result of plan spectrum analysis over the second variant of Q1-Q6 in JOB.
Dashed and straight lines in the figure are the distributions of $P_d$ and $P_d^*$ plans, respectively, which show the number of plans (on y-axis) for different runtime value cutoffs, e.g., 100ms, 200ms, ..., 1000ms (on x-axis).

\begin{landscape}
  \begin{table}
  \setlength{\tabcolsep}{4.0pt}
  \def\arraystretch{1.0}%
  \hspace{-1.2cm}
    \begin{tabular}{|c|c|c|c|c|c|c|c|c|c|c|c|c|c|c|c|c|c|c|c|}
    \hline
      & \textbf{Q1a} & \textbf{Q1b} & \textbf{Q1c} & \textbf{Q1d} & \textbf{Q2a} & \textbf{Q2b} & \textbf{Q2c} & \textbf{Q2d} & \textbf{Q3a} & \textbf{Q3b} & \textbf{Q3c} & \textbf{Q4a} & \textbf{Q4b} & \textbf{Q4c} & \textbf{Q5a} & \textbf{Q5b} & \textbf{Q5c} & \textbf{Q6a} & \textbf{Q6b} \\
      \hline 
      {\texttt{DuckDB}} & 234.2 & 331.8 & 196.8 & 327.6 & 207.0 & 208.2 & 156.0 & 239.0 & 1491.4 & 491.8 & 1551.8 & 216.0 & 193.8 & 216.8 & 96.0 & 177.0 & 1628.4 & 885.4 & 878.4 \\ 
      \hline 
      \multirow{2}{*}{\texttt{GRainDB}} & 34.2 & 3.0 & 13.8 & 6.0 & 154.0 & 143.0 & 17.2 & 227.2 & 328.0 & 502.0 & 473.6 & 114.2 & 73.6 & 144.4 & 116.4 & 146.0 & 425.4 & 57.0 & 97.0 \\ 
      & \textbf{6.8x} & \textbf{110.6x} & \textbf{14.3x} & \textbf{54.6x} & \textbf{1.3x} & \textbf{1.5x} & \textbf{9.1x} & \textbf{1.1x} & \textbf{4.5x} & \textbf{1.0x} & \textbf{3.3x} & \textbf{1.9x} & \textbf{2.6x} & \textbf{1.5x} & \textbf{0.8x} & \textbf{1.2x} & \textbf{3.8x} & \textbf{15.5x} & \textbf{9.1x} \\ 
      \hline 

      & \textbf{Q6c} & \textbf{Q6d} & \textbf{Q6e} & \textbf{Q6f} & \textbf{Q7a} & \textbf{Q7b} & \textbf{Q7c} & \textbf{Q8a} & \textbf{Q8b} & \textbf{Q8c} & \textbf{Q8d} & \textbf{Q9a} & \textbf{Q9b} & \textbf{Q9c} & \textbf{Q9d} & \textbf{Q10a} & \textbf{Q10b} & \textbf{Q10c} & \textbf{Q11a} \\
      \hline
      {\texttt{DuckDB}} & 857.2 & 863.4 & 849.4 & 1601.0 & 879.8 & 764.4 & 1383.6 & 1307.4 & 1267.4 & 1707.6 & 1097.6 & 1933.2 & 1538.4 & 1906.8 & 3538.4 & 1404.2 & 1248.4 & 1396.4 & 264.2 \\ 
      \hline 
      \multirow{2}{*}{\texttt{GRainDB}} & 416.8 & 430.6 & 419.0 & 1727.6 & 203.4 & 166.8 & 864.6 & 349.8 & 208.6 & 1944.6 & 973.2 & 554.8 & 494.2 & 881.4 & 3104.4 & 614.2 & 526.2 & 1436.8 & 44.4 \\ 
      & \textbf{2.1x} & \textbf{2.0x} & \textbf{2.0x} & \textbf{0.9x} & \textbf{4.3x} & \textbf{4.6x} & \textbf{1.6x} & \textbf{3.7x} & \textbf{6.1x} & \textbf{0.9x} & \textbf{1.1x} & \textbf{3.5x} & \textbf{3.1x} & \textbf{2.2x} & \textbf{1.1x} & \textbf{2.3x} & \textbf{2.4x} & \textbf{1.0x} & \textbf{6.0x} \\ 
      \hline 

      & \textbf{Q11b} & \textbf{Q11c} & \textbf{Q11d} & \textbf{Q12a} & \textbf{Q12b} & \textbf{Q12c} & \textbf{Q13a} & \textbf{Q13b} & \textbf{Q13c} & \textbf{Q13d} & \textbf{Q14a} & \textbf{Q14b} & \textbf{Q14c} & \textbf{Q15a} & \textbf{Q15b} & \textbf{Q15c} & \textbf{Q15d} & \textbf{Q16a} & \textbf{Q16b} \\
      {\texttt{DuckDB}} & 220.0 & 307.2 & 307.6 & 684.0 & 643.2 & 859.0 & 981.2 & 398.2 & 326.8 & 679.8 & 1644.2 & 1683.4 & 1695.2 & 987.8 & 984.4 & 903.0 & 612.4 & 652.4 & 2316.0 \\ 
      \hline 
      \multirow{2}{*}{\texttt{GRainDB}} & 97.8 & 254.4 & 142.6 & 242.4 & 16.0 & 415.6 & 336.8 & 208.0 & 145.6 & 614.6 & 202.2 & 158.8 & 270.8 & 283.4 & 87.6 & 317.4 & 309.0 & 177.6 & 2025.6 \\ 
      & \textbf{2.2x} & \textbf{1.2x} & \textbf{2.2x} & \textbf{2.8x} & \textbf{40.2x} & \textbf{2.1x} & \textbf{2.9x} & \textbf{1.9x} & \textbf{2.2x} & \textbf{1.1x} & \textbf{8.1x} & \textbf{10.6x} & \textbf{6.3x} & \textbf{3.5x} & \textbf{11.2x} & \textbf{2.8x} & \textbf{2.0x} & \textbf{3.7x} & \textbf{1.1x} \\ 
      \hline 

      & \textbf{Q16c} & \textbf{Q16d} & \textbf{Q17a} & \textbf{Q17b} & \textbf{Q17c} & \textbf{Q17d} & \textbf{Q17e} & \textbf{Q17f} & \textbf{Q18a} & \textbf{Q18b} & \textbf{Q18c} & \textbf{Q19a} & \textbf{Q19b} & \textbf{Q19c} & \textbf{Q19d} & \textbf{Q20a} & \textbf{Q20b} & \textbf{Q20c} & \textbf{Q21a} \\
      {\texttt{DuckDB}} & 1215.4 & 1110.0 & 1164.8 & 704.4 & 595.0 & 726.4 & 2134.2 & 1400.4 & 1797.0 & 2433.0 & 3565.8 & 2632.4 & 2395.4 & 2913.8 & 2979.0 & 1118.4 & 1399.2 & 1102.0 & 1629.2 \\ 
      \hline 
      \multirow{2}{*}{\texttt{GRainDB}} & 1378.8 & 1026.8 & 486.0 & 309.2 & 206.4 & 325.6 & 2921.4 & 978.0 & 612.6 & 525.2 & 1952.2 & 419.6 & 226.0 & 857.0 & 2159.4 & 1071.6 & 993.6 & 1065.0 & 54.2 \\ 
      & \textbf{0.9x} & \textbf{1.1x} & \textbf{2.4x} & \textbf{2.3x} & \textbf{2.9x} & \textbf{2.2x} & \textbf{0.7x} & \textbf{1.4x} & \textbf{2.9x} & \textbf{4.6x} & \textbf{1.8x} & \textbf{5.4x} & \textbf{10.6x} & \textbf{3.4x} & \textbf{1.4x} & \textbf{1.0x} & \textbf{1.4x} & \textbf{1.0x} & \textbf{30.1x} \\ 
      \hline 

      & \textbf{Q21b} & \textbf{Q21c} & \textbf{Q22a} & \textbf{Q22b} & \textbf{Q22c} & \textbf{Q22d} & \textbf{Q23a} & \textbf{Q23b} & \textbf{Q23c} & \textbf{Q24a} & \textbf{Q24b} & \textbf{Q25a} & \textbf{Q25b} & \textbf{Q25c} & \textbf{Q26a} & \textbf{Q26b} & \textbf{Q26c} & \textbf{Q27a} & \textbf{Q27b} \\
      {\texttt{DuckDB}} & 718.6 & 1653.0 & 1471.8 & 1425.0 & 1881.8 & 1710.2 & 866.6 & 838.6 & 859.6 & 2554.8 & 2551.4 & 2318.8 & 2138.0 & 3584.6 & 1074.6 & 816.4 & 1142.6 & 761.4 & 717.6 \\ 
      \hline 
      \multirow{2}{*}{\texttt{GRainDB}} & 52.2 & 49.2 & 864.0 & 433.0 & 387.0 & 412.6 & 296.8 & 100.6 & 280.0 & 788.8 & 297.4 & 1376.6 & 176.4 & 1834.4 & 733.4 & 424.0 & 1145.0 & 44.4 & 36.6 \\ 
      & \textbf{13.8x} & \textbf{33.6x} & \textbf{1.7x} & \textbf{3.3x} & \textbf{4.9x} & \textbf{4.1x} & \textbf{2.9x} & \textbf{8.3x} & \textbf{3.1x} & \textbf{3.2x} & \textbf{8.6x} & \textbf{1.7x} & \textbf{12.1x} & \textbf{2.0x} & \textbf{1.5x} & \textbf{1.9x} & \textbf{1.0x} & \textbf{17.1x} & \textbf{19.6x} \\ 
      \hline 

      & \textbf{Q27c} & \textbf{Q28a} & \textbf{Q28b} & \textbf{Q28c} & \textbf{Q29a} & \textbf{Q29b} & \textbf{Q29c} & \textbf{Q30a} & \textbf{Q30b} & \textbf{Q30c} & \textbf{Q31a} & \textbf{Q31b} & \textbf{Q31c} & \textbf{Q32a} & \textbf{Q32b} & \textbf{Q32c} & \textbf{Q33a} & \textbf{Q33b} & \\
      {\texttt{DuckDB}} & 1595.4 & 2068.4 & 993.0 & 1882.8 & 2742.6 & 2205.6 & 2506.0 & 2198.0 & 2287.6 & 3189.0 & 2523.8 & 2432.6 & 3457.8 & 126.0 & 275.2 & 350.0 & 336.2 & 574.6 & \\ 
      \hline 
      \multirow{2}{*}{\texttt{GRainDB}} & 46.6 & 240.2 & 199.4 & 264.6 & 266.6 & 227.0 & 259.0 & 673.0 & 233.8 & 1501.2 & 612.0 & 296.0 & 817.8 & 8.2 & 173.6 & 178.6 & 149.4 & 223.8 & \\ 
      & \textbf{34.2x} & \textbf{8.6x} & \textbf{5.0x} & \textbf{7.1x} & \textbf{10.3x} & \textbf{9.7x} & \textbf{9.7x} & \textbf{3.3x} & \textbf{9.8x} & \textbf{2.1x} & \textbf{4.1x} & \textbf{8.2x} & \textbf{4.2x} & \textbf{15.4x} & \textbf{1.6x} & \textbf{2.0x} & \textbf{2.3x} & \textbf{2.6x} & \\ 
      \hline 
    \end{tabular}
  \vspace{10pt}
  \captionsetup{justification=centering}
  \caption{Runtime (in ms) of DuckDB and GRainDB on all 113 queries in JOB.}
  \label{tab:job-full}
  \end{table}

  \begin{table}
  \bgroup
  \setlength{\tabcolsep}{2.5pt}
  \def\arraystretch{1.0}%
  \hspace{-1.0cm}
    \begin{tabular}{ |c|c|c|c|c|c|c|c|c|c|c|c|c|c|c|c|c|c|c|c|c|c|c|c|c|c| }
      \hline
      & \textbf{Q1} & \textbf{Q2} & \textbf{Q3} & \textbf{Q4} & \textbf{Q5} & \textbf{Q6} & \textbf{Q7} & \textbf{Q8} & \textbf{Q9} & \textbf{Q10} & \textbf{Q11} & \textbf{Q12} & \textbf{Q13} & \textbf{Q14} & \textbf{Q15} & \textbf{Q16} & \textbf{Q17} & \textbf{Q18} & \textbf{Q19} & \textbf{Q20} & \textbf{Q21} & \textbf{Q22} \\ 
      \hline       
      {\texttt{DuckDB}} & 8830.2 & 1324.0 & 2757.0 & 2817.2 & 3616.4 & 742.8 & 6863.8 & 2347.5 & 8633.0 & 6531.8 & 386.3 & 2639.8 & 6365.0 & 836.5 & 1856.6 & 1624.8 & 4684.8 & 13063.0 & 7279.6 & 1802.0 & 11446.6 & 1265.8 \\ 
      \hline  
      \multirow{2}{*}{\texttt{GRainDB}}& 9450.6 & 519.0 & 1515.0 & 2830.8 & 2881.8 & 683.8 & 8217.8 & 1598.5 & 7492.5 & 4158.0 & 344.0 & 3247.8 & 6736.8 & 867.0 & 1832.8 & 1798.4 & 4219.4 & 13296.4 & 8418.0 & 1817.2 & 10236.8 & 1233.4 \\ 
      & \textbf{0.9x} & \textbf{2.6x} & \textbf{1.8x} & \textbf{1.0x} & \textbf{1.3x} & \textbf{1.1x} & \textbf{0.8x} & \textbf{1.5x} & \textbf{1.2x} & \textbf{1.6x} & \textbf{1.1x} & \textbf{0.8x} & \textbf{0.9x} & \textbf{1.0x} & \textbf{1.0x} & \textbf{0.9x} & \textbf{1.1x} & \textbf{1.0x} & \textbf{0.9x} & \textbf{1.0x} & \textbf{1.1x} & \textbf{1.0x} \\ 
      \hline
  \end{tabular}\vspace{0.2 cm}
  \egroup
  \captionsetup{justification=centering}
  \caption{Runtime (in ms) of DuckDB and GRainDB on all 22 queries in TPC-H.}
  \label{tab:tpch-full}
  \vspace{-5px}
  \end{table}

  \begin{table}
  \bgroup
  \setlength{\tabcolsep}{1.0pt}
  \def\arraystretch{0.8}%
  \hspace{-1.2cm}
    \begin{tabular}{ |c|c|c|c|c|c|c|c|c|c|c|c|c|c|c|c|c|c|c|c|c|c|c|c|c|c|c|c| }
      \hline
      & \textbf{IS1} & \textbf{IS2} & \textbf{IS3} & \textbf{IS4} & \textbf{IS5} & \textbf{IS6} & \textbf{IS7} & \textbf{IC1-1} & \textbf{IC1-2} & \textbf{IC1-3} & \textbf{IC2} & \textbf{IC3-1} & \textbf{IC3-2} & \textbf{IC4} & \textbf{IC5-1} & \textbf{IC5-2} & \textbf{IC6-1} & \textbf{IC6-2} & \textbf{IC7} & \textbf{IC8} & \textbf{IC9-1} & \textbf{IC9-2} & \textbf{IC11-1} & \textbf{IC11-2} & \textbf{IC12} \\ 
      \hline      
      {\texttt{DuckDB}} & 0.8 & 524.8 & 36.6 & 0.2 & 4.5 & 148.0 & 989.0 & 38.0 & 72.0 & 110.5 & 926.0 & 1177.8 & 4647.0 & 402.0 & 636.0 & 3125.0 & 244.6 & 471.2 & 1186.8 & 1017.0 & 441.8 & 1312.6 & 35.8 & 68.4 & 788.4 \\ 
      \hline  
      \multirow{2}{*}{\texttt{GRainDB}}& 1.2 & 19.6 & 3.4 & 0.2 & 0.6 & 5.0 & 11.0 & 4.0 & 6.4 & 38.2 & 134.8 & 119.4 & 1665.0 & 54.0 & 174.0 & 2768.0 & 13.0 & 22.0 & 33.2 & 14.0 & 113.6 & 752.0 & 2.8 & 9.0 & 234.8 \\ 
      & \textbf{0.7x} & \textbf{26.8x} & \textbf{10.8x} & \textbf{1.0x} & \textbf{7.5x} & \textbf{29.6x} & \textbf{90.0x} & \textbf{9.5x} & \textbf{11.2x} & \textbf{2.9x} & \textbf{6.9x} & \textbf{9.9x} & \textbf{2.8x} & \textbf{7.4x} & \textbf{3.7x} & \textbf{1.1x} & \textbf{18.8x} & \textbf{21.4x} & \textbf{35.7x} & \textbf{72.6x} & \textbf{3.9x} & \textbf{1.8x} & \textbf{12.8x} & \textbf{7.6x} & \textbf{3.4x} \\ 
      \hline
      \multirow{2}{*}{\texttt{GRainDB-JM}} & 1.1 & 19 & 3.2 & 0.2 & 0.6 & 5.4 & 10.0 & 6.0 & 25.0 & 68.8 & 349.2 & 119.0 & 1706.3 & 54.0 & 198.2 & 3082.2 & 13.6 & 61.6 & 29.0 & 15.4 & 113.6 & 686.0 & 3.8 & 38.6 & 253.0 \\ 
      & \textbf{0.7x} & \textbf{27.6x} & \textbf{11.4x} & \textbf{1.0x} & \textbf{7.5x} & \textbf{27.4x} & \textbf{98.9x} & \textbf{6.3x} & \textbf{2.9x} & \textbf{1.6x} & \textbf{2.7x} & \textbf{9.9x} & \textbf{2.7x} & \textbf{7.4x} & \textbf{3.2x} & \textbf{1.0x} & \textbf{18.0x} & \textbf{7.6x} & \textbf{40.9x} & \textbf{66.0x} & \textbf{3.9x} & \textbf{1.9x} & \textbf{9.4x} & \textbf{1.8x} & \textbf{3.1x} \\ 
      \hline  
      \multirow{2}{*}{\texttt{GRainDB-JM-RSJ}} & 0.8 & 407.8 & 34.6 & 0.2 & 0.6 & 5.0 & 1020.8 & 33.2 & 62.8 & 107.0 & 1004.0 & 1195.0 & 2122.4 & 345.0 & 620.0 & 3195.2 & 244.2 & 299.2 & 1185.8 & 1016.0 & 430.2 & 804.4 & 34.2 & 68.0 & 640.2 \\ 
      & \textbf{1.0x} & \textbf{1.3x} & \textbf{1.1x} & \textbf{1.0x} & \textbf{7.5x} & \textbf{29.6x} & \textbf{1.0x} & \textbf{1.1x} & \textbf{1.1x} & \textbf{1.0x} & \textbf{0.9x} & \textbf{1.0x} & \textbf{2.2x} & \textbf{1.2x} & \textbf{1.0x} & \textbf{1.0x} & \textbf{1.0x} & \textbf{1.6x} & \textbf{1.0x} & \textbf{1.0x} & \textbf{1.0x} & \textbf{1.6x} & \textbf{1.0x} & \textbf{1.0x} & \textbf{1.2x} \\ 
      \hline  
    \end{tabular}\vspace{0.2 cm}
    \newline
  \egroup
  \captionsetup{justification=centering}
  \caption{Runtime (in ms) of GRainDB on each query in SNB-M under different optimizations. DuckDB implements no optimizations. GRainDB-JM-RSJ only implements RID materialization. GRainDB-JM in addition implements reverse semijoins. GRainDB in addition implements join merging.}
  \label{tab:snb-m-ablation}
  \vspace{-5px}
  \end{table}
\end{landscape}

\end{appendix}